\documentclass[twocolumn,aps,prb,reprint,superscriptaddress,longbibliography,nofootinbib]{revtex4-2}
\usepackage{graphicx}
\usepackage{multirow}
\usepackage{xr-hyper}
\usepackage{hyperref}
\usepackage{dsfont} 
\usepackage{soul}
\hypersetup{
     colorlinks   = true,
     citecolor    = blue
}
\usepackage{physics}

\usepackage{blkarray}
\usepackage[dvipsnames]{xcolor}
\usepackage[utf8]{inputenc}
\usepackage[vietnamese, english]{babel}
\usepackage{chngcntr}
\usepackage{xcolor}
\usepackage{braket}
\usepackage{sidecap}
\usepackage{lipsum}
\usepackage{bbold}
\usepackage{array}
\usepackage{comment}

\begin{document}

\title{Band Renormalization, Quarter Metals, and Chiral Superconductivity \\ in Rhombohedral Tetralayer Graphene}

\author{Guillermo Parra-Mart\'inez}
\thanks{These authors contributed equally.}
\affiliation{IMDEA Nanoscience, C/ Faraday 9, 28049 Madrid, Spain}
\author{Alejandro Jimeno-Pozo}
\thanks{These authors contributed equally.}
\affiliation{IMDEA Nanoscience, C/ Faraday 9, 28049 Madrid, Spain}
\author{\foreignlanguage{vietnamese}{Võ Tiến Phong}}
\affiliation{Department of Physics, Florida State University, Tallahassee, FL, 32306, U.S.A.}
\affiliation{National High Magnetic Field Laboratory, Tallahassee, FL, 32310, U.S.A}
\author{H\'ector Sainz-Cruz}
\affiliation{IMDEA Nanoscience, C/ Faraday 9, 28049 Madrid, Spain}
\author{Daniel Kaplan}
\affiliation{Center for Materials Theory, Department of Physics and Astronomy,
Rutgers University, 136 Frelinghuysen Rd., Piscataway, NJ 08854, USA}
\author{Peleg Emanuel}
\affiliation{Department of Condensed Matter Physics, Weizmann Institute of Science, Rehovot 7610001, Israel}
\author{Yuval Oreg}
\affiliation{Department of Condensed Matter Physics, Weizmann Institute of Science, Rehovot 7610001, Israel}
\author{Pierre A. Pantale\'on}
\email{pierre.pantaleon@imdea.org}
\affiliation{IMDEA Nanoscience, C/ Faraday 9, 28049 Madrid, Spain}
\author{Jos\'e \'Angel Silva-Guill\'en}
\email{joseangel.silva@imdea.org}
\affiliation{IMDEA Nanoscience, C/ Faraday 9, 28049 Madrid, Spain}
\author{Francisco Guinea}
\affiliation{IMDEA Nanoscience, C/ Faraday 9, 28049 Madrid, Spain}
\affiliation{Donostia International Physics Center, Paseo Manuel de Lardiz\'{a}bal 4, 20018 San Sebastián, Spain}

\begin{abstract}
Recently, exotic superconductivity emerging from a spin-and-valley-polarized metallic phase has been discovered in rhombohedral tetralayer graphene. To explain this observation, we study the role of electron-electron interactions in driving flavor symmetry breaking, using the Hartree-Fock (HF) approximation, and in stabilizing superconductivity mediated by repulsive interactions. Though mean-field HF correctly predicts the isospin flavors and reproduces the experimental phase diagram, it overestimates the band renormalization near the Fermi energy and suppresses superconducting instabilities. To address this, we introduce a physically motivated scheme that includes internal screening in the HF calculation. Using this formalism, we find superconductivity arising from the spin-valley polarized phase for a range of electric fields and electron dopings.  Our findings reproduce the experimental observations and reveal a $p$-wave, finite-momentum, time-reversal-symmetry-broken superconducting state, encouraging further investigation into exotic phases in graphene multilayers.
\end{abstract}

\maketitle

{\it Introduction \textendash} Narrow-band multilayer graphene systems, both with and without moir\'{e} structures, host an abundance of sought-after strongly-correlated states such as integer and fractional quantum anomalous Hall effects and unconventional superconductivity~\cite{Cetal18b, Yankowitz2019, Lu2019_SC_TBG, Stepanov2020_TBG, Oh2021UnconvSC,Park2021_SC_TTG, Hao2021_SC_TTG,Kim2022evidence,Liu2022Isospin,Uri2023Superconductivity,Park2022_multi, Zhang2022_promotionSC,Su2023Superconductivity, Zhou2021SuperRTG, patterson_superconductivity_2024,zhou2022isospin, zhang2023spin, holleis2023SC,han_signatures_2024,choi_electric_2024,lu2024fractional}. These materials are appealing both from a theoretical perspective, due to the richness of their phase diagrams, and from an experimental one, due to their \textit{in situ} tunability by applying displacement fields and varying the carrier density~\cite{Andrei2020Graphene,Balents2020Superconductivity}. 
Recently, superconductivity has been observed in rhombohedral tetralayer graphene (RTLG) both upon electron and hole doping~\cite{han_signatures_2024,choi_electric_2024}.
Interestingly, on the electron side ~\cite{han_signatures_2024}, there are strong indications that the highest critical temperature pocket ($T_\mathrm{c}\sim200$~mK) emerges from a quarter-metal phase, where both spins and valleys are polarized. 
This suggests a highly exotic, possibly chiral, superconducting state that breaks time-reversal symmetry and possesses a finite center-of-mass momentum order parameter.

Prior to the experiment reported in Ref. \cite{han_signatures_2024}, superconductivity has already been observed in rhombohedral multilayer graphene under different experimental conditions \cite{Zhou2021SuperRTG,zhou2022isospin,zhang2023spin, holleis2023SC, patterson_superconductivity_2024}. These observations consequently inspired a wealth of theoretical studies exploring the rich landscape of normal and superconducting phases in Bernal and rhombohedral graphene~\cite{ghazaryan_unconventional_2021, you2022kohn, chatterjee_inter-valley_2022, cea2022superconductivity, Cea2023Superconductivity,JimenoPozo2023, ZiyanLi2023, Dong2023Multilayer, qin_functional_2023, shavit_inducing_2023, dong_superconductivity_2024, long_evolution_2024,  vinas_bostrom_phonon-mediated_2024, fischer_spin_2024, vituri2024incommensurateintervalleycoherentstates, Braz2024_magneticABC}. In particular, many of these studies suggest that superconductivity can arise from purely repulsive long-range Coulomb interactions that are screened within the Random Phase Approximation (RPA). However, in  these previous theoretical and experimental studies, superconductivity emerging from a fully-polarized quarter metal has not been considered. As we shall demonstrate in this work, properly analyzing this novel finite-momentum superconducting phase requires careful accounting for electron-electron renormalization of the band structure, which, in turn, requires the employment of appropriate theoretical techniques beyond standard mean-field Hartree-Fock theory often used in previous works.

Superconductivity is exponentially sensitive to the topology and density of states (DOS) at the Fermi level, both of which are strongly affected by electron-electron interactions. On one hand, trigonal warping, the imbalance between states at $\mathbf{k}$ and $-\mathbf{k}$ in a single valley, can be detrimental to Cooper pairing. On the other, a large DOS is required to reproduce critical temperatures in the range experimentally measured. For accurate modeling, these two competing factors must be simultaneously considered. A straightforward application of Hartree-Fock theory does \textit{not} work since it suppresses the DOS at the Fermi energy~\cite{monkhorst_hartree-fock_1979}, eliminating the Cooper instability entirely. While this effect arises from the divergence of the Coulomb interaction at $q = 0$, even a double-gated Coulomb interaction, which is regular at $q = 0$, can overestimate the reduction in DOS, potentially leading to unrealistically low values of $T_c$. To mitigate this, internal screening in the Coulomb potential must be taken into account.

In this Letter, we develop a new method to address the challenges aforementioned by accounting for the effects of band renormalization both with gate screening and with internal screening to capture the correct Fermi surface topology and DOS; we name this method screened Hartree-Fock (sHF). Once properly renormalized, the electronic bands are then used to calculate superconducting properties. For the normal states, we perform  HF calculations to obtain a phase diagram at high displacement fields. Our results are in good agreement with experimental findings~\cite{han_signatures_2024}, and we verify that the normal state surrounding SC1 of~\cite{han_signatures_2024} is indeed a quarter metal. Finally, we use our new approach to obtain the superconducting critical temperature and order parameter. We find that superconductivity is strongly influenced by residual trigonal warping, with a leading $p$-wave order parameter emerging over a range of electron doping and displacement fields that align with experimental observations, as shown in Fig.~\ref{fig:Figura2}. Furthermore, the critical temperatures obtained with this approach match the experimental results in Ref.~\cite{han_signatures_2024}. This contrasts with the significantly different values found when interactions are not included. The $p$-wave order parameter that we obtain is in general agreement with other recent theoretical works on tetralayer graphene~\cite{Chou_Intravalley_tetralayer_2024, yang_topological_2024, qin_chiral_2024,Geier_isospin_tetralayer_2024,Yoon2025Quarter,jahin_enhanced_2024,wang_chiral_2024,Li2024_spontaneous}. However, these other studies have not fully considered the effects of Coulomb renormalization of electronic bands.  In this work, we emphasize that band renormalization cannot be neglected since it is crucial to the existence, or lack thereof, of superconductivity in this system.

{\it Screened Hartree Fock Scheme \textendash} Hartree-Fock theory is a variational method that minimizes total energy. Hence, it is a standard tool for determining the normal ground state energy, using which one can extract the normal-state phase diagram. However, as mentioned before, it suffers a known deficiency of overestimating the reduction in DOS at the Fermi energy, resulting in an unrealistic suppression of $T_\mathrm{c}$ (see Secs. 4 and 5 of Supplementary Information (SI) \cite{SM}). 
In this section, we formulate a method that overcomes this shortcoming in order to correctly describe both normal and superconducting states measured in the experiment of Ref.~\cite{han_signatures_2024}.
Our approach consists of two interdependent calculations: (1) for the normal-state phase diagram, we use standard Hartree-Fock theory to determine phase boundaries\footnote{This rests on the assumption that although standard Hartree-Fock theory may not get the correct DOS at the Fermi level, it is sufficient for the comparison of relative total energies of various competing ground state candidates. Therefore, when determining the phase diagram of the normal states, we continue to use standard HF. It is only when the wavefunctions are needed to compute superconducting properties that those obtained from standard HF no longer suffice. In those cases, we turn to our new formalism of including internal screening in HF theory.}, and (2) in regions where superconductivity is expected to emerge, we implement our screened Hartree-Fock theory to obtain the wavefunctions using which we determine superconducting properties.

We start by briefly describing our calculations. More detailed information about the method is included in the SI section. After building a continuum model for RTLG (see Sec. I of SI), we can write the two-particle Coulomb interaction as 
\begin{equation}
\label{eq:four-fermion Coulomb}
    \hat{\mathcal{V}}_\mathcal{C} = \frac{1}{2\Omega}  \sum_{\substack{\mathbf{k}, \mathbf{p}, \mathbf{q} , \alpha, \alpha'}} \mathcal{V}_0(\mathbf{q}) \hat{c}^\dagger_{\alpha, \mathbf{k}  + \mathbf{q} }\hat{c}^\dagger_{\alpha', \mathbf{p} - \mathbf{q}}\hat{c}_{\alpha', \mathbf{p}}\hat{c}_{\alpha, \mathbf{k} },
\end{equation}
where $\hat{c}$ and $\hat{c}^\dagger$ denote annihilation and creation fermionic operators respectively. To account for gate screening, we use the double-gated screened Coulomb potential given by
\begin{equation}
\mathcal{V}_{0}(\mathbf{q}) = \frac{2\pi e^2}{\epsilon_c |\mathbf{q}| }\tanh{\left(d_{g} |\mathbf{q}|\right)},
\label{eq:dualgate}
\end{equation}
where $e$ is the electron charge, $\epsilon_c=10$ is the dielectric constant associated with encapsulating the system in hexagonal boron nitride (hBN), and $d_{g}=20$ nm is the distance to the gates. The four-fermion term in \eqref{eq:four-fermion Coulomb} can be decomposed into two-fermion terms using Wick's theorem, leading to a Hartree and a Fock contribution to the Hamiltonian. This mean-field Hamiltonian is then solved self-consistently starting with three initial ansatz: a quarter metal that occupies only one isospin\footnote{Throughout, the term isospin is used to refer generically to both spin and valley.}, a half metal that occupies two isospins equally, and a symmetric state that occupies all four isospins. The lowest-energy solution among the three candidates is chosen as the ground state at a particular filling and displacement field\footnote{In this work, we have not considered more general ground states. However, we note that they are possible, including valley-coherent ground states, for example. Driven by experimental constraints, we focus here primarily on the quarter-metal phase and its competition with the half-metal and full-metal phases. We leave other possibilities to future work. }. More details can be found in Sec. II of SI. 

\begin{figure}[t!] 
\begin{center}
\includegraphics[width=1\linewidth]{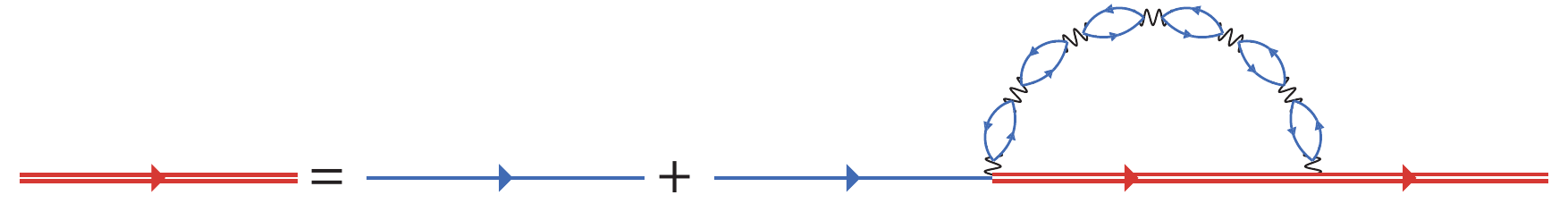}    
\end{center}
\caption{
\textbf{Screened HF diagram.} Diagrammatic representation of the screened Hartree-Fock procedure used for the calculations of superconductivity. Double (single) arrowed lines correspond to interacting (bare) Green functions, wavy lines correspond to interactions mediated by the bare Coulomb potential and the bubbles correspond to electron-hole pairs.}
\label{fig:Figura1a}
\end{figure}

Next, we tackle the superconductivity calculations within a diagrammatic method in the spirit of the Kohn-Luttinger (KL) mechanism~\cite{Kohn1965, Chubukov1993KL2D}.
Based on the HF results, we consider the emergence of superconductivity in the quarter-metal polarized phase.
To reiterate, to get correct values for DOS, we introduce screening from the bare bands and recalculate the HF potential. This procedure is diagrammatically represented in Fig.~\ref{fig:Figura1a}. More details are presented in Sec. III of SI. Concretely, we use the zero-frequency susceptibility of the bare bands at $T=0$, $\Pi_0(\mathbf{q})$, to compute the internally-screened Coulomb potential 
\begin{equation}
   \mathcal{V}_\mathrm{RPA}^{0}(\mathbf{q}) = \frac{\mathcal{V}_{0}(\mathbf{q})}{1 - {\Pi_0}(\mathbf{q})\mathcal{V}_{0}(\mathbf{q})}.
    \label{eq:V0RPAmain}
\end{equation}
We then perform a mean-field calculation by self-consistently determining the HF Hamiltonian, crucially using  ${\cal V}_{\rm RPA}^{0}(\bf q)$ and not $\mathcal{V}_0(\mathbf{q}).$  From this, we obtain the energies and wave functions, $\widetilde{\epsilon}_{\bf q}$ and $\widetilde{\psi}_{\bf q}$. These solutions are subsequently used to calculate a temperature-dependent susceptibility, $\widetilde{\Pi}(\mathbf{q})$, which is, in turn, employed to calculate the modified temperature-dependent screened potential
\begin{equation}
    \widetilde{\mathcal{V}}(\mathbf{q}) = \frac{\mathcal{V}_{0}(\mathbf{q})}{1 - \widetilde{\Pi}(\mathbf{q})\mathcal{V}_{0}(\mathbf{q})}.
    \label{eq:VcHFRPAmain}
\end{equation}
To avoid double counting, we use ${\cal V}_0(\mathbf{q})$ in Eq.~(\ref{eq:VcHFRPAmain})  rather than ${\cal V}_{\rm RPA}^{0}(\mathbf{q})$.
The temperature-dependent screened potential $\widetilde {\cal V}(\mathbf{q})$ is finally used in the gap equation for the kernel $\Gamma(\mathbf{k},\mathbf{q})$:
\begin{equation}
    \Delta(\mathbf{k}) = \sum_{\mathbf{q}} \Gamma(\mathbf{k},\mathbf{q}) \Delta(\mathbf{q}).
\end{equation}
The critical temperature and order parameter are obtained when the largest eigenvalue of $\Gamma$ reaches unity.

\begin{figure}[t!] 
\includegraphics[width=1\linewidth]{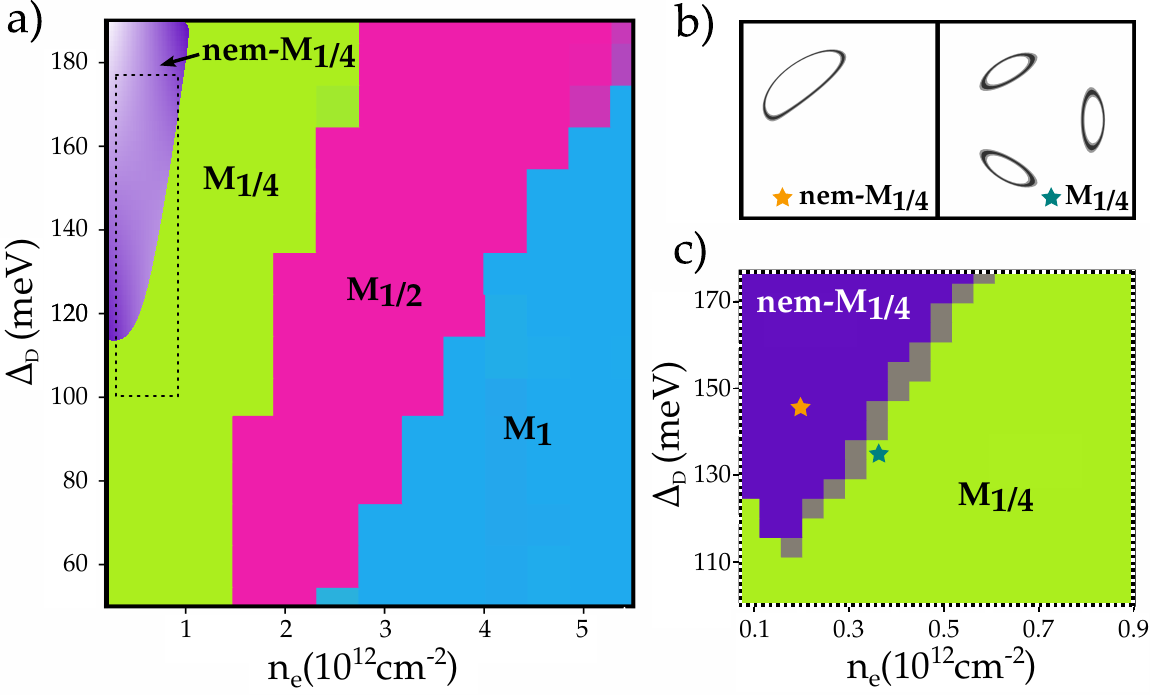}
\caption{
\textbf{Hartree-Fock phase diagram.} 
(a) Self-consistently calculated HF phase diagram, showing M$_{1/4}$, M$_{1/2}$ and M$_1$ phases as a function of the external displacement field ($\Delta_{\mathrm{D}}$) and electronic density ($n_e$). 
The purple region at low densities and high displacement fields highlights schematically where the nem-M$_{1/4}$ is expected to emerge. In (c), we show the calculated phase boundary between M$_{1/4}$ and nem-M$_{1/4}$ in the region indicated by a dashed rectangle in (a). (b) Representative Fermi surfaces at the same point in parameter space indicated by the yellow and blue stars in (c)  solutions with and without $\mathcal{C}_3$ symmetry breaking.}
\label{fig:Figura1}
\end{figure}

\begin{figure}[t] 
\begin{center}
\includegraphics[width=0.43\textwidth]{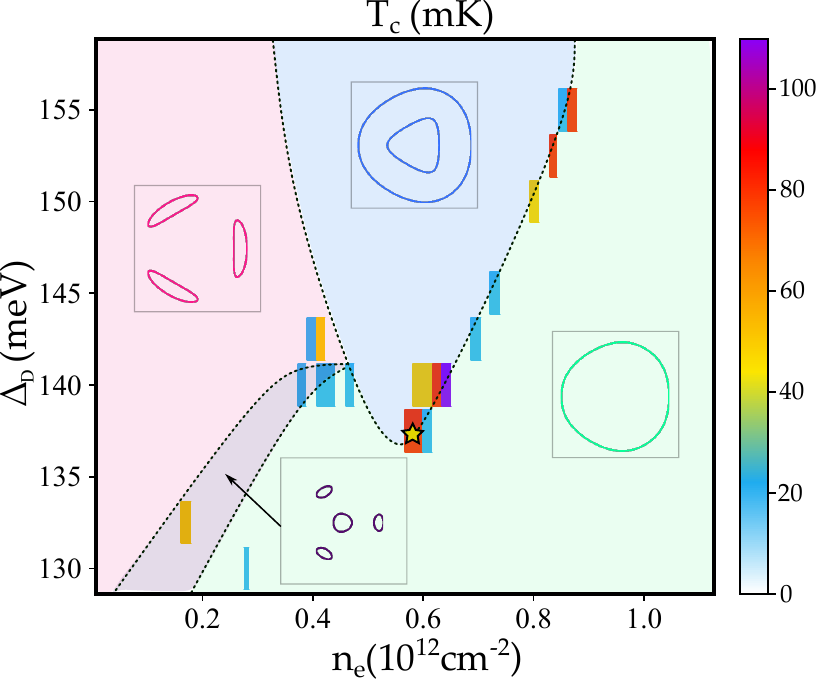}    
\end{center}
\caption{\textbf{Superconductivity phase diagram.} Superconducting critical temperature, $T_c$, as a function of displacement and electron filling in the M$_{1/4}$ phase. The colored regions indicate different Fermi surface shapes.
Black dashed lines mark Lifshitz transitions, between different shapes of the fermi surfaces (see Sec. I of SI). A representative Fermi surface is displayed within each region. 
The yellow star indicates the parameters used for the calculation of the superconducting order parameter shown in Fig.~\ref{fig:Figura3}(a-b).}
\label{fig:Figura2} 
\end{figure}

{\it Normal and Superconducting Phase Diagrams \textendash} The HF calculation described above produces the phase diagram shown in Fig.~\ref{fig:Figura1}(a). We identify four different metallic phases:
\begin{enumerate}
    \item The ${\cal C}_3$-symmetry-broken quarter-metal phase, nem-M$_{1/4}$, is characterized by a Fermi surface with only one pocket in one spin-valley flavor, see Fig.~\ref{fig:Figura1}(b), and appears at low densities and high displacement fields.
    \item The ${\cal C}_3$-symmetric quarter-metal phase, M$_{1/4}$, occupies a single isospin species and arises at higher densities and lower displacement fields compared to the nem-M$_{1/4}$ phase.
    \item The spin-polarized half-metal phase, M$_{1/2}$, occupies both valleys of a single spin flavor and emerges at higher densities compared to the M$_{1/4}$ phase.
    \item The symmetric, unpolarized phase, M$_1$, occupies all four isospins equally and emerges at the highest densities.
\end{enumerate}
The resulting phase diagram, including the presence of the M$_{1/4}$ phase at low densities, aligns well with the experimental observations reported in Ref.~\cite{han_signatures_2024}, where the quarter-metal phase is flanked diagonally by other phases. Interestingly, as shown in Fig.~\ref{fig:Figura1}(b) and Fig.~\ref{fig:Figura1}(c), allowing for broken ${\cal C}_3$ symmetry in the HF calculations, we find the additional nem-M$_{1/4}$ phase at low densities. Fig.~\ref{fig:Figura1}(b) illustrates a representative Fermi surface transition from three pockets to a single pocket. 

These results can be understood as follows: 
in the absence of interactions, the bands near the valley momenta are very flat and exhibit significant trigonal deformations.
Beyond a crossover momentum, away from the center of the valley, the bands become more dispersive and isotropic.
Since exchange interactions occur between occupied states with the same spin and valley indices, at low densities, exchange effects
dominate and Stoner instabilities~\cite{stoner1938collective} are favored, leading to the selection of the M$_{1/4}$ phase as the ground state.
The tendency of exchange to minimize Fermi surface area (see Sec. V in SI) then leads to spontaneous breaking of $\mathcal{C}_3$ and the emergence of the nem-M$_{1/4}$ phase as shown in Fig.~\ref{fig:Figura1}(b) and (c).
As a result, the Fermi surface becomes more isotropic, and trigonal effects are suppressed.
As the electronic density is increased, the dispersion is no longer negligible and the competition with the exchange interaction leads first to the spin-polarized (M$_{1/2}$) state and, at higher densities, to an unpolarized symmetric (M$_1$) phase.

Now, we turn our attention to the superconducting phases found in electron-doped RTLG~\cite{han_signatures_2024}.
Fig.~\ref{fig:Figura2} is the most important result of our work. 
We display the superconducting critical temperature of electron-doped RTLG as a function of electron filling and displacement field. 
It should be noted that the entire range of displacement fields and doping of Fig.~\ref{fig:Figura2} is in the range where the quarter-metal (M$_{1/4}$) state appears, as shown in Fig.~\ref{fig:Figura1}. 
The superconducting sleeve that we find qualitatively matches the principal superconducting dome (so-called SC1) observed in Ref.~\cite{han_signatures_2024}
and closely follows the line of transition between annular and circular Fermi surfaces (see Sec. I and Fig. S1 in SI).
These Lifshitz transitions and their associated high DOS have previously been observed to be highly beneficial for superconductivity in numerous systems~\cite{pantaleon_superconductivity_2023} and are possibly crucial for the analysis of SC1 in electron-doped RTLG~\cite{Geier_isospin_tetralayer_2024}.
These transitions are present and occur smoothly in the bare bands as well as in the RPA-screened HF bands.  However, in the non-screened HF bands, the chemical potential bypasses the Lifshitz transitions, effectively eliminating them and suppressing the formation of any superconducting state (see Secs. 4 and 5 in SI for further details). 

\begin{figure}[t] 
\begin{center}
\includegraphics[width=1\linewidth]{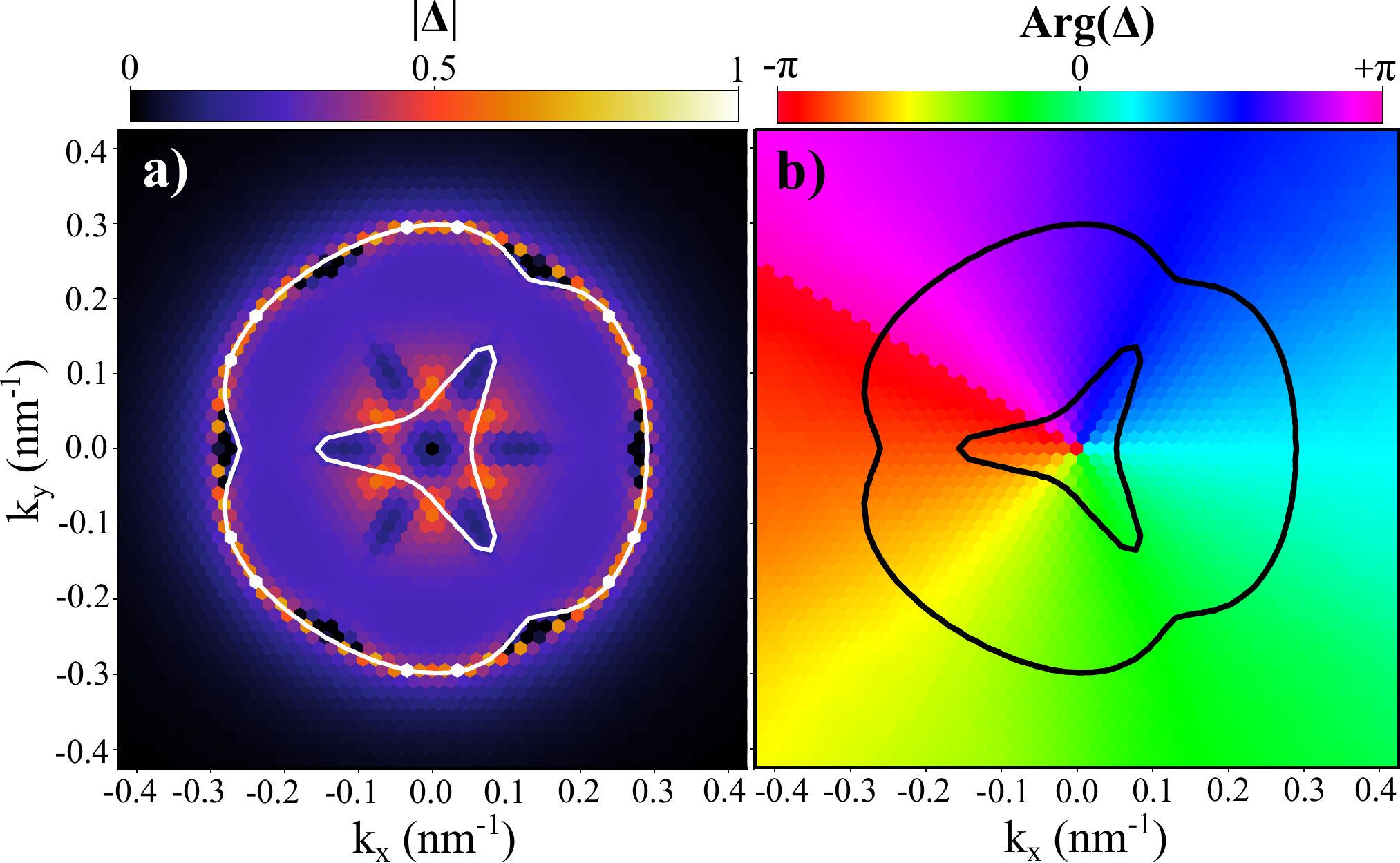}     
\end{center}
\caption{\textbf{Superconducting order parameter}. Order parameter of the leading eigenvalue in the screened Hartree-Fock superconductivity calculations. (a) Normalized magnitude of the order parameters and (b) its corresponding phase. The Fermi surface is indicated by solid white and black lines in each panel, respectively. Calculations were made at values $\Delta_{\mathrm{D}}=137.5$~meV and $\text{n}_e=6.01\times10^{11}\text{cm}^{-2}$ with $T_c~=~80$~mK, marked in Fig.~\ref{fig:Figura2} as a yellow star.} 
\label{fig:Figura3}
\end{figure}

The superconducting order parameter is shown in Fig.~\ref{fig:Figura3}, for a representative state, indicated by a yellow star in Fig.~\ref{fig:Figura2}. 
We find that the order parameter exhibits a $p$-wave symmetry, is nodal and complex (see Sec. XI in SI)\footnote{For clarity, it is worth emphasizing that by $p$-wave, we mean that the order parameter exhibits strong $p$-wave character. In other words, its phase in momentum space winds around the center-of-mass momentum once every $2\pi.$ However, unlike the usual $p$-wave order parameter, ours exhibits a strong dependence on the radial coordinate of momentum as well, as clearly demonstrated in Fig. \ref{fig:Figura3}. More precisely, by $p$-wave, we mean that the order parameter can be written as $\Delta(\mathbf{k}) \approx \Delta(|\mathbf{k}|) e^{\pm i \theta_\mathbf{k}}$}. 
Notably, when superimposing the corresponding Fermi surface (continuous lines in Fig.~\ref{fig:Figura3}), we find that the order parameter is not pinned to the Fermi surface. 
This feature has also been found in twisted systems~\cite{long_evolution_2024}. 
The $p$-wave order parameter, which commonly occurs at smaller electron density, closely matches results from previous studies~\cite{Geier_isospin_tetralayer_2024,Chou_Intravalley_tetralayer_2024,qin_chiral_2024,yang_topological_2024}.

{\it Conditions for Superconductivity \textendash} We recall that in our approach, for the emergence of a superconducting state, we consider a pairing mechanism based on the screened Coulomb interaction, $\widetilde{\mathcal{V}} ( \mathbf{q} )$, which is repulsive for all values of $\mathbf{q}$. 
Superconductivity may arise in the zero-frequency limit, $\omega \rightarrow 0$, if two key conditions are met. 
First, the order parameter $\Delta(\mathbf{k})$ changes sign as a function of $\mathbf{k}$. 
Second, for sufficiently large values of $\mathbf{q},$ the screened interaction must satisfy the inequality $\widetilde{\mathcal{V}} ( \mathbf{q} ) > \widetilde{\mathcal{V}} ( \mathbf{q} = 0 ) > 0$ (see section 3 of SI). 
These criteria allow for superconductivity when the values of $\Delta(\mathbf{k})$ and $\Delta(\mathbf{k} + \mathbf{q})$ have opposite signs
for a sufficient number of combinations of $\mathbf{k}$ and $\mathbf{q}$. 
This criterion is satisfied in our calculations (see Sec. I in SI). 
The RPA calculation performed here leads to an electron polarizability such that $\lim_{\mathbf{q} \rightarrow \mathbf{0} , \omega \rightarrow 0} \widetilde{\Pi} (\mathbf{q},i\omega) = D (\epsilon_F)$, so that $\lim_{\mathbf{q} \rightarrow \mathbf{0} , \omega \rightarrow 0} \widetilde{\mathcal{V}} (\mathbf{q},i\omega)=D^{-1} (\epsilon_F)$,  where $D (\epsilon_F)$ is the density of states at the Fermi energy. Superconductivity arises when, at least for some values of $\mathbf{q}$, $\widetilde{\Pi}(\mathbf{q}\ne0,0) \le \widetilde{\Pi}(\mathbf{q}=0,0)$~\cite{kagan_possibility_1988,Chubukov1993KL2D}. 
The value of $\widetilde{\Pi} (\mathbf{q},0)$ depends on the form factors $|\langle \mathbf{k} + \mathbf{q} | \mathbf{k} \rangle |^2$ (band indices omitted). 
Hence, superconductivity is enhanced when $| \langle  \mathbf{k} + \mathbf{q} | \mathbf{k} \rangle |^2 \ll | \langle  \mathbf{k} | \mathbf{k} \rangle |^2 = 1$. 
This difference increases in the presence of trigonal warping, while it is suppressed by the exchange interaction. This analysis is related to the quantum metric of the band~\cite{shavit_quantum_2024-1,jahin_enhanced_2024}\footnote{It is worth noting that in moire lattices, in addition to the geometric effects discussed in~\cite{shavit_quantum_2024-1,jahin_enhanced_2024,guerci_topological_2024}, the RPA dielectric function and the screened potential are matrices whose elements depend on moiré reciprocal lattice vectors. The resulting Umklapp processes can give a significant contribution to the calculation of the critical temperature~\cite{cea21Coulomb,long_evolution_2024}.}. As a consequence, when the calculation of superconductivity is carried out starting from the bare bands, the critical temperatures are generally overestimated (see Sec. V and Fig. S6 in SI) , whereas when interactions are included, as we have done here, the results are in better agreement with experiments.

On the other hand, the bands in the valley polarized phases show trigonal deformations around the center of the valley. 
This implies that $\epsilon_{\mathbf{k}} \ne \epsilon_{-\mathbf{k}}$. 
As a result, a BCS weak coupling analysis of the pairing does not lead to the well-known Cooper instability, and it is not guaranteed that a finite superconducting critical temperature can be defined irrespective of the strength of the pairing interaction. The situation is reminiscent of a conventional superconductor in the presence of pair-breaking defects~\cite{AG59,ambegaokar_theory_1965,noauthor_gapless_1969}. 
In addition, pairing is favored between electrons with $\epsilon_{\mathbf{k}} \approx \epsilon_{- \mathbf{k}}$, so that the order parameter $\Delta(-\mathbf{k})$ needs not to be restricted to momenta near the Fermi surface. This effect explains the de-pinning of $\Delta(\mathbf{k})$ from the Fermi surface shown in Fig.~\ref{fig:Figura3} and its approximate ${\cal C}_6$ symmetry.

{\it Conclusions \textendash} The origin of superconductivity and other strongly correlated phases in graphene stacks remains a formidable enigma. However, ever-increasing evidence points to an electronic pairing mechanism for the superconducting phases. Unconventional, electron-mediated mechanisms, in which the screened Coulomb interaction acts as the pairing glue, have previously demonstrated reasonable success in reproducing experimental observations in both twisted~\cite{Gonzlez2019, Roy2019, Goodwin2019, cea21Coulomb, Lewandowski2021, Sharma2020, Samajdar2020SC, cea21Coulomb,Pahlevanzadeh2021DMFT, Crepel2022Unconventional, Cea2023Superconductivity,Gonzlez2023TTG,long_evolution_2024,guerci_topological_2024} and non-twisted stacks~\cite{cea2022superconductivity, you2022kohn, Cea2023Superconductivity, JimenoPozo2023, ZiyanLi2023} with a minimal number of tunable parameters. Electron-mediated mechanisms have also been invoked to describe the superconducting quarter-metal states observed in rhombohedral tetralayer graphene~\cite{Geier_isospin_tetralayer_2024,Chou_Intravalley_tetralayer_2024, yang_topological_2024, qin_chiral_2024}. Here, we take a further step by first analyzing the stability of the quarter-metal phase as a function of the electron doping and external potential via the HF approximation, and then, solving the linearized gap equation within the interacting model, which accounts for the sHF renormalized band structure. It is worth noting that the HF phase diagram that we find, which contains a nematic quarter metal, an ordinary quarter metal, a half metal, and a full metal phase, is in good agreement with the experiment of Ref.~\cite{han_signatures_2024}.
This allows us to predict the parent state for the superconducting phase.
We would like to emphasize that this is in contrast to the non-interacting scenario where the parent state is assumed but not actually calculated and, moreover, $T_\mathrm{c}$ is overestimated by an order of magnitude.
Furthermore, the superconducting phenomenology we obtain also matches the experiment both in terms of the critical temperature and the electronic densities at which the superconducting region appears. As the predicted superconducting state is simultaneously finite-momentum (occurring within a single pocket near either $\mathbf{K}$ or $\mathbf{K}'$ valleys) and intrinsically time-reversal and inversion symmetry breaking, we expect a sizable diode effect related to the critical current intensities $I_{L}/I_{R}$, where $L,R$ refer to the direction of the applied current relative to the principal axes of the Fermi surface pockets~\cite{yuan2022supercurrent}. Nematicity may also be accessed by a multi-terminal probe or optical response~\cite{Kaplan2025}, similar to recent experiments on transport in topological magnets~\cite{wang2023quantum,gao2023quantum}. It is possible that in the finite-momentum state, any fluctuations, such as those related to the spontaneous breaking of $\mathbf{K},\mathbf{K}'$ degeneracy, may act as pair-breaking~\cite{fulde_superconductivity_1964, larkin_nonuniform_64} excitations. This fact may explain the telegraphic noise in $R_{xx}$ observed in Ref.~\cite{han_signatures_2024}.

The order parameters revealed by our calculations are similar to the ones predicted when starting from the bare band structure~\cite{Geier_isospin_tetralayer_2024,Chou_Intravalley_tetralayer_2024, yang_topological_2024}: it shows predominantly a $p$-wave symmetry, with suppressed contribution of higher harmonics across electronic density and displacement field. The precise structure of the order parameter can be determined through photoconductivity measurements~\cite{Kaplan2025}; any reduction of symmetry below $\mathcal{C}_6$, such as for a $p$ wave, would lead to a dramatic increase of photoconductivity in response to normal-incidence light. In all scenarios, the photoconductivity is non-vanishing due to the ground state being inversion and time-reversal symmetry broken. 

Our work shows that including interactions is crucial for understanding and correctly describing the experimental results in graphene multilayers, paving the way towards the understanding of the microscopic phenomena behind the formation of such exotic phases.

{\it Acknowledgments \textendash} We thank Cyprian Lewandowski, Oskar Vafek, and Saul A. Herrera for fruitful discussions. IMDEA Nanociencia acknowledges support from the ‘Severo Ochoa’ Programme for Centres of Excellence in R\&D (CEX2020-001039-S/AEI/10.13039/501100011033). The IMDEA Nanociencia team acknowledges support from NOVMOMAT, project PID2022-142162NB-I00 funded by MICIU/AEI/10.13039/501100011033 and by FEDER, UE as well as financial support through the (MAD2D-CM)-MRR MATERIALES AVANZADOS-IMDEA-NC.G.P.-M. is supported by Comunidad de Madrid through the PIPF2022 programme (grant number PIPF-2022TEC-26326). D.K., Y.O., P.E., A.J.-P. and F.G. acknowledge that this research was supported in part by grant NSF PHY-2309135 to the Kavli Institute for Theoretical Physics (KITP). J.A. S.-G. has received financial support through the ``Ram\'on y Cajal'' Fellowship program, grant RYC2023-044383-I financed by MICIU/AEI/10.13039/501100011033 and FSE+. This research was funded in part by the DFG Collaborative Research Center (CRC) 183, and by ISF grants No. 1914/24 and No. 2478/24. D.K. is supported by an Abrahams postdoctoral fellowship of the Center for Materials Theory, Rutgers University and the Zuckerman STEM fellowship. V.T.P. is supported by C. Lewandowski's start-up funds from Florida State University and the National High Magnetic Field Laboratory. The National High Magnetic Field Laboratory is supported by the National Science Foundation through NSF/DMR-2128556 and the State of Florida. We thankfully acknowledge RES resources provided by BSC in MareNostrum5 to FI-2024-3-0029, FI-2024-3-0030 and FI-2025-1-0032.


%

\onecolumngrid
\newpage

\setcounter{equation}{0}
\setcounter{figure}{0}
\setcounter{table}{0}
\setcounter{page}{1}
\setcounter{section}{0}

\renewcommand{\theequation}{S\arabic{equation}}
\renewcommand{\thefigure}{S\arabic{figure}}

\begin{center}
\Large Supplementary Information for \\
Band Renormalization, Quarter Metals, and Chiral Superconductivity\\ in Rhombohedral Tetralayer Graphene
\end{center}
\begin{center}
\normalsize Guillermo Parra-Mart\'inez, Alejandro Jimeno-Pozo, \foreignlanguage{vietnamese}{Võ Tiến Phong},
H\'ector Sainz-Cruz, Daniel Kaplan, Peleg Emanuel, Yuval Oreg, Pierre A. Pantale\'on, Jos\'e \'Angel Silva-Guill\'en and Francisco Guinea
\end{center}

\tableofcontents
\clearpage
\section{Continuum model of Rhombohedral Tetralayer Graphene} \label{SM: Electronic}

\begin{figure}[ht] 
\begin{center}
\includegraphics[width=0.5\linewidth]{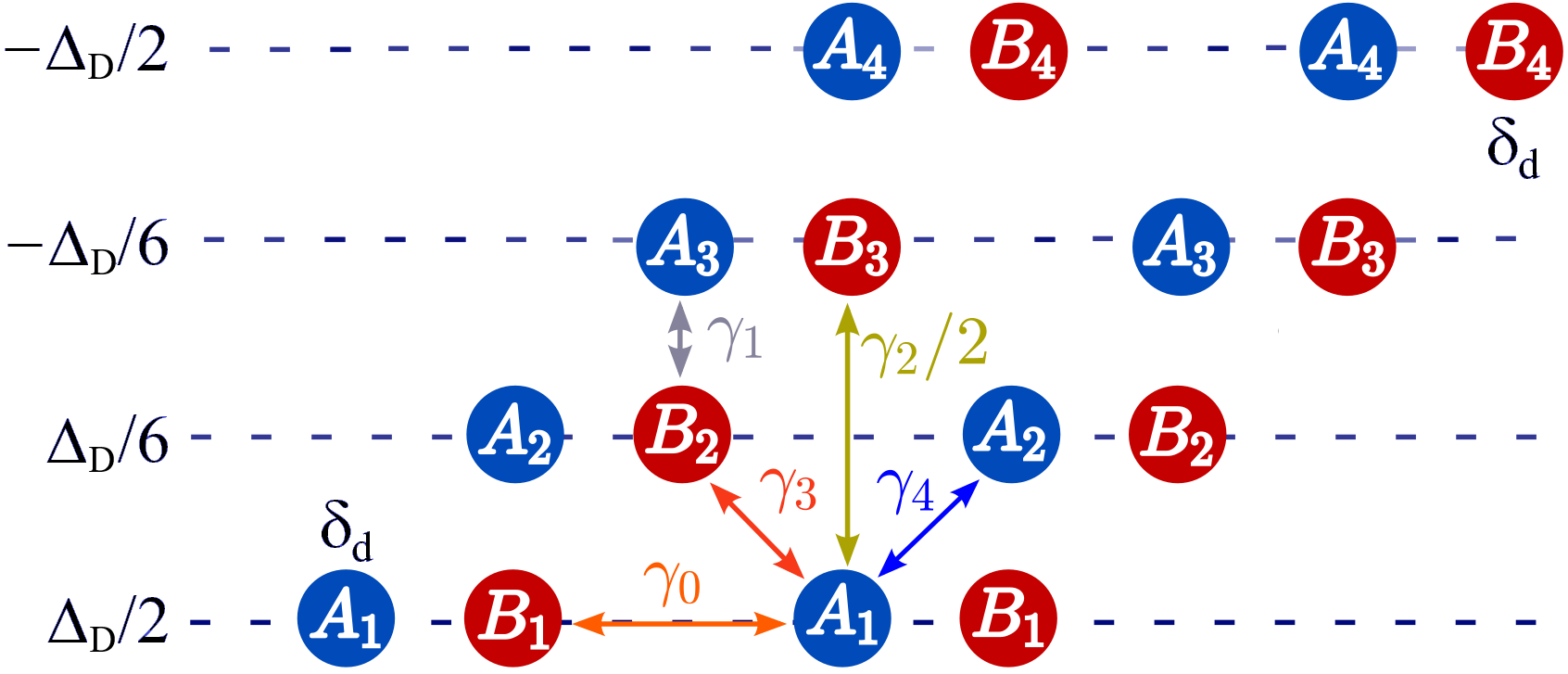}    
\end{center}
\caption{\textbf{Lattice structure of ABCA tetralayer graphene with a representation of the hopping parameters between carbon atoms}. The displacement field, $\Delta_{\mathrm{D}}$, acting in each layer is also indicated. $\delta_d$ is the on-site energy at the corresponding site.}
\label{fig:FiguraSM1}
\end{figure}

Based on models for the electronic structure of graphite~\cite{McClure1956,dresselhaus1981intercalation}, the bands of rhombohedral multilayers have been extensively studied since the isolation of graphene~\cite{guinea_electronic_2006,min_chiral_2008,koshino_trigonal_2009,heikkila_dimensional_2011,ho_evolution_2016,slizovskiy_films_2019}.
In Fig.~\ref{fig:FiguraSM1}, we display the lattice structure of ABCA tetralayer graphene, where each unit cell consists of eight carbon atoms, two per layer, connected through hopping amplitudes $\gamma_{i}$. The continuum Hamiltonian is given by \eqref{eq:Hamil},
\begin{equation}
    \mathcal{H} = \mqty(\Delta_{\mathrm{D}}/2+\delta_d & v_{0}\pi^{\dagger} & v_{4}\pi^{\dagger} & v_{3}\pi & 0 & \gamma_{2}/2 & 0 & 0\\ v_{0}\pi & \Delta_{\mathrm{D}}/2 & \gamma_{1} & v_{4}\pi^{\dagger} & 0 & 0 & 0 & 0\\ v_{4}\pi & \gamma_{1} & \Delta_{\mathrm{D}}/6 & v_{0}\pi^{\dagger} & v_{4}\pi^{\dagger} & v_{3}\pi & 0 & \gamma_{2}/2\\ v_{3}\pi^{\dagger} & v_{4}\pi & v_{0}\pi & \Delta_{\mathrm{D}}/6 & \gamma_{1} & v_{4}\pi^{\dagger} & 0 & 0 \\ 0 & 0 & v_{4}\pi & \gamma_{1} & -\Delta_{\mathrm{D}}/6 & v_{0}\pi^{\dagger} & v_{4}\pi^{\dagger} & v_{3}\pi \\ \gamma_{2}/2 & 0 & v_{3}\pi^{\dagger} & v_{4}\pi & v_{0}\pi & -\Delta_{\mathrm{D}}/6 & \gamma_{1} & v_{4}\pi^{\dagger} \\ 0 & 0 & 0 & 0 & v_{4}\pi & \gamma_{1} & -\Delta_{\mathrm{D}}/2 & v_{0}\pi^{\dagger}\\ 0 & 0 & \gamma_{2}/2 & 0 & v_{3}\pi^{\dagger} & v_{4}\pi & v_{0}\pi & -\Delta_{\mathrm{D}}/2+\delta_d),
    \label{eq:Hamil}
\end{equation}
where $\pi = \xi k_{x} +\mathrm{i}k_{y}$ with $\xi=\pm 1$ the valley index and $v_{i}=\frac{\sqrt{3}}{2}a\gamma_{i}$ with $a=2.46$ \r{A} the lattice constant of graphene \cite{ghazaryan_multilayer_2023}. The parameter $\Delta_{\mathrm{D}}$ takes into account an external displacement field and $\delta_d$ encodes an on-site potential which is only present at sites $A_1$ and $B_4$ since these two atoms do not have a nearest neighbor on the adjacent layer. The parameters are given in Table \ref{tab: Tight-binding parameters}.

\begin{table}[h!]
\begin{center}
\begin{tabular}{cccccc} 
 \hline
 \hline
$\gamma_0$ & $\gamma_1$ & $\gamma_2$ & $\gamma_3$ & $\gamma_4$ & $\delta_d$ \\
\hline
$3.1$ & $0.38$ & $-0.015$ & $-0.29$ & $-0.141$ & $-0.0105$ \\
\hline
\hline
\end{tabular}

\caption{\textbf{Model parameters.} Values taken from Refs.~\cite{Zhou2021HalfMetRTG,Zhang2010Band} and are quoted in electronvolts.}
\label{tab: Tight-binding parameters}
\end{center}
\end{table}

\section{Hartree-Fock Formalism for Isospin-Polarized Phases}  \label{SM: HF RPA Formalism}

\begin{figure}[t!] 
\begin{center}
    \includegraphics[width=0.7\linewidth]{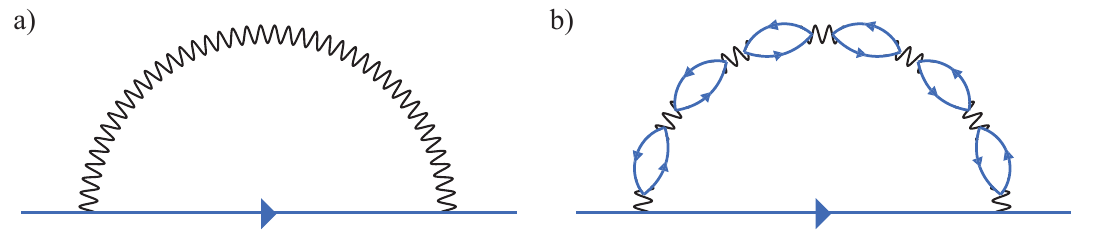}
\end{center}
\caption{\textbf{Sketch of possible self energy diagrams which modify the noninteracting electron bands}. \textbf{(a)} Fock diagram for the HF phase diagram calculations. \textbf{(b)} RPA renormalized Fock diagram for the superconductivity calculations.}
\label{fig:FiguraSM2}
\end{figure}

In this section, we write the interacting Hamiltonian and provide details on the Hartree-Fock calculation. 
We work with plane-wave basis states $\langle{\mathbf{r}}\ket{\psi_{\alpha,\mathbf{k}} } = \Omega^{-1/2}e^{i \mathbf{k} \cdot \mathbf{r}} \ket{\alpha}$, where $\alpha$ is a multi-index label that includes valley $\xi$, spin $s$, sublattice $\sigma$, and layer $\ell$, and $\Omega$ is the area of the entire system. We completely ignore intervalley scattering in the non-interacting limit; so these basis states can be taken to be orthonormal $\langle{\psi_{\alpha, \mathbf{k}}}\ket{\psi_{\alpha',\mathbf{k}'}} = \langle{\alpha} \ket{\alpha'} \Omega^{-1}\int_{\mathrm{\Omega}}d^2\mathbf{r} e^{-i \left( \mathbf{k} - \mathbf{k}' \right) \cdot \mathbf{r}}  =  \delta_{\alpha, \alpha'} \delta_{\mathbf{k}, \mathbf{k}'}$. The single-particle kinetic energy is $\hat{\mathcal{H}}_0 = \sum_{\substack{\mathbf{k} , \alpha, \alpha' }} \hat{c}_{\alpha,\mathbf{k}}^\dagger  \mathbb{K}^{\alpha, \alpha'} (\mathbf{k}) \hat{c}_{\alpha',\mathbf{k}},$ where $\mathbb{K}^{\alpha, \alpha'} (\mathbf{k}) = \bra{\psi_{\alpha,\mathbf{k}} } \hat{\mathcal{H}}_0\ket{\psi_{\alpha',\mathbf{k}} }$ is the non-interacting Hamiltonian matrix and $\hat{c}_{\alpha,\mathbf{k}}^\dagger $ creates $\ket{\psi_{\alpha,\mathbf{k} }}.$ The two-particle Coulomb interaction in Eq.(1) of the main text can be decomposed as 
\begin{equation}
    \begin{split}
        \hat{\mathcal{V}}_C &\approx \frac{1}{\Omega} \sum_{ \substack{ \mathbf{k} ,\alpha}}  \left[ \mathcal{V}_0\left( \mathbf{0}\right)\sum_{\mathbf{p} , \alpha'}\langle \hat{c}^\dagger_{\alpha', \mathbf{p} } \hat{c}_{\alpha', \mathbf{p} } \rangle \right]\hat{c}^\dagger_{\alpha, \mathbf{k} }\hat{c}_{\alpha, \mathbf{k} }- \frac{1}{\Omega} \sum_{\substack{ \mathbf{k} ,\alpha, \alpha' }}  \left[\sum_{ \mathbf{p} }\mathcal{V}_0\left( \mathbf{p}- \mathbf{k} \right) \langle \hat{c}^\dagger_{\alpha', \mathbf{p} } \hat{c}_{\alpha, \mathbf{p} } \rangle \right]\hat{c}^\dagger_{\alpha, \mathbf{k}} \hat{c}_{\alpha', \mathbf{k}}.
    \end{split}
\end{equation}
In this form of the Coulomb potential, we have assumed that the final Slater determinant state preserves lattice translational symmetries. Hence, the Hartree and Fock terms are both diagonal in $\mathbf{k}.$   To simplify notation, we introduce the following Hartree and Fock matrices 
\begin{equation}
    \begin{split}
        \mathbb{H}^{\alpha \alpha'} &= +\frac{\delta_{\alpha, \alpha'}}{\Omega} \mathcal{V}_0(\mathbf{0})\sum_{\mathbf{p} ,  \beta} \mathbb{D}^{\beta \beta}(\mathbf{p}), \\
        \mathbb{F}^{\alpha \alpha'}(\mathbf{k}) &= \mathbin{-} \frac{1}{\Omega} \sum_{\mathbf{p}}\mathcal{V}_0\left(\mathbf{p}  - \mathbf{k} \right) \mathbb{D}^{\alpha \alpha'}(\mathbf{p}).
    \end{split}
\end{equation}
Here, we have defined a density matrix $\mathbb{D}^{\alpha \alpha '}(\mathbf{k}) = \langle \hat{c}^\dagger_{\alpha',\mathbf{k}} \hat{c}_{\alpha, \mathbf{k}} \rangle;$ we draw attention to the order of the indices in our definition of the density matrix\footnote{Another common convention is to define $\mathbb{D}^{\alpha' \alpha }(\mathbf{k}) = \langle \hat{c}^\dagger_{\alpha',\mathbf{k}} \hat{c}_{\alpha, \mathbf{k}} \rangle.$ In this case, the Fock term becomes $\mathbb{F}^{\alpha \alpha'}(\mathbf{k}) = - \frac{1}{\Omega} \sum_{\mathbf{p}}\mathcal{V_\mathcal{C}}\left(\mathbf{p}  - \mathbf{k} \right) \mathbb{D}^{\alpha' \alpha}(\mathbf{p}).$ The Fock energy is then $\frac{1}{2}\sum_{\alpha,\alpha',\mathbf{k}}\mathbb{F}^{\alpha \alpha'}(\mathbf{k})\mathbb{D}^{\alpha \alpha'}(\mathbf{k}) = \frac{1}{2} \mathrm{Tr} \left[\mathbb{F} \mathbb{D}^\mathrm{T}\right]$}. Explicitly, the matrix elements can be calculated from the coefficients of the one-particle eigenstates $\mathbb{D}^{\alpha \alpha'}(\mathbf{k}) = \sum_{n \in \mathrm{occupied}}   \phi_n^\alpha(\mathbf{k})\phi_{n}^{\alpha',*}(\mathbf{k}),$ where the eigenstates are $\ket{\psi_{n, \mathbf{k}}} = \sum_{\alpha} \phi_n^\alpha(\mathbf{k}) \ket{\psi_{\alpha,\mathbf{k}} }$ and $n$ labels band index. Normalization requires $\sum_{\alpha}\left|\phi_n^\alpha(\mathbf{k})\right|^2 = 1.$ The mean-field Hamiltonian now includes both the non-interacting part and the new Hartree and Fock terms
\begin{equation}
\label{eq: Hartree-Fock Hamiltonian}
    \hat{\mathcal{H}}^\mathrm{HF} = \sum_{\substack{\mathbf{k} , \alpha, \alpha' }}  \hat{c}_{\alpha,\mathbf{k}}^\dagger  \left[ \mathbb{K}^{\alpha \alpha'}(\mathbf{k}) + \mathbb{H}^{\alpha \alpha'} + \mathbb{F}^{\alpha \alpha'}(\mathbf{k})  \right] \hat{c}_{\alpha',\mathbf{k}}.
\end{equation}
As usual, we observe that both the Hartree and Fock terms require knowledge of the occupied wavefunctions; in particular, the Fock term is highly nonlocal. On the other hand, the Hartree term is just proportional to the identity matrix and depends only on the overall number density since $N_\mathrm{total} = \Omega^{-1}\sum_{\mathbf{p} ,  \beta} \mathbb{D}^{\beta \beta}(\mathbf{p})$ is just the total electron number (this relies on the assumption that the Coulomb potential carries no internal indices like layer). We must solve this Hamiltonian iteratively until convergence. In our calculations, we include all eight bands with a cutoff only in momentum, and we do not substrate any reference density from the density matrix. The eigenvalues of $\hat{\mathcal{H}}^\mathrm{HF} \ket{\psi^\mathrm{HF}_{n}} = \epsilon^\mathrm{HF}_{n} \ket{\psi^\mathrm{HF}_{n}}$ are \textit{not} real energies but are merely  Lagrange multipliers. Nonetheless,  they are helpful in finding the energy of the Hartree-Fock ground state $\ket{\Omega^\mathrm{HF}}$ as well as the energy of the first excited state due to Koopmans theorem. The total energy of a Hartree-Fock state is: 
\begin{equation}
\begin{split}
    \mathcal{E}_{\Omega} &= \bra{\Omega^\mathrm{HF}} \left( \hat{\mathcal{H}_0} + \hat{\mathcal{V}}_\mathrm{C} \right) \ket{\Omega^\mathrm{HF}} =\mathrm{Tr} \left[\left(\mathbb{K} + \frac{\mathbb{H}}{2} + \frac{\mathbb{F}}{2}   \right) \mathbb{D} \right]\\
\end{split}
\end{equation} 
where the trace is over both internal indices and momentum. Factors of $\frac{1}{2}$ in this expression are important to note. The absolute value of the energy of the ground state $\mathcal{E}_{\Omega}$ can be arbitrarily shifted. Therefore, it is more meaningful to compare it to the energy of a definite reference state. At a particular filling, we choose the reference state to be the symmetric ground state with equal population in all valley-spin flavors, denoted by $\bar{\mathcal{E}}_\Omega.$ Thus, the quantity to be extremized is then $\Delta \mathcal{E} = \mathcal{E}_\Omega - \bar{\mathcal{E}}_\Omega.$
In our calculation, we include only the Fock  term (schematically shown in Fig.~\ref{fig:FiguraSM2}(a)) since the Hartree term only leads to a rigid shift of the band structure and, therefore, it does not contribute to the energy difference landscape.

\section{Kohn-Luttinger sHF formalism for superconductivity }  \label{SM: HFRPA Superconductivity}

\begin{figure}[ht] 
\begin{center}
    \includegraphics[width=0.7\linewidth]{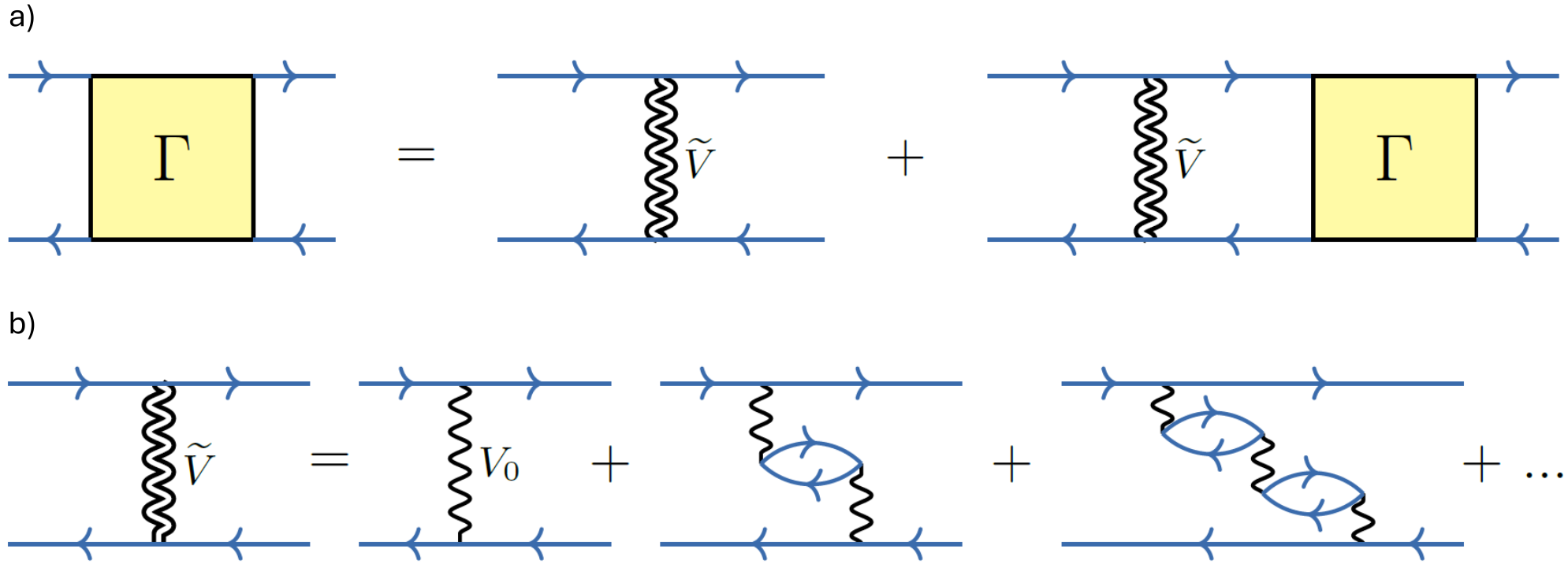}
\end{center}
\caption{\textbf{Diagrammatic representation} of \textbf{(a)} Bethe-Salpeter equation and \textbf{(b)} RPA screened interaction that defines the kernel $\Gamma$ according to  \eqref{eq:kernelSupp}.}
\label{fig:FiguraSM3}
\end{figure}

We consider the emergence of superconductivity in the quarter-metal (non-nematic) polarized phase. We adopt a diagrammatic method based on the Kohn-Luttinger (KL) mechanism~\cite{Kohn1965, Chubukov1993KL2D}, which has previously shown good agreement with experimental results for both twisted~\cite{cea21Coulomb,phong2021band, long_evolution_2024,guerci_topological_2024} and untwisted systems~\cite{Ghazaryan2021Anular, cea2022superconductivity, JimenoPozo2023, pantaleon_superconductivity_2023, ZiyanLi2023, Dong2023spin, Dong2023Multilayer} and it has also been employed in highly doped monolayer graphene~\cite{Herrera2024Topological}. 
In contrast to the usual KL mechanism which considers different second-order processes to screen the Coulomb potential, our approach focuses on a single process (electron-hole excitations, bubble diagram) which is summed to infinite order via RPA, see Fig.~\ref{fig:FiguraSM3}(b).

\begin{figure*}[t!]
\begin{center}
\includegraphics[width=0.9\textwidth]{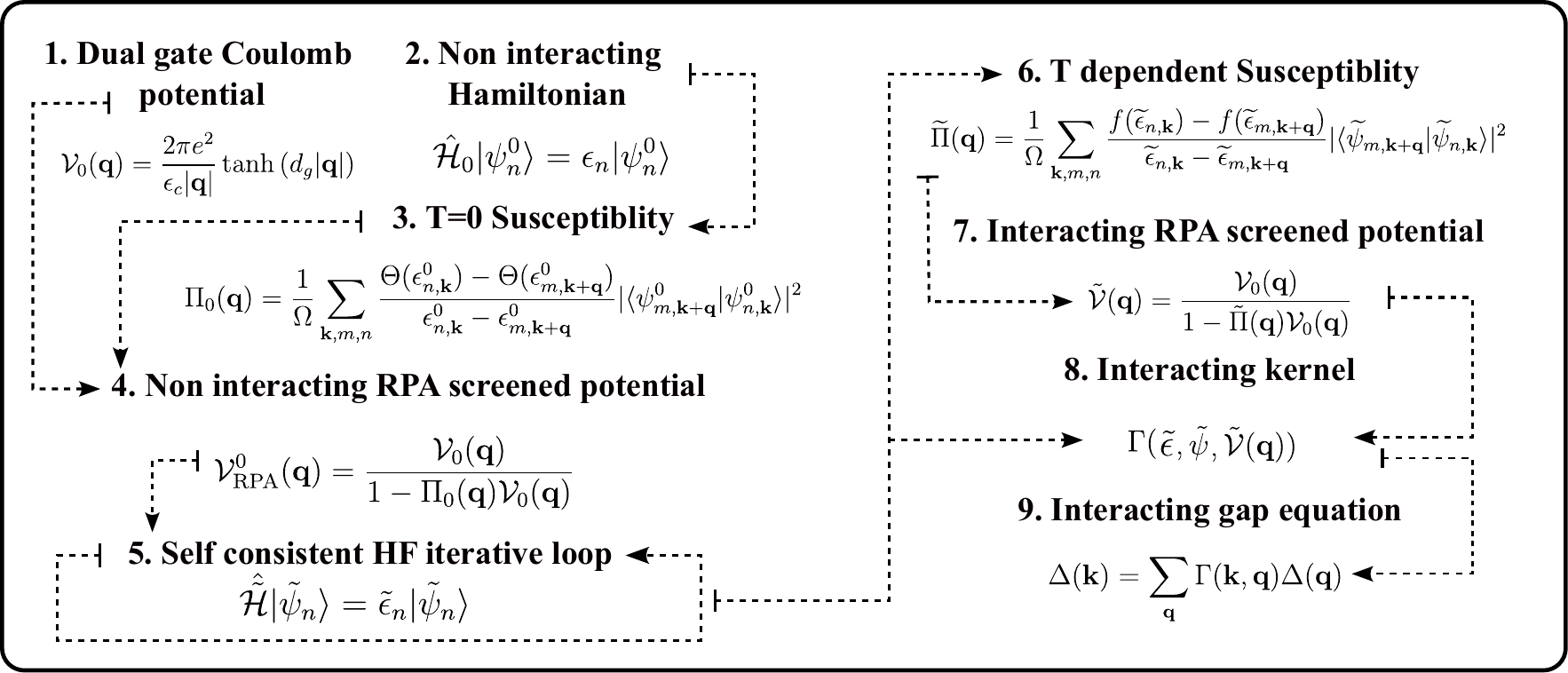}
\end{center}
\caption{\textbf{Workflow scheme of the screened Hartree-Fock plus superconductivity calculation for the results shown in orange}{Fig. 3 and Fig. 4 of the main text}.} 
\label{fig:FiguraSM4} 
\end{figure*}

The  Hartree-Fock approximation has been widely used to study the phase diagram of graphene stacks as it provides a reliable estimate of the total energies of the different phases. However, it also leads to an increase in the bandwidth and a reduction of the Fermi-level DOS. This effect is most extreme for the unscreened Coulomb interaction, where the density of states vanishes logarithmically in two and three dimensions~\cite{monkhorst_hartree-fock_1979}. This nonphysical feature of the HF bands significantly influences the results of calculations involving excited states. In particular, it suppresses the Cooper instability which allows for the existence of superconductivity in metals in the weak coupling regime. Two common circumventions to this problem involve either using a Coulomb interaction screened by external metallic gates or imposing finite grids with cutoffs for small momenta. Neither one of these workarounds is ideal since they both depend on extrinsic factors. The latter method is especially pathological since it may affect the accuracy of the computed results due to finite-size effects. The former method is at least physically motivated since dual gates are, in fact, used in experiments to tune carrier density and displacement field. Because of that, we do use a gate-screened potential in this work. Unfortunately, this does not completely solve the problem because even with a gate-screened potential, the Fermi level DOS is still suppressed even if it is not completely quenched. Therefore, out of necessity, we use a combined approach that includes both metallic gate screening and intrinsic internal screening; this approach is summarized in Fig.~\ref{fig:FiguraSM4} and schematically shown in Figs. \ref{fig:FiguraSM2} and \ref{fig:FiguraSM3}. In what follows, we explain this approach in detail.

To start, we employ the non-interacting energies and wave functions, $\epsilon^0$  and $\psi^0$, to construct the zero-frequency susceptibility at $T=0,$ 

\begin{equation}
    \Pi_0(\vb{q}) = \frac{1}{\Omega}\sum_{\vb{k},m,n} \frac{\Theta(\epsilon^{0}_{n,\vb{k}}) - \Theta(\epsilon^{0}_{m,\vb{k}+\vb{q}})}{\epsilon^{0}_{n,\vb{k}} - \epsilon^{0}_{m,\vb{k}+\vb{q}}} \abs{\langle\psi^{0}_{m,\vb{k}+\vb{q}}|\psi^{0}_{n, \vb{k}}\rangle}^{2},
    \label{eq:SusBare}
\end{equation}
where $\Theta(\epsilon^{0}_{n,\vb{k}})$ is the Heaviside step function and   $\epsilon^0_{n,\vb{k}}=E^0_{n,\vb{k}}-\mu$ with $\mu$ being the Fermi energy, $E^0_{n,\vb{k}}$ are the eigenvalues and $\psi^{0}_{n,\vb{k}}$ the eigenvectors of the non-interacting Hamiltonian, \eqref{eq:Hamil}, and  $\Omega$ is the area of the system. 

Next, we incorporate a screened interaction at $T=0$, to compute the electronic self-energy, analogous to the $GW$ method used in band structure calculations:

\begin{equation}
   \mathcal{V}^0_\mathrm{RPA}(\vb{q}) = \frac{\mathcal{V}_{0}(\vb{q})}{1 - {\Pi_0}(\vb{q})\mathcal{V}_{0}(\vb{q})},
    \label{eq:VcHFRPAsupp}
\end{equation}
where $\mathcal{V}_{0}$ is the dual-gate Coulomb potential given by Eq.(2) of the main text. 
We then perform a mean-field calculation by solving HF self-consistently with the screened potential, $\mathcal{V}^0_\mathrm{RPA}$,and obtain $\widetilde{\epsilon}$ and $\widetilde{\psi}$.
These solutions are then used to calculate a temperature-dependent susceptibility, $\widetilde{\Pi}$,
\begin{equation}
    \widetilde{\Pi}(\vb{q}) = \frac{1}{\Omega}\sum_{\vb{k},m,n} \frac{f(\widetilde{\epsilon}_{n,\vb{k}}) - f(\widetilde{\epsilon}_{m,\vb{k}+\vb{q}})}{\widetilde{\epsilon}_{n,\vb{k}} - \widetilde{\epsilon}_{m,\vb{k}+\vb{q}}} \abs{\langle\widetilde{\psi}_{m,\vb{k}+\vb{q}}|\widetilde{\psi}_{n, \vb{k}}\rangle}^{2},
    \label{eq:tildepi}
\end{equation}
where $m,n$ are band indexes, $f$ is the Fermi-Dirac distribution and $\widetilde{\epsilon}_{n,\vb{k}}=\widetilde{E}_{n,\vb{k}}-\mu$ with $\mu$ being the Fermi energy, $\widetilde{E}_{n,\vb{k}}$ are the eigenvalues, $\widetilde{\psi}_{n,\vb{k}}$ the eigenvectors of the interacting system.

Building upon the resulting sHF band structure, $\widetilde{\epsilon}$ and $\widetilde{\psi}$, we proceed with the analysis of superconductivity where the pairing potential responsible for Cooper pair formation is the dual gate Coulomb potential, Eq.(2) of the main text, 
effectively screened by electron-hole excitations of the interacting system given by

\begin{equation}
    \widetilde{\mathcal{V}}(\vb{q}) = \frac{\mathcal{V}_{0}(\vb{q})}{1 - \widetilde{\Pi}(\vb{q})\mathcal{V}_{0}(\vb{q})}.
    \label{eq:VcHFRPA}
\end{equation}
It is important to note that this approach avoids double-counting the screening effects for the superconducting states. 

Finally, we solve the linearized gap equation to characterize the superconducting state. To compute the critical temperature ($T_{\mathrm{c}}$) and order parameter we solve the linearized gap equation given by

\begin{equation}
\label{eq:initial_gap_eq}
    \Delta_{ij}(\vb{k}) = -\frac{k_{B}T}{\Omega}\sum_{\vb{q},\mathrm{i}\omega}\widetilde{\mathcal{V}}(\vb{k}-\vb{q}) \sum_{i',j'} \mathcal{G}_{0}^{i,i'}(\vb{q},\mathrm{i}\omega)\mathcal{G}_{0}^{j,j'}(-\vb{q},-\mathrm{i}\omega) \Delta_{i',j'}(\vb{q}),
\end{equation}
where $\mathcal{G}_{0}$ is the electron Green function.  
Then, the electron Green function takes the form,
\begin{equation}
\label{eq:green_func}
    \mathcal{G}_{0}^{i,j}(\vb{k},\pm\mathrm{i}\omega)=\sum_{n}\frac{\widetilde{\phi}^{i}_{n}(\vb{k})\widetilde{\phi}^{j,*}_{n}(\vb{k})}{\mathrm{i}\omega \mp \widetilde{\epsilon}_{n,\vb{k}}},
\end{equation} 
where the index $n,i,j$ labels the band, sublattice and layer, respectively. We shall recall that $\widetilde{\epsilon}_{n,\vb{k}} = E_{n,\vb{k}}-\mu$ and $\widetilde{\phi}_{n}(\vb{k})$ correspond to the energies and eigenfunctions of the interacting system, i.e. $\widetilde{\epsilon}_{n,\vb{k}}$ and $\widetilde{\psi}_{n,\vb{k}}$.
Introducing the definition in \eqref{eq:green_func} into \eqref{eq:initial_gap_eq} leads to
\begin{equation}
\begin{split}
    \Delta_{ij}(\vb{k}) = \frac{k_{B}T}{\Omega}\sum_{\vb{q},\mathrm{i}\omega}\widetilde{\mathcal{V}}(\vb{k}-\vb{q}) &\sum_{i',j'}  \sum_{n,m}\frac{\widetilde{\phi}_{n}^{i}(\vb{q})\widetilde{\phi}_{n}^{i',*}(\vb{q})}{\mathrm{i}\omega - \widetilde{\epsilon}_{n,\vb{q}}}\times   \frac{\widetilde{\phi}_{m}^{j}(-\vb{q})\widetilde{\phi}_{m}^{j',*}(-\vb{q})}{\mathrm{i}\omega + \widetilde{\epsilon}_{m,-\vb{q}}} \Delta_{i',j'}(\vb{q}) .    
\end{split}
\end{equation}
The Matsubara sum can be carried out using
\begin{equation}
    \sum_{\mathrm{i}\omega} \frac{1}{\mathrm{i}\omega - \widetilde{\epsilon}_{n,\vb{q}}} \frac{1}{\mathrm{i}\omega + \widetilde{\epsilon}_{m,-\vb{q}}} = \frac{1}{k_{B}T}\frac{f(-\widetilde{\epsilon}_{m,-\vb{q}}) - f(\widetilde{\epsilon}_{n,\vb{q}})}{\widetilde{\epsilon}_{n,\vb{q}} + \widetilde{\epsilon}_{m,-\vb{q}}},
\end{equation}
which leads to
\begin{equation}
    \begin{split}
    \Delta_{ij}(\vb{k}) &= -\frac{1}{\Omega}\sum_{\vb{q}}\widetilde{\mathcal{V}}(\vb{k}-\vb{q}) \sum_{i',j'}  \sum_{n,m} \widetilde{\phi}_{n}^{i}(\vb{q})\widetilde{\phi}_{n}^{i',*}(\vb{q}) \widetilde{\phi}_{m}^{j}(-\vb{q})\widetilde{\phi}_{m}^{j',*}(-\vb{q})  \frac{f(-\widetilde{\epsilon}_{m,-\vb{q}}) - f(\widetilde{\epsilon}_{n,\vb{q}})}{\widetilde{\epsilon}_{n,\vb{q}} + \widetilde{\epsilon}_{m,-\vb{q}}} \Delta_{i',j'}(\vb{q}),\\
    &= -\frac{1}{\Omega}\sum_{\vb{q}} \widetilde{\mathcal{V}}(\vb{k}-\vb{q}) \sum_{n,m} \frac{f(-\widetilde{\epsilon}_{m,-\vb{q}}) - f(\widetilde{\epsilon}_{n,\vb{q}})}{\widetilde{\epsilon}_{n,\vb{q}} + \widetilde{\epsilon}_{m,-\vb{q}}} \widetilde{\phi}_{n}^{i}(\vb{q}) \widetilde{\phi}_{m}^{j}(-\vb{q})  \sum_{i',j'} \widetilde{\phi}_{n}^{i',*}(\vb{q}) \widetilde{\phi}_{m}^{j',*}(-\vb{q}) \Delta_{i',j'}(\vb{q}).
    \end{split}
\end{equation}
We now project the above equation onto the band basis by defining
\begin{equation}
    \Delta_{nm}(\vb{k}) = \sum_{ij} \widetilde{\phi}_{n}^{i,*}(\vb{k}) \widetilde{\phi}_{m}^{j,*}(-\vb{k}) \sqrt{\frac{f(-\widetilde{\epsilon}_{m,-\vb{k}}) - f(\widetilde{\epsilon}_{n,\vb{k}})}{\widetilde{\epsilon}_{n,\vb{k}} + \widetilde{\epsilon}_{m,-\vb{k}}}} \Delta_{i,j}(\vb{k}),
\end{equation}
and obtain
\begin{equation}
    \begin{split}
        \Delta_{n_{1}m_{1}}(\vb{k}) =& -\frac{1}{\Omega}\sum_{\vb{q}, n, m} \widetilde{\mathcal{V}}(\vb{k}-\vb{q})  \sqrt{\frac{f(-\widetilde{\epsilon}_{m_{1},-\vb{k}}) - f(\widetilde{\epsilon}_{n_{1},\vb{k}})}{\widetilde{\epsilon}_{n_{1},\vb{k}} + \widetilde{\epsilon}_{m_{1},-\vb{k}}}}  \sqrt{\frac{f(-\widetilde{\epsilon}_{m,-\vb{q}}) - f(\widetilde{\epsilon}_{n,\vb{q}})}{\widetilde{\epsilon}_{n,\vb{q}} + \widetilde{\epsilon}_{m,-\vb{q}}}} \times \\
        &\sum_{ij}\widetilde{\phi}_{n}^i(\vb{q})\widetilde{\phi}_{n_{1}}^{i,*}(\vb{k}) \widetilde{\phi}_{m}^j(-\vb{q})\widetilde{\phi}_{m_{1}}^{j,*}(-\vb{k}) \Delta_{nm}(\vb{q}).
    \end{split}
\end{equation}
Therefore, the gap equation may be written as an eigenvalue problem
\begin{equation}
    \Delta_{n_{1}m_{1}}(\vb{k}) = \sum_{\vb{q}, n, m} \Gamma_{n_{1}m_{1},n,m}(\vb{k},\vb{q}) \Delta_{nm}(\vb{q}),
\end{equation}
where the kernel is given by
\begin{equation}
    \begin{split}
        \Gamma_{n_{1}m_{1},n,m}(\vb{k},\vb{q}) = -\frac{\widetilde{\mathcal{V}}(\vb{k}-\vb{q})}{\Omega} & \sqrt{\frac{f(-\widetilde{\epsilon}_{m_{1},-\vb{k}}) - f(\widetilde{\epsilon}_{n_{1},\vb{k}})}{\widetilde{\epsilon}_{n_{1},\vb{k}} + \widetilde{\epsilon}_{m_{1},-\vb{k}}}}\sqrt{\frac{f(-\widetilde{\epsilon}_{m,-\vb{q}}) - f(\widetilde{\epsilon}_{n,\vb{q}})}{\widetilde{\epsilon}_{n,\vb{q}} + \widetilde{\epsilon}_{m,-\vb{q}}}}  \langle \widetilde{\psi}_{n_{1},\vb{k}}|\widetilde{\psi}_{n,\vb{q}}\rangle \langle\widetilde{\psi}_{m_{1},-\vb{k}}|\widetilde{\psi}_{m,-\vb{q}}\rangle.
    \end{split}
    \label{eq:kernelSupp}
\end{equation}
In the calculations, we fix the phase of the wave functions so that the first component ($A_{1}$) is real. Hence, the kernel is then given by
\begin{equation}
    \begin{split}
        \Gamma_{n_{1}m_{1},n,m}(\vb{k},\vb{q}) =& -\frac{\widetilde{\mathcal{V}}(\vb{k}-\vb{q})}{\Omega}  \sqrt{\frac{f(-\widetilde{\epsilon}_{m_{1},-\vb{k}}) - f(\widetilde{\epsilon}_{n_{1},\vb{k}})}{\widetilde{\epsilon}_{n_{1},\vb{k}} + \widetilde{\epsilon}_{m_{1},-\vb{k}}}}  \sqrt{\frac{f(-\widetilde{\epsilon}_{m,-\vb{q}}) - f(\widetilde{\epsilon}_{n,\vb{q}})}{\widetilde{\epsilon}_{n,\vb{q}} + \widetilde{\epsilon}_{m,-\vb{q}}}}\ \times \\
        &e^{-\mathrm{i}\left(-\theta_{k}+\theta_{q}-\theta_{-k}+\theta_{-q}\right)}  \langle \widetilde{\psi}_{n_{1},\vb{k}}|\widetilde{\psi}_{n,\vb{q}}\rangle \langle\widetilde{\psi}_{m_{1},-\vb{k}}|\widetilde{\psi}_{m,-\vb{q}}\rangle.
    \end{split}
\end{equation}
By diagonalizing the kernel as a function of the temperature and the Fermi energy, we identify the onset of superconductivity as when its leading eigenvalue takes a value of $1$. 
We would like to emphasize that this involves recalculating the screened Coulomb potential, $\widetilde{\mathcal{V}}(\vb{q})$, as a function of the temperature within the RPA (\eqref{eq:VcHFRPA}), using the energies and wave functions of the interacting system, $\widetilde{\epsilon}_{n,\vb{k}}$ and $\widetilde{\psi}_{n,\vb{k}}$.

\section{Fermi Surface Topology} \label{SM: FS Topology}

\begin{figure}[ht] 
\begin{center}
 \includegraphics[width=0.8\textwidth]{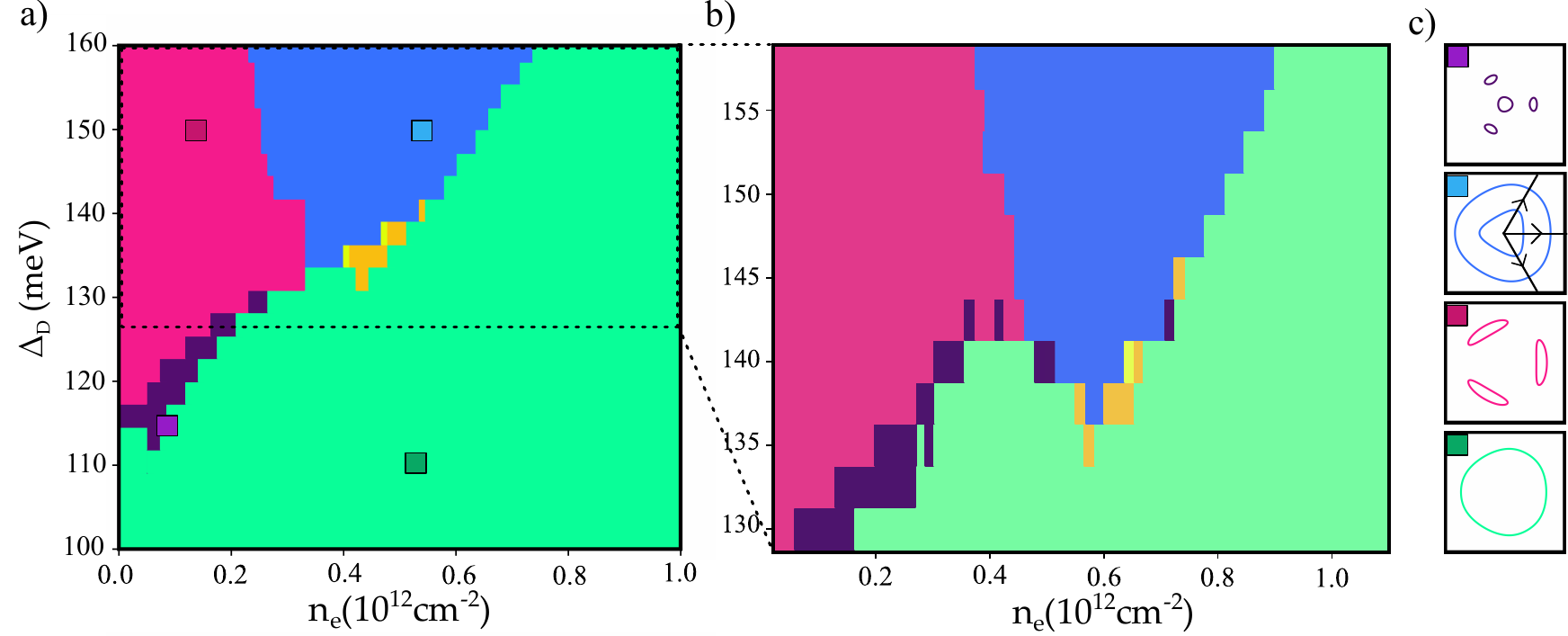}   
\end{center}
\caption{\textbf{Topology of the Fermi surfaces} as a function of the external electric field and the electronic density for \textbf{(a)} the noninteracting model and \textbf{(b)} interacting model (sHF). \textbf{(c)} Representative Fermi surfaces at external potential, $\Delta_{D}$, and electron density, $n_{e}$, pointed out by the colored squares in (a). Yellow and orange regions in panels (a) and (b) are intermediate regions where the Fermi surface is between annular and circular.}
\label{fig:FiguraSM5}
\end{figure}

Fig. ~\ref{fig:FiguraSM5} illustrates the variation in the Fermi surface topology as a function of the displacement field and electronic density for the noninteracting case,  Fig.~\ref{fig:FiguraSM5}(a), and the sHF case, Fig.~\ref{fig:FiguraSM5}(b). The color scheme represents specific Fermi surface shapes, as indicated in Fig.~\ref{fig:FiguraSM5}(c). In order to differentiate the different topologies we make use of $\mathcal{C}_3$ symmetry of the energy eigenvalues and numerically obtain the number of times the Fermi surface crosses three high-symmetry axes in the Brillouin zone, shown in black arrows in Fig.~\ref{fig:FiguraSM5}(c). This method provides an scalar number that we use to distinguish the different Fermi surfaces and hence the different colors. Orange and yellow regions are intermediate shapes between annular and circular Fermi surfaces that appear when pockets merge with each other right at the transition. Therefore, the boundaries surrounding each colored region correspond to Lifshitz transitions. 

In the green region of both panels, the Fermi surface consists of a single, quasi-circular pocket slightly distorted by trigonal warping. As the displacement field increases and at different densities, a series of Lifshitz transitions occurs, leading to distinct Fermi surface topologies. In the transition from green to purple, the Fermi surface changes from a quasi-circular pocket to four pockets. Further increasing the displacement field results in a second transition where the Fermi surface evolves from four pockets to three enlarged pockets. At intermediate densities (with respect to the density window in each panel), the blue region corresponds to an annular Fermi surface. In the transition from blue to green, some regions, colored in orange and not shown in Fig.~\ref{fig:FiguraSM5}(c), represent intermediate topologies between circular and annular. For very large electronic densities, the circular Fermi surface becomes dominant. 

\section{Results for Superconductivity in the Non-Interacting Model} \label{SM: SC NonInteracting}

\begin{figure}[t] 
\begin{center}
\includegraphics[width=0.95\textwidth]{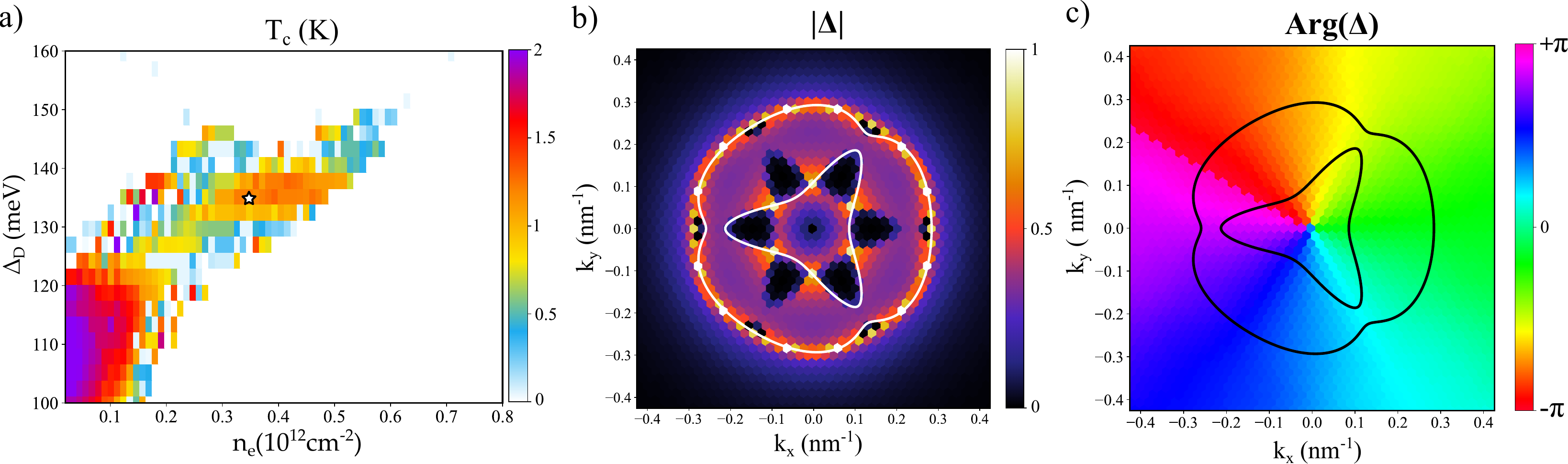}    
\end{center}
\caption{\textbf{Superconductivity in the non-interacting system.} \textbf{(a)} Critical temperature as a function of the displacement field and electronic density. \textbf{(b)} Order parameter in the non-interacting model at external potential $\Delta_{\mathrm{D}}=135$ meV and electron density $\text{n}_{e}=0.38 \times 10^{12}\text{cm}^{-2}$ marked by a star in (a). In this case, the order parameter is complex featuring a winding phase.}
\label{fig:FiguraSM6} 
\end{figure}

Fig.~\ref{fig:FiguraSM6} presents the superconductivity calculations for the noninteracting single-valley model. We followed the standard procedure used in our previous works~\cite{cea2022superconductivity, Cea2023Superconductivity, JimenoPozo2023,  ZiyanLi2023, Herrera2024Topological}. These non-interacting results align with those reported in earlier studies~\cite{Chou_Intravalley_tetralayer_2024, yang_topological_2024, Geier_isospin_tetralayer_2024, qin_chiral_2024}. Specifically, the variation in the critical temperature shown in Fig.~\ref{fig:FiguraSM6}(a) is consistent with the findings of Yang et al.~\cite{yang_topological_2024}. Similarly, the $p$-wave intravalley order parameter depicted in Fig.~\ref{fig:FiguraSM6}(b) closely matches the results reported by Geier et al.~\cite{Geier_isospin_tetralayer_2024}. The maximum critical temperature is about $2.5$ K with a dielectric constant $\epsilon_{c} =6$. Note that this large critical temperature is obtained for a single flavor calculation as in Refs.~\cite{JimenoPozo2023,ZiyanLi2023}. Note that, as described in the main text and in the following sections, a full self-consistent Fock calculations is required to obtain the polarized phases and the corresponding superconducting states. 

\section{Results for the sHF Screened Potential}  \label{SM: HF RPA Screened Potential}

\begin{figure}[ht] 
\begin{center}
\includegraphics[width=0.5\textwidth]{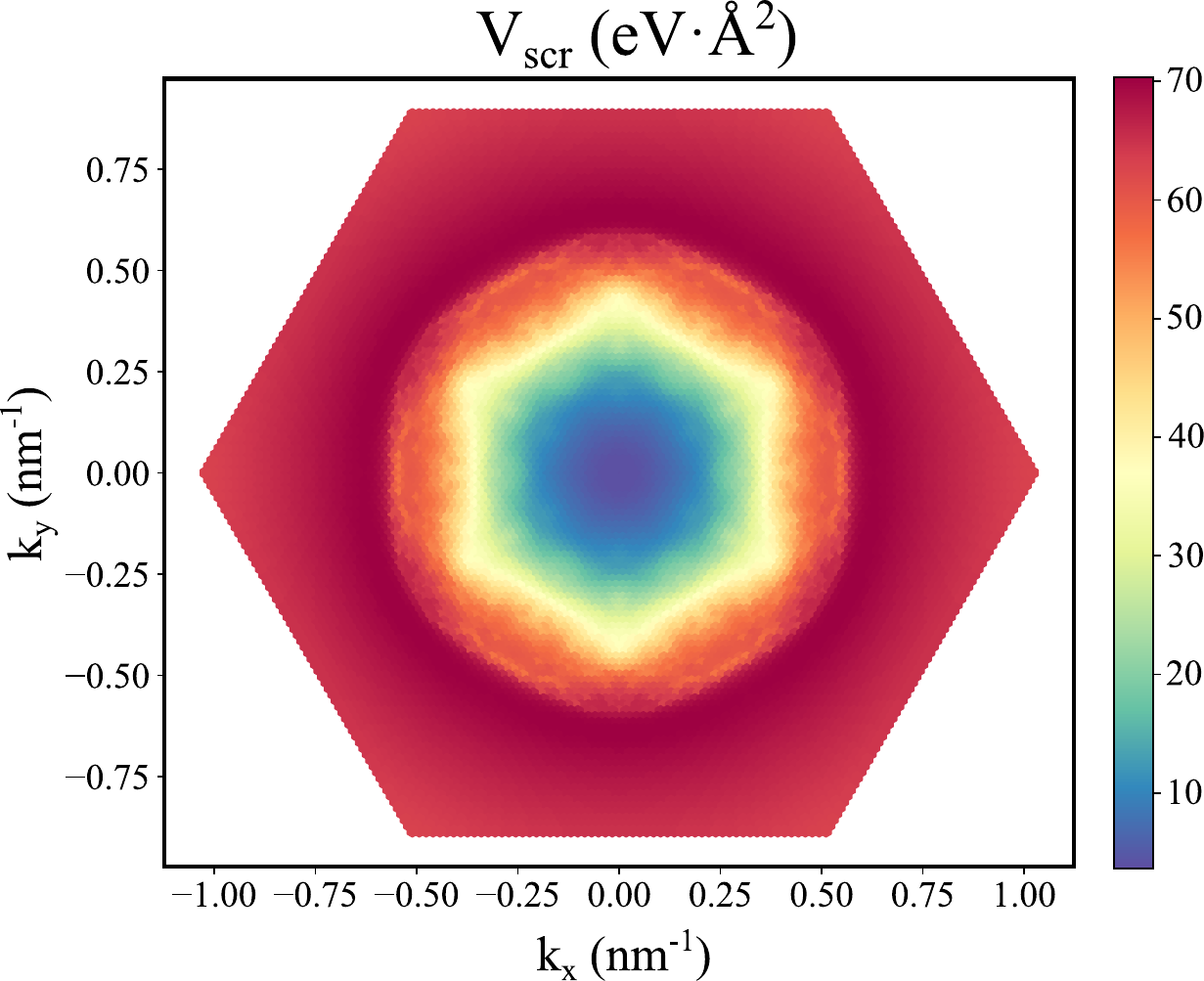}
\end{center}
\caption{\textbf{Screened Coulomb potential} for the superconducting state indicated by a star in Fig. 3 of the main text with corresponding order parameter shown in Fig. 4 of the main text.}
\label{fig:FiguraSM7}
\end{figure}

In Fig.~\ref{fig:FiguraSM7}, we show the screened potential obtained with Eq. (4) of the main text. The Coulomb potential is screened by the interacting HF bands. The potential has a minimum at the center and rapidly increases for large momenta. As described in the main text and in our previous works for twisted~\cite{cea21Coulomb,phong2021band, long_evolution_2024} and non-twisted systems~\cite{Ghazaryan2021Anular, cea2022superconductivity, JimenoPozo2023, Pantaleon2023ReviewSC, ZiyanLi2023}, this condition is favorable for the emergence of a superconducting state. 
We have checked that our sHF calculations are converged with a discretization of the grid in reciprocal space of $N_k\sim12000$ $\vb{k}$-points.

\section{Band Dispersion and Interactions at Low Densities} \label{SM: Band Low}

It is instructive to expand the band structure of the conduction band at low momenta, $\vb{k} \rightarrow \mathbf{0}$, which determines the properties at low densities. Following~\cite{koshino_trigonal_2009}, we write an effective $2 \times 2$ Hamiltonian:
\begin{align}
   {\cal H}^{eff}_{\vb{k}} &= \left( \begin{array}{cc} V_{E} & X ( \vb{k} )
   \\
  X^* ( \vb{k} )&- V_{E} \end{array} \right),
\end{align}
where $V_{E}$ is a potential difference between the two effective layers and
\begin{align}
    X ( \vb{k} ) &\approx
    \frac{v_F \gamma_2}{\gamma_1 } | \vb{k} | e^{i \phi_{\vb{k}}}
   + 3 \frac{v_3^2}{\gamma_1} | \vb{k} |^2 e^{- 2 i \phi_{\vb{k}}} 
   + \frac{v_F^2 v_3}{\gamma_1^2} | \vb{k} |^3 e^{ i \phi_{\vb{k}}} + \frac{v_F^4}{\gamma_1^3} | \vb{k} |^4 e^{4 i \phi_{\vb{k}}},
\end{align}
and $v_F = (3 \gamma_0 a ) /2$ and $v_3 = ( 3 \gamma_3 a ) /2$ and $a$ is the distance between nearest neighbor sites. We can rewrite $X ( \vb{ k} )$ as:
\begin{align}
    X ( \vb{k} ) &= \alpha | \vb{k} | e^{i \phi_{\vb{k}}} + \alpha \frac{| \vb{k} |^2}{k_0} e^{- 2 i \phi_{\vb{k}}} + 3 \alpha \frac{| \vb{k} |^3}{k_0 k_1} e^{ i \phi_{\vb{k}}}
    + \alpha \frac{| \vb{k} |^4}{k_0 k_1^2} e^{4 i \phi_{\vb{k}}},
\end{align}
where:
\begin{align}
    \alpha &= \frac{v_F \gamma_2}{\gamma_1}, \nonumber \\
    k_0 &= \frac{v_F \gamma_2}{v_3^2}, \nonumber \\
    k_1 &= \frac{v_3 \gamma_1}{v_F^2}.
\end{align}
Using the parameters in~\cite{koshino_trigonal_2009} and $V_{E} = 25$ meV, we find:
\begin{align}
    \frac{k_1}{k_0} &= \frac{v_3^3 \gamma_1}{v_F^3 \gamma_2} \approx 0.02
\end{align}
The fact that $k_1 \ll k_0$ implies that the function $X (\vb{k} )$ shows a crossover:
\begin{align}
    X ( \vb{k} ) &\approx \left\{ \begin{array}{lr} 
    \frac{v_f \gamma_2}{\gamma_1} | \vb{k} | e^{i \phi_{\vb{k}}} &| \vb{k} | \ll k_c, \\
    \frac{v_F^4}{\gamma_1^3} | \vb{k} |^4 e^{4 i \phi_{\vb{k}}} &k_c \ll | \vb{k} |,
    \end{array}
    \right.
\end{align}
where $k_c \approx  ( k_0 k_1^2 )^{1/3}$. There is a region around $| \vb{k} | \sim k_c$ with a more complicated dispersion, and where trigonal effects are relevant. The density associated to $k_c$, $\rho_c = k_c^2 / ( 4 \pi )$ (assuming that a single flavor is occupied) is $\rho \approx 3.7 \times 10^{11} {\rm cm}^{-2}$. For $| \vb{k} | \ll k_c$ the band dispersion is approximately given by:
  \begin{align}
     \epsilon_{\vb{k}} &\approx \sqrt{V_{E}^2 +  \left( \frac{v_F \gamma_2}{\gamma_1} \right)^2 | \vb{k} |^2} \approx V_{E} +  \frac{1}{2 V_{E}} \left( \frac{v_F \gamma_2}{\gamma_1} \right)^2 | \vb{k} |^2 + \cdots
  \end{align}
Thus, at low densities,$\rho \ll \rho_c$, the band is quadratic, and the system resembles a two-dimensional electron gas. We assume that the electron-electron interactions can be described by a Coulomb potential with effective coupling $e^2 / \epsilon_{c}$. Then, we can quantify the relation between the kinetic and interaction energy by a dimensionless parameter $r_s$. We obtain
\begin{align}
    r_s &= \frac{\epsilon_{c}}{e^2} \frac{2}{V_{E}} \left( \frac{\gamma_0 \gamma_2}{\gamma_1} \right)^2 \sqrt{\frac{\pi}{\rho}}, 
\end{align}
where for $\rho = \rho_c$ we obtain $r_s = 26$, so that the system is in the strongly correlated regime.   

The two-dimensional electron gas at low values of $r_s$ has been extensively studied. It is not clear, however, if a polarized phase exists in this regime. Numerical studies favor a transition from a paramagnetic metal to a Wigner crystal~\cite{loos_uniform_2016}. For completeness, we recall aspects of the Hartree-Fock solution: The size and shape of the Fermi surface do not change in a self-consistent Hartree-Fock analysis. It is a circle of radius $k_F = 2 \sqrt{\pi \rho}$, where $\rho$ is the density, and we consider that a single electron flavor is occupied. We assume that, at low densities, changes in the wave functions can be neglected, so that form factors are equal to 1. Then, the kinetic and the Fock energies per unit area can be written as:
\begin{align}
E_{kin} &= \frac{1}{2 \pi} 
\left( \frac{v_F \gamma_2}{\gamma_1} \right)^2 \frac{1}{2 V_{E}}\int_0^{2 \pi} d \theta \int_0^{k_F} dk k \times k^2 = \alpha k_F^4, \nonumber \\
    E_{ex} &= - \frac{e^2 k_F^3}{8 \epsilon_{c}} \int_0^1 u du \int_0^1 u' d u' \int_0^{2 \pi} \frac{u u'd\theta}{\sqrt{u^2 + u'^2 + 2 u u' \cos ( \theta )}}  = -\frac{e^2 k_F^3}{24 \epsilon_{c}} = - \beta k_F^3.
\end{align}
For a system where $N$ flavors are partially filled, the total energy is:
\begin{align}
    E_{tot}^N = \alpha \frac{N}{N^{4/2}} k_F^4 -\beta \frac{N}{N^{3/2}} k_F^3
\end{align}
where, as before, the value of $k_F$ is referred to the density when a single flavor is occupied.
The system will be spontaneously polarized when $E^1_{tot} \le E^N_{tot}$. This happens for:
\begin{align}
    k_F &\le \frac{\sqrt{N}}{\sqrt{N} + 1} \frac{\beta}{\alpha}
\end{align}
For $N=4$ we find that flavor polarization is favored for densities such that $\rho \le 7.9 \times 10^{12} {\rm cm}^{-1}$. At these densities, the expansion of the dispersion to second order in momentum is not valid. This result implies that flavor polarization will be favored throughout the entire region where the expansion is justified.

\begin{figure}[h]
\begin{center}
\includegraphics[width=0.45\linewidth]{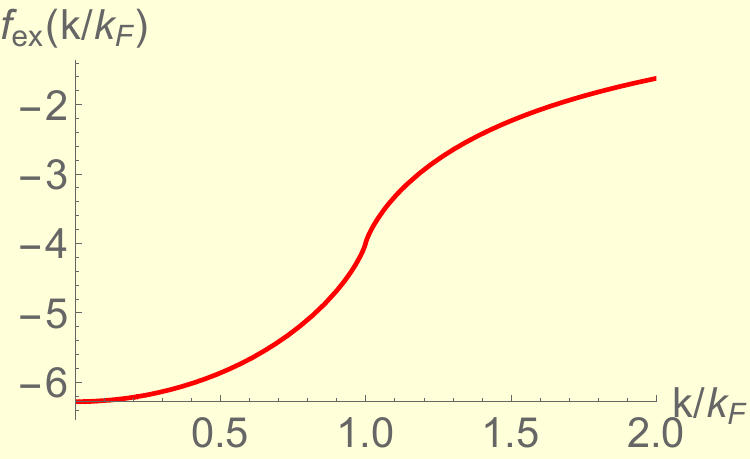}     
\end{center}
 \caption{Function $f_{ex} ( k / k_F )$ which determines the dispersion of the Hartree-Fock bands at low densities, see text.}
 \label{fig:FiguraSM8}
\end{figure}

The contribution of the Fock term to the energy bands is:
\begin{align}
    \epsilon_{\vb{k}}^{ex} &= \frac{e^2 k_F}{2 \epsilon_{c}} f_{ex} \left( \frac{| \vb{k} |}{k_F} \right)
\end{align}
where:
\begin{align}
    f_{ex} ( x ) &= -x \int_0^{2 \pi} d \theta \int_0^{1/x} d x' \frac{x'}{\sqrt{1 + x'^2 + 2 x' \cos ( \theta )}}
\end{align}
where:
\begin{align}
    f_{ex} ( 0 ) &= - 2 \pi \nonumber \\
    f_{ex} ( 1 ) &= -4 \nonumber \\
    \lim_{x \rightarrow \infty} f_{ex} ( x ) &= 0
\end{align}
The function $f_{ex} ( x )$ is plotted in Fig.~\ref{fig:FiguraSM8}. The value of $\partial f_{ex} ( x ) / \partial x$ diverges logarithmically as $x \rightarrow 1$. As a result, the density of states of the Hartree-Fock bands goes to zero as $| \vb{k} | \rightarrow k_F$. This implies that there is no Cooper instability, so that a weak coupling BCS analysis does not lead to pairing, irrespective of the mechanism.

It is worth noting that this logarithmic singularity in the velocity at the Fermi surface is a general property of the HF approximation. Let us consider a point $\vb{k}$ close to the Fermi surface. The most singular part of the integral which describes the exchange contribution to the Fock self-energy of a state with momentum $\vb{k}$ comes from the occupied states $ \vb{k}' \in {\cal S}_F$ such that $| \vec{k} - \vec{k}' | \ll k_F$ where ${\cal S}_F$ is the Fermi surface, and $k_F$ is a scale associated to the size of the Fermi surface. In order to take into account the most singular contribution to the self-energy, we approximate the Fermi surface near $\vb{k}$ by a straight line along the $k_y$ axis, between points $- \Lambda_1 \le k_y \le \Lambda_2$, where $\Lambda_1$ and $\lambda_2$ are cutoffs of order $k_F$. We consider a point $\vb{k} = ( k_x , 0 )$. Then the contribution of the region around $\vb{k}$ to the self-energy can be written as:
\begin{align}
      \delta \epsilon_F ( k_x , 0 )  &\approx 
      - \frac{e^2}{2 \pi \epsilon_{c}} \int_0^{\Lambda_x} d {k'}_x  \int_{-\Lambda_1}^{\Lambda_2} \frac{d {k'}_y}{\sqrt{( k_x - {k'}_x)^2 + {k'}_y^2}} \propto \nonumber \\
      &\propto
     - \frac{e^2}{2 \pi \epsilon_{c}}  \int_0^{\Lambda_x} d {k'}_x  
      \left[ \sinh^{-1} \left( \frac{\Lambda_2}{| k_x - {k'}_x|} \right)
      + \sinh^{-1} \left( \frac{\Lambda_1}{| k_x - {k'}_x|} \right)
      \right] \sim \nonumber \\
      &\sim - \frac{e^2 | k_x |}{2 \pi \epsilon_{c}} 
       \left[ \log \left( \frac{\Lambda_2}{|k_x|} \right)
      + \log \left( \frac{\Lambda_1}{|k_x|} \right)
      \right] + \cdots
\end{align}
where $\Lambda_x \sim k_F$, and, for sufficiently close $\vb{k} , \vb{k}'$ we can assume $| \langle \vb{k} | \vb{k}' \rangle |^2 \approx 1$. This generic presence of a singularity in the velocity at the Fermi surface makes the Hartree-Fock bands unreliable for the calculation of excited states.

\section{Smoothening Fermi Surface with Non-Screened Hartree Fock Bands}  \label{SM: FS Smooth }

A common feature of the Fock term is the smoothing of the Fermi surface. To see this, note that the Fock energy may be written as $- \frac{1}{2}\sum_{\mathbf{k}, \mathbf{q}}\mathcal{V}\left(q\right) |\Lambda_{\mathbf{k}} \left(\mathbf{q}\right)|^2 f_{\mathbf{k}} f_{\mathbf{k + q}}$ with $\Lambda$ the form factor. Typically, the interaction $\mathcal{V}\left(q\right)$ is maximal at $q = 0$ and so are the form factors, since $\Lambda_{\mathbf{k}} \left(0\right) = 1$. Hence, the Fock energy is dominated by small $q$ terms. Let us simplify the problem by considering these terms only. Suppose some state $\mathbf{k}$ is occupied, i.e., $f_{\mathbf{k}} \approx 1$ To minimize the Fock energy, states in its vicinity should be occupied as well. Alternatively, the Fermi surface, which is the interface between occupied and unoccupied regions, should be minimized. The total number of occupied states, {\it the volume} of the Fermi sea, is set by the density, $n = \int \frac{d^d k}{\left(2\pi\right)^d} f_{\mathbf{k}}$. Hence, the minimization of the Fock energy is equivalent to the construction a shape of minimal surface with a given volume. Therefore, we conclude that non-screened exchange terms tend to produce smoother, more circularly symmetric Fermi surfaces.

\section{Trigonal Warping and Gapless Superconductivity}

\begin{figure}[h]
\begin{center}
\begin{tabular}{lcr}
\includegraphics[width=0.15\linewidth]{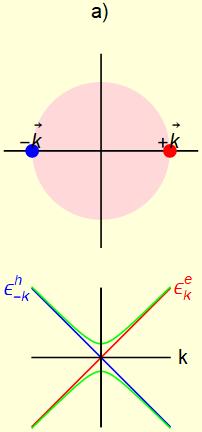} 
&
\includegraphics[width=0.15\linewidth]{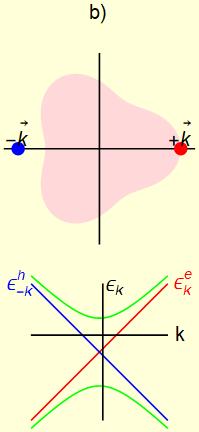} 
&
\includegraphics[width=0.15\linewidth]{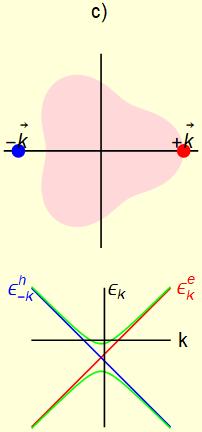}
\\
\end{tabular}
\end{center}
 \caption{\textbf{Effects of trigonal warping on superconductivity.} Top: Sketch of the Fermi surface in systems (a) without, and (b) and (c) with trigonal distortion. Bottom: Quasiparticle dispersion in the normal (red and blue lines) and superconducting phase (green line). (a) No trigonal distortion. Isotropic Fermi surface. Gapped quasiparticles in the superconducting phase. (b)  Trigonal distortion. Anisotropic Fermi surface. Large superconducting gap. Gapped quasiparticles in the superconducting phase. (c) Trigonal distortion. Anisotropic Fermi surface. Small superconducting gap. Gapless quasiparticles in the superconducting phase.}
 \label{fig:FiguraSM9}
\end{figure}

After a self-consistent calculation of the superconducting gap, $\Delta ( \vb{k} )$, the calculation of the quasiparticle bands using the Bogoliubov-de Gennes equations is reduced to the calculation of the eigenvalues of a $2 \times 2$ matrix:
\begin{align}
    {\cal H}_{\vb{k}}^{BdG} &= \left( \begin{array}{cc} \epsilon^e_{\vb{k}} &\Delta ( \vb{k} ) \\ \Delta^* ( \vb{k} ) &- \epsilon_{-\vb{k}}^h
    \end{array}
    \right)
    \label{hamilBdG}
\end{align}
where $\epsilon^e_{\vb{k}}$ and $\epsilon^h_{-\vb{k}}$ are the energies of electrons and holes in the normal phase measured with respect to the Fermi energy $\mu$. In the absence of trigonal warping, the values of $\vb{k}$ for which $\epsilon^e_{\vb{k}} = 0$ and $\epsilon^h_{\vb{k}}= 0$ are the same. As a result, the quasiparticle energies in the superconducting phase show a gap equal to $2 | \Delta ( \vb{k}) |$, as shown in Fig.~\ref{fig:FiguraSM9}(a). In the presence of trigonal warping, the values $\vb{k}_e$ and $\vb{k}_h$ where $\epsilon^e_{\vb{k}_e} = 0$ and $\epsilon^h_{\vb{k}_h} = 0$ are not the same, as shown in Fig.~\ref{fig:FiguraSM9}(b) and (c). For sufficiently small values of $| \Delta (\vb{k}) |$, the quasiparticle spectrum in the superconducting phase is gapless, as sketched in Fig.~\ref{fig:FiguraSM9}.

\section{Superconductivity Due to Electron-Phonon Coupling}  \label{SM: SC Electron Phonon }

We analyze further possible mechanisms for superconductivity in quarter filled rhombohedral graphene tetralayers. We study here the coupling to optical phonons, as the electron-phonon coupling to acoustical phonons goes to zero as $\sqrt{ | \vb{q} |} $ for low momenta $\vb{q} \rightarrow \mathbf{0}$. Simple tight binding arguments suggest that the phonons more strongly coupled to electrons are the LO and TO phonons at $\Gamma$ and the TO phonons at $K$. In the tight binding the electron-phonon coupling has the same value for these three modes~\cite{castro_neto_electron-phonon_2007,basko_theory_2008}. The TO phonons at $K$ induce intervalley transitions, which do not exist in a valley polarized phase. In graphene, band structure calculations give a value for the dimensionless coupling between electrons and optical phonons~\cite{calandra_electron-phonon_2007} of $\lambda ( n ) = 5.05 \times 10^{-9} \sqrt{n} \, {\rm }$ cm where $n$ is the electron density. We can get an estimate of the value of this parameter in a rhombohedral tetralayer by multiplying this value by 2/3, in order to take into account the absence of $K$ phonons, and replacing the density of states of graphene by the density of states of the parabolic bands which approximate the small $\vb{q}$ dispersion in the tetralayer, as discussed previously. As the density of states of a 2D parabolic band is constant, we get $\lambda \approx 0.1$, independent of density.

\section{Fourier Expansion of the Order Parameter}  \label{SM: OP Harmonics}

\begin{figure}[h]
\begin{center}
   \includegraphics[width=1\linewidth]{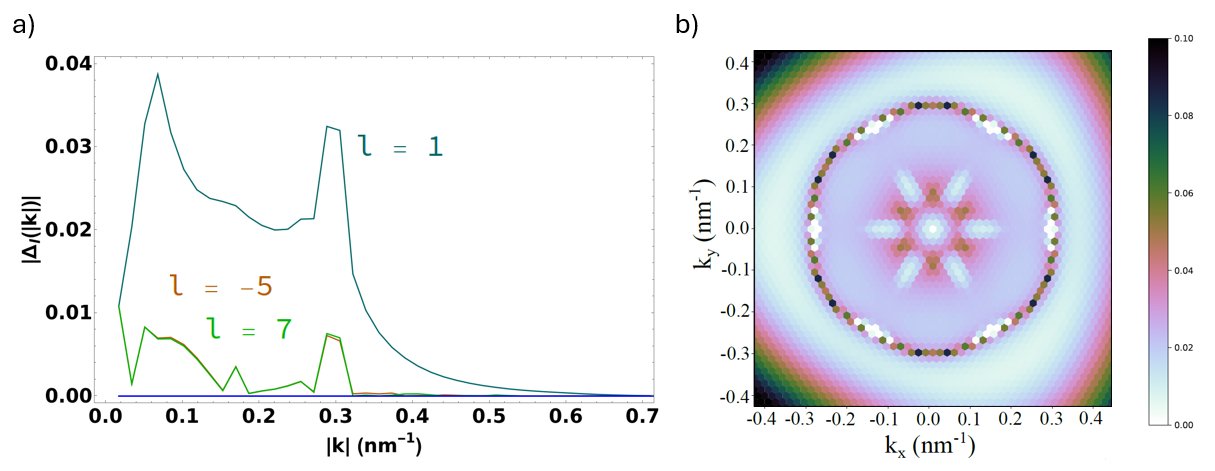}  
\end{center}
 \caption{\textbf{Fourier analysis of order parameter.} (a) Fourier coefficients for the order parameter as a function of the norm of the k-points in the grid. (b) Quantity $\mathcal{Q}(\mathbf{k})$ calculated from the order parameter. Here, the order parameter refers to that in Fig. 4 of the main text.}
 \label{fig:op_analysis}
\end{figure}

We can extract the symmetry of the order parameter by performing a Fourier decomposition of it as
\begin{equation}
    \Delta(\vb{k}) = \sum_{l} \Delta_{l}(|\vb{k}|) e^{\mathrm{i} l \theta_{\vb{k}}},
\end{equation}
with $\theta_{\vb{k}} = \text{atan2}(k_{y}/k_{x})$. The Fourier coefficients are given by
\begin{equation}
    \Delta_{l}(|\vb{k}|) = \frac{1}{N_{|\vb{k}|}}\sum_{|\vb{k}|=|\vb{k}'|} \Delta(\vb{k}') e^{-\mathrm{i} l \theta_{\vb{k}'}},
\end{equation}
where $N_{|\mathbf{k}|}$ is the number of points included in the contour of constant $|\vb{k}|$.
The results of performing such an analysis to the complex order parameter shown in Fig. 4 of the main text are given in Fig.~\ref{fig:op_analysis}(a). First, we find that the largest component for any $|\mathbf{k}|$ is $l = +1,$ which confirms the $p$-wave nature of the order parameter seen as a counterclockwise phase winding shown in Fig. 4(b) of the main text. However, unlike in a strictly $p$-wave superconductor, the order parameter contains \textit{radial} modulation in $|\mathbf{k}|$ and peaks around $|\mathbf{k}| \approx 0.065$ nm$^{-1}$ and 
$|\mathbf{k}| \approx 0.3$ nm$^{-1}$. Furthermore, additional Fourier harmonics beyond $l = +1$ are also present due to the \textit{angular} modulation of the magnitude of the order parameter. In particular, when we factor out the largest angular dependence $\Delta(\vb{k}) = e^{i\theta_\mathbf{k}} \left[\sum_{l} \Delta_{l}(|\vb{k}|) e^{\mathrm{i} (l-1) \theta_{\vb{k}}}\right],$ we find that, up to an overall $|\mathbf{k}|$-independent phase, the term in the bracket is approximately real. This requires $\Delta_{l}(|\mathbf{k}|) \approx \Delta_{-l+2}(|\mathbf{k}|)^*.$ Consequently, we find that the next largest Fourier components contributing to the order parameter are $l = -5, 7,$ and they have approximately equal amplitude as required. Also, we notice that these two components peak at $|\mathbf{k}| \approx 0.065$ nm$^{-1}$ and $|\mathbf{k}| \approx 0.3$ nm$^{-1}$ too because at those values of $|\mathbf{k}|$, the order parameter has an approximate six-fold symmetry in its magnitude. In the region between  $|\mathbf{k}| \approx 0.2$ nm$^{-1}$ and $|\mathbf{k}| \approx 0.3$ nm$^{-1},$ the  $l = -5,7$ components are suppressed relative to the $l = +1$ component because the order parameter has much less variation in its amplitude. The $l = -2,4$ components are allowed by $\mathcal{C}_3$ symmetry, but they are greatly suppressed due to the emergent $\mathcal{C}_6$ symmetry. Numerically, we find that the $l = -2,4$ components are much smaller than the other $l = -5,1,7$ components but they are definitely much larger than machine precision of zero, confirming that $\mathcal{C}_6$ is \textit{not} actually a symmetry of the system and is only approximate. On the other hand, the other components that are prohibited by $\mathcal{C}_3,$ we find, are numerically exactly zero to machine precision.

Finally, in order to reveal the nodal character of the order parameter, we also compute the following quantity:
\begin{equation}
    \mathcal{Q}(\vb{k}) = \sqrt{(E_{\mathbf{k}}-\mu)^{2} + \Delta(\mathbf{k})^{2}}.
\end{equation}
This quantity has the form of the quasiparticle spectrum. However, since we have only computed the order parameter at $T_c,$ $\Delta(\mathbf{k})$ is not actually the gap parameter but should be proportional to it. Because of that, we can infer the nodes of the quasiparticle spectrum from the nodes of $\mathcal{Q}(\vb{k}).$ This quantity $\mathcal{Q}(\vb{k})$ computed with $\Delta(\mathbf{k})$ in Fig. 4 of the main text is shown in Fig.~\ref{fig:op_analysis}(b). We note that several approximate zeroes appear in $\mathcal{Q}(\vb{k})$ at $\vb{k}$ on the Fermi surface related by $\mathcal{C}_{6}$ symmetry. These nodes, of course, correspond to the nodes in $\Delta(\mathbf{k}).$ Therefore, as far as we can ascertain numerically on a finite mesh, the order parameter shown in Fig. 4 of the main text corresponds to a gapless superconductor.

\begin{thebibliography}{100}%
\makeatletter
\providecommand \@ifxundefined [1]{%
 \@ifx{#1\undefined}
}%
\providecommand \@ifnum [1]{%
 \ifnum #1\expandafter \@firstoftwo
 \else \expandafter \@secondoftwo
 \fi
}%
\providecommand \@ifx [1]{%
 \ifx #1\expandafter \@firstoftwo
 \else \expandafter \@secondoftwo
 \fi
}%
\providecommand \natexlab [1]{#1}%
\providecommand \enquote  [1]{``#1''}%
\providecommand \bibnamefont  [1]{#1}%
\providecommand \bibfnamefont [1]{#1}%
\providecommand \citenamefont [1]{#1}%
\providecommand \href@noop [0]{\@secondoftwo}%
\providecommand \href [0]{\begingroup \@sanitize@url \@href}%
\providecommand \@href[1]{\@@startlink{#1}\@@href}%
\providecommand \@@href[1]{\endgroup#1\@@endlink}%
\providecommand \@sanitize@url [0]{\catcode `\\12\catcode `\$12\catcode `\&12\catcode `\#12\catcode `\^12\catcode `\_12\catcode `\%12\relax}%
\providecommand \@@startlink[1]{}%
\providecommand \@@endlink[0]{}%
\providecommand \url  [0]{\begingroup\@sanitize@url \@url }%
\providecommand \@url [1]{\endgroup\@href {#1}{\urlprefix }}%
\providecommand \urlprefix  [0]{URL }%
\providecommand \Eprint [0]{\href }%
\providecommand \doibase [0]{https://doi.org/}%
\providecommand \selectlanguage [0]{\@gobble}%
\providecommand \bibinfo  [0]{\@secondoftwo}%
\providecommand \bibfield  [0]{\@secondoftwo}%
\providecommand \translation [1]{[#1]}%
\providecommand \BibitemOpen [0]{}%
\providecommand \bibitemStop [0]{}%
\providecommand \bibitemNoStop [0]{.\EOS\space}%
\providecommand \EOS [0]{\spacefactor3000\relax}%
\providecommand \BibitemShut  [1]{\csname bibitem#1\endcsname}%
\let\auto@bib@innerbib\@empty
\bibitem [{\citenamefont {Cao}\ \emph {et~al.}(2018)\citenamefont {Cao}, \citenamefont {Fatemi}, \citenamefont {Fang}, \citenamefont {Watanabe}, \citenamefont {Taniguchi}, \citenamefont {Kaxiras},\ and\ \citenamefont {Jarillo-Herrero}}]{Cetal18b}%
  \BibitemOpen
  \bibfield  {author} {\bibinfo {author} {\bibfnamefont {Y.}~\bibnamefont {Cao}}, \bibinfo {author} {\bibfnamefont {V.}~\bibnamefont {Fatemi}}, \bibinfo {author} {\bibfnamefont {S.}~\bibnamefont {Fang}}, \bibinfo {author} {\bibfnamefont {K.}~\bibnamefont {Watanabe}}, \bibinfo {author} {\bibfnamefont {T.}~\bibnamefont {Taniguchi}}, \bibinfo {author} {\bibfnamefont {E.}~\bibnamefont {Kaxiras}},\ and\ \bibinfo {author} {\bibfnamefont {P.}~\bibnamefont {Jarillo-Herrero}},\ }\bibfield  {title} {\bibinfo {title} {Unconventional superconductivity in magic-angle graphene superlattices},\ }\href {https://doi.org/10.1038/nature26160} {\bibfield  {journal} {\bibinfo  {journal} {Nature}\ }\textbf {\bibinfo {volume} {556}},\ \bibinfo {pages} {43} (\bibinfo {year} {2018})}\BibitemShut {NoStop}%
\bibitem [{\citenamefont {Yankowitz}\ \emph {et~al.}(2019)\citenamefont {Yankowitz}, \citenamefont {Chen}, \citenamefont {Polshyn}, \citenamefont {Zhang}, \citenamefont {Watanabe}, \citenamefont {Taniguchi}, \citenamefont {Graf}, \citenamefont {Young},\ and\ \citenamefont {Dean}}]{Yankowitz2019}%
  \BibitemOpen
  \bibfield  {author} {\bibinfo {author} {\bibfnamefont {M.}~\bibnamefont {Yankowitz}}, \bibinfo {author} {\bibfnamefont {S.}~\bibnamefont {Chen}}, \bibinfo {author} {\bibfnamefont {H.}~\bibnamefont {Polshyn}}, \bibinfo {author} {\bibfnamefont {Y.}~\bibnamefont {Zhang}}, \bibinfo {author} {\bibfnamefont {K.}~\bibnamefont {Watanabe}}, \bibinfo {author} {\bibfnamefont {T.}~\bibnamefont {Taniguchi}}, \bibinfo {author} {\bibfnamefont {D.}~\bibnamefont {Graf}}, \bibinfo {author} {\bibfnamefont {A.~F.}\ \bibnamefont {Young}},\ and\ \bibinfo {author} {\bibfnamefont {C.~R.}\ \bibnamefont {Dean}},\ }\bibfield  {title} {\bibinfo {title} {Tuning superconductivity in twisted bilayer graphene},\ }\href {https://doi.org/10.1126/science.aav1910} {\bibfield  {journal} {\bibinfo  {journal} {Science}\ }\textbf {\bibinfo {volume} {363}},\ \bibinfo {pages} {1059} (\bibinfo {year} {2019})}\BibitemShut {NoStop}%
\bibitem [{\citenamefont {Lu}\ \emph {et~al.}(2019)\citenamefont {Lu}, \citenamefont {Stepanov}, \citenamefont {Yang}, \citenamefont {Xie}, \citenamefont {Aamir}, \citenamefont {Das}, \citenamefont {Urgell}, \citenamefont {Watanabe}, \citenamefont {Taniguchi}, \citenamefont {Zhang}, \citenamefont {Bachtold}, \citenamefont {MacDonald},\ and\ \citenamefont {Efetov}}]{Lu2019_SC_TBG}%
  \BibitemOpen
  \bibfield  {author} {\bibinfo {author} {\bibfnamefont {X.}~\bibnamefont {Lu}}, \bibinfo {author} {\bibfnamefont {P.}~\bibnamefont {Stepanov}}, \bibinfo {author} {\bibfnamefont {W.}~\bibnamefont {Yang}}, \bibinfo {author} {\bibfnamefont {M.}~\bibnamefont {Xie}}, \bibinfo {author} {\bibfnamefont {M.~A.}\ \bibnamefont {Aamir}}, \bibinfo {author} {\bibfnamefont {I.}~\bibnamefont {Das}}, \bibinfo {author} {\bibfnamefont {C.}~\bibnamefont {Urgell}}, \bibinfo {author} {\bibfnamefont {K.}~\bibnamefont {Watanabe}}, \bibinfo {author} {\bibfnamefont {T.}~\bibnamefont {Taniguchi}}, \bibinfo {author} {\bibfnamefont {G.}~\bibnamefont {Zhang}}, \bibinfo {author} {\bibfnamefont {A.}~\bibnamefont {Bachtold}}, \bibinfo {author} {\bibfnamefont {A.~H.}\ \bibnamefont {MacDonald}},\ and\ \bibinfo {author} {\bibfnamefont {D.~K.}\ \bibnamefont {Efetov}},\ }\bibfield  {title} {\bibinfo {title} {Superconductors, orbital magnets and correlated states in magic-angle bilayer graphene},\ }\href
  {https://doi.org/10.1038/s41586-019-1695-0} {\bibfield  {journal} {\bibinfo  {journal} {Nature}\ }\textbf {\bibinfo {volume} {574}},\ \bibinfo {pages} {653} (\bibinfo {year} {2019})}\BibitemShut {NoStop}%
\bibitem [{\citenamefont {Stepanov}\ \emph {et~al.}(2020)\citenamefont {Stepanov}, \citenamefont {Das}, \citenamefont {Lu}, \citenamefont {Fahimniya}, \citenamefont {Watanabe}, \citenamefont {Taniguchi}, \citenamefont {Koppens}, \citenamefont {Lischner}, \citenamefont {Levitov},\ and\ \citenamefont {Efetov}}]{Stepanov2020_TBG}%
  \BibitemOpen
  \bibfield  {author} {\bibinfo {author} {\bibfnamefont {P.}~\bibnamefont {Stepanov}}, \bibinfo {author} {\bibfnamefont {I.}~\bibnamefont {Das}}, \bibinfo {author} {\bibfnamefont {X.}~\bibnamefont {Lu}}, \bibinfo {author} {\bibfnamefont {A.}~\bibnamefont {Fahimniya}}, \bibinfo {author} {\bibfnamefont {K.}~\bibnamefont {Watanabe}}, \bibinfo {author} {\bibfnamefont {T.}~\bibnamefont {Taniguchi}}, \bibinfo {author} {\bibfnamefont {F.~H.~L.}\ \bibnamefont {Koppens}}, \bibinfo {author} {\bibfnamefont {J.}~\bibnamefont {Lischner}}, \bibinfo {author} {\bibfnamefont {L.}~\bibnamefont {Levitov}},\ and\ \bibinfo {author} {\bibfnamefont {D.~K.}\ \bibnamefont {Efetov}},\ }\bibfield  {title} {\bibinfo {title} {Untying the insulating and superconducting orders in magic-angle graphene},\ }\href {https://doi.org/10.1038/s41586-020-2459-6} {\bibfield  {journal} {\bibinfo  {journal} {Nature}\ }\textbf {\bibinfo {volume} {583}},\ \bibinfo {pages} {375} (\bibinfo {year} {2020})}\BibitemShut {NoStop}%
\bibitem [{\citenamefont {Oh}\ \emph {et~al.}(2021)\citenamefont {Oh}, \citenamefont {Nuckolls}, \citenamefont {Wong}, \citenamefont {Lee}, \citenamefont {Liu}, \citenamefont {Watanabe}, \citenamefont {Taniguchi},\ and\ \citenamefont {Yazdani}}]{Oh2021UnconvSC}%
  \BibitemOpen
  \bibfield  {author} {\bibinfo {author} {\bibfnamefont {M.}~\bibnamefont {Oh}}, \bibinfo {author} {\bibfnamefont {K.~P.}\ \bibnamefont {Nuckolls}}, \bibinfo {author} {\bibfnamefont {D.}~\bibnamefont {Wong}}, \bibinfo {author} {\bibfnamefont {R.~L.}\ \bibnamefont {Lee}}, \bibinfo {author} {\bibfnamefont {X.}~\bibnamefont {Liu}}, \bibinfo {author} {\bibfnamefont {K.}~\bibnamefont {Watanabe}}, \bibinfo {author} {\bibfnamefont {T.}~\bibnamefont {Taniguchi}},\ and\ \bibinfo {author} {\bibfnamefont {A.}~\bibnamefont {Yazdani}},\ }\bibfield  {title} {\bibinfo {title} {Evidence for unconventional superconductivity in twisted bilayer graphene},\ }\href {https://doi.org/10.1038/s41586-021-04121-x} {\bibfield  {journal} {\bibinfo  {journal} {Nature}\ }\textbf {\bibinfo {volume} {600}},\ \bibinfo {pages} {240–245} (\bibinfo {year} {2021})}\BibitemShut {NoStop}%
\bibitem [{\citenamefont {Park}\ \emph {et~al.}(2021)\citenamefont {Park}, \citenamefont {Cao}, \citenamefont {Watanabe}, \citenamefont {Taniguchi},\ and\ \citenamefont {Jarillo-Herrero}}]{Park2021_SC_TTG}%
  \BibitemOpen
  \bibfield  {author} {\bibinfo {author} {\bibfnamefont {J.~M.}\ \bibnamefont {Park}}, \bibinfo {author} {\bibfnamefont {Y.}~\bibnamefont {Cao}}, \bibinfo {author} {\bibfnamefont {K.}~\bibnamefont {Watanabe}}, \bibinfo {author} {\bibfnamefont {T.}~\bibnamefont {Taniguchi}},\ and\ \bibinfo {author} {\bibfnamefont {P.}~\bibnamefont {Jarillo-Herrero}},\ }\bibfield  {title} {\bibinfo {title} {Tunable strongly coupled superconductivity in magic-angle twisted trilayer graphene},\ }\href {https://doi.org/10.1038/s41586-021-03192-0} {\bibfield  {journal} {\bibinfo  {journal} {Nature}\ }\textbf {\bibinfo {volume} {590}},\ \bibinfo {pages} {249} (\bibinfo {year} {2021})}\BibitemShut {NoStop}%
\bibitem [{\citenamefont {Hao}\ \emph {et~al.}(2021)\citenamefont {Hao}, \citenamefont {Zimmerman}, \citenamefont {Ledwith}, \citenamefont {Khalaf}, \citenamefont {Najafabadi}, \citenamefont {Watanabe}, \citenamefont {Taniguchi}, \citenamefont {Vishwanath},\ and\ \citenamefont {Kim}}]{Hao2021_SC_TTG}%
  \BibitemOpen
  \bibfield  {author} {\bibinfo {author} {\bibfnamefont {Z.}~\bibnamefont {Hao}}, \bibinfo {author} {\bibfnamefont {A.~M.}\ \bibnamefont {Zimmerman}}, \bibinfo {author} {\bibfnamefont {P.}~\bibnamefont {Ledwith}}, \bibinfo {author} {\bibfnamefont {E.}~\bibnamefont {Khalaf}}, \bibinfo {author} {\bibfnamefont {D.~H.}\ \bibnamefont {Najafabadi}}, \bibinfo {author} {\bibfnamefont {K.}~\bibnamefont {Watanabe}}, \bibinfo {author} {\bibfnamefont {T.}~\bibnamefont {Taniguchi}}, \bibinfo {author} {\bibfnamefont {A.}~\bibnamefont {Vishwanath}},\ and\ \bibinfo {author} {\bibfnamefont {P.}~\bibnamefont {Kim}},\ }\bibfield  {title} {\bibinfo {title} {Electric field{\textendash}tunable superconductivity in alternating-twist magic-angle trilayer graphene},\ }\href {https://doi.org/10.1126/science.abg0399} {\bibfield  {journal} {\bibinfo  {journal} {Science}\ }\textbf {\bibinfo {volume} {371}},\ \bibinfo {pages} {1133} (\bibinfo {year} {2021})}\BibitemShut {NoStop}%
\bibitem [{\citenamefont {Kim}\ \emph {et~al.}(2022)\citenamefont {Kim}, \citenamefont {Choi}, \citenamefont {Lewandowski}, \citenamefont {Thomson}, \citenamefont {Zhang}, \citenamefont {Polski}, \citenamefont {Watanabe}, \citenamefont {Taniguchi}, \citenamefont {Alicea},\ and\ \citenamefont {Nadj-Perge}}]{Kim2022evidence}%
  \BibitemOpen
  \bibfield  {author} {\bibinfo {author} {\bibfnamefont {H.}~\bibnamefont {Kim}}, \bibinfo {author} {\bibfnamefont {Y.}~\bibnamefont {Choi}}, \bibinfo {author} {\bibfnamefont {C.}~\bibnamefont {Lewandowski}}, \bibinfo {author} {\bibfnamefont {A.}~\bibnamefont {Thomson}}, \bibinfo {author} {\bibfnamefont {Y.}~\bibnamefont {Zhang}}, \bibinfo {author} {\bibfnamefont {R.}~\bibnamefont {Polski}}, \bibinfo {author} {\bibfnamefont {K.}~\bibnamefont {Watanabe}}, \bibinfo {author} {\bibfnamefont {T.}~\bibnamefont {Taniguchi}}, \bibinfo {author} {\bibfnamefont {J.}~\bibnamefont {Alicea}},\ and\ \bibinfo {author} {\bibfnamefont {S.}~\bibnamefont {Nadj-Perge}},\ }\bibfield  {title} {\bibinfo {title} {Evidence for unconventional superconductivity in twisted trilayer graphene},\ }\href {https://doi.org/10.1038/s41586-022-04715-z} {\bibfield  {journal} {\bibinfo  {journal} {Nature}\ }\textbf {\bibinfo {volume} {606}},\ \bibinfo {pages} {494} (\bibinfo {year} {2022})}\BibitemShut {NoStop}%
\bibitem [{\citenamefont {Liu}\ \emph {et~al.}(2022)\citenamefont {Liu}, \citenamefont {Zhang}, \citenamefont {Watanabe}, \citenamefont {Taniguchi},\ and\ \citenamefont {Li}}]{Liu2022Isospin}%
  \BibitemOpen
  \bibfield  {author} {\bibinfo {author} {\bibfnamefont {X.}~\bibnamefont {Liu}}, \bibinfo {author} {\bibfnamefont {N.~J.}\ \bibnamefont {Zhang}}, \bibinfo {author} {\bibfnamefont {K.}~\bibnamefont {Watanabe}}, \bibinfo {author} {\bibfnamefont {T.}~\bibnamefont {Taniguchi}},\ and\ \bibinfo {author} {\bibfnamefont {J.~I.~A.}\ \bibnamefont {Li}},\ }\bibfield  {title} {\bibinfo {title} {Isospin order in superconducting magic-angle twisted trilayer graphene},\ }\href {http://dx.doi.org/10.1038/s41567-022-01515-0} {\bibfield  {journal} {\bibinfo  {journal} {Nature Physics}\ }\textbf {\bibinfo {volume} {18}},\ \bibinfo {pages} {522–527} (\bibinfo {year} {2022})}\BibitemShut {NoStop}%
\bibitem [{\citenamefont {Uri}\ \emph {et~al.}(2023)\citenamefont {Uri}, \citenamefont {de~la Barrera}, \citenamefont {Randeria}, \citenamefont {Rodan-Legrain}, \citenamefont {Devakul}, \citenamefont {Crowley}, \citenamefont {Paul}, \citenamefont {Watanabe}, \citenamefont {Taniguchi}, \citenamefont {Lifshitz}, \citenamefont {Fu}, \citenamefont {Ashoori},\ and\ \citenamefont {Jarillo-Herrero}}]{Uri2023Superconductivity}%
  \BibitemOpen
  \bibfield  {author} {\bibinfo {author} {\bibfnamefont {A.}~\bibnamefont {Uri}}, \bibinfo {author} {\bibfnamefont {S.~C.}\ \bibnamefont {de~la Barrera}}, \bibinfo {author} {\bibfnamefont {M.~T.}\ \bibnamefont {Randeria}}, \bibinfo {author} {\bibfnamefont {D.}~\bibnamefont {Rodan-Legrain}}, \bibinfo {author} {\bibfnamefont {T.}~\bibnamefont {Devakul}}, \bibinfo {author} {\bibfnamefont {P.~J.~D.}\ \bibnamefont {Crowley}}, \bibinfo {author} {\bibfnamefont {N.}~\bibnamefont {Paul}}, \bibinfo {author} {\bibfnamefont {K.}~\bibnamefont {Watanabe}}, \bibinfo {author} {\bibfnamefont {T.}~\bibnamefont {Taniguchi}}, \bibinfo {author} {\bibfnamefont {R.}~\bibnamefont {Lifshitz}}, \bibinfo {author} {\bibfnamefont {L.}~\bibnamefont {Fu}}, \bibinfo {author} {\bibfnamefont {R.~C.}\ \bibnamefont {Ashoori}},\ and\ \bibinfo {author} {\bibfnamefont {P.}~\bibnamefont {Jarillo-Herrero}},\ }\bibfield  {title} {\bibinfo {title} {Superconductivity and strong interactions in a tunable moir{\'{e}} quasicrystal},\ }\href
  {https://doi.org/10.1038/s41586-023-06294-z} {\bibfield  {journal} {\bibinfo  {journal} {Nature}\ }\textbf {\bibinfo {volume} {620}},\ \bibinfo {pages} {762} (\bibinfo {year} {2023})}\BibitemShut {NoStop}%
\bibitem [{\citenamefont {Park}\ \emph {et~al.}(2022)\citenamefont {Park}, \citenamefont {Cao}, \citenamefont {Xia}, \citenamefont {Sun}, \citenamefont {Watanabe}, \citenamefont {Taniguchi},\ and\ \citenamefont {Jarillo-Herrero}}]{Park2022_multi}%
  \BibitemOpen
  \bibfield  {author} {\bibinfo {author} {\bibfnamefont {J.~M.}\ \bibnamefont {Park}}, \bibinfo {author} {\bibfnamefont {Y.}~\bibnamefont {Cao}}, \bibinfo {author} {\bibfnamefont {L.-Q.}\ \bibnamefont {Xia}}, \bibinfo {author} {\bibfnamefont {S.}~\bibnamefont {Sun}}, \bibinfo {author} {\bibfnamefont {K.}~\bibnamefont {Watanabe}}, \bibinfo {author} {\bibfnamefont {T.}~\bibnamefont {Taniguchi}},\ and\ \bibinfo {author} {\bibfnamefont {P.}~\bibnamefont {Jarillo-Herrero}},\ }\bibfield  {title} {\bibinfo {title} {Robust superconductivity in magic-angle multilayer graphene family},\ }\href {https://doi.org/10.1038/s41563-022-01287-1} {\bibfield  {journal} {\bibinfo  {journal} {Nature Materials}\ }\textbf {\bibinfo {volume} {21}},\ \bibinfo {pages} {877} (\bibinfo {year} {2022})}\BibitemShut {NoStop}%
\bibitem [{\citenamefont {Zhang}\ \emph {et~al.}(2022)\citenamefont {Zhang}, \citenamefont {Polski}, \citenamefont {Lewandowski}, \citenamefont {Thomson}, \citenamefont {Peng}, \citenamefont {Choi}, \citenamefont {Kim}, \citenamefont {Watanabe}, \citenamefont {Taniguchi}, \citenamefont {Alicea}, \citenamefont {von Oppen}, \citenamefont {Refael},\ and\ \citenamefont {Nadj-Perge}}]{Zhang2022_promotionSC}%
  \BibitemOpen
  \bibfield  {author} {\bibinfo {author} {\bibfnamefont {Y.}~\bibnamefont {Zhang}}, \bibinfo {author} {\bibfnamefont {R.}~\bibnamefont {Polski}}, \bibinfo {author} {\bibfnamefont {C.}~\bibnamefont {Lewandowski}}, \bibinfo {author} {\bibfnamefont {A.}~\bibnamefont {Thomson}}, \bibinfo {author} {\bibfnamefont {Y.}~\bibnamefont {Peng}}, \bibinfo {author} {\bibfnamefont {Y.}~\bibnamefont {Choi}}, \bibinfo {author} {\bibfnamefont {H.}~\bibnamefont {Kim}}, \bibinfo {author} {\bibfnamefont {K.}~\bibnamefont {Watanabe}}, \bibinfo {author} {\bibfnamefont {T.}~\bibnamefont {Taniguchi}}, \bibinfo {author} {\bibfnamefont {J.}~\bibnamefont {Alicea}}, \bibinfo {author} {\bibfnamefont {F.}~\bibnamefont {von Oppen}}, \bibinfo {author} {\bibfnamefont {G.}~\bibnamefont {Refael}},\ and\ \bibinfo {author} {\bibfnamefont {S.}~\bibnamefont {Nadj-Perge}},\ }\bibfield  {title} {\bibinfo {title} {Promotion of superconductivity in magic-angle graphene multilayers},\ }\href {https://doi.org/10.1126/science.abn8585} {\bibfield
  {journal} {\bibinfo  {journal} {Science}\ }\textbf {\bibinfo {volume} {377}},\ \bibinfo {pages} {1538} (\bibinfo {year} {2022})}\BibitemShut {NoStop}%
\bibitem [{\citenamefont {Su}\ \emph {et~al.}(2023)\citenamefont {Su}, \citenamefont {Kuiri}, \citenamefont {Watanabe}, \citenamefont {Taniguchi},\ and\ \citenamefont {Folk}}]{Su2023Superconductivity}%
  \BibitemOpen
  \bibfield  {author} {\bibinfo {author} {\bibfnamefont {R.}~\bibnamefont {Su}}, \bibinfo {author} {\bibfnamefont {M.}~\bibnamefont {Kuiri}}, \bibinfo {author} {\bibfnamefont {K.}~\bibnamefont {Watanabe}}, \bibinfo {author} {\bibfnamefont {T.}~\bibnamefont {Taniguchi}},\ and\ \bibinfo {author} {\bibfnamefont {J.}~\bibnamefont {Folk}},\ }\bibfield  {title} {\bibinfo {title} {Superconductivity in twisted double bilayer graphene stabilized by {WSe$_{2}$}},\ }\href {http://dx.doi.org/10.1038/s41563-023-01653-7} {\bibfield  {journal} {\bibinfo  {journal} {Nature Materials}\ }\textbf {\bibinfo {volume} {22}},\ \bibinfo {pages} {1332–1337} (\bibinfo {year} {2023})}\BibitemShut {NoStop}%
\bibitem [{\citenamefont {Zhou}\ \emph {et~al.}(2021{\natexlab{a}})\citenamefont {Zhou}, \citenamefont {Xie}, \citenamefont {Taniguchi}, \citenamefont {Watanabe},\ and\ \citenamefont {Young}}]{Zhou2021SuperRTG}%
  \BibitemOpen
  \bibfield  {author} {\bibinfo {author} {\bibfnamefont {H.}~\bibnamefont {Zhou}}, \bibinfo {author} {\bibfnamefont {T.}~\bibnamefont {Xie}}, \bibinfo {author} {\bibfnamefont {T.}~\bibnamefont {Taniguchi}}, \bibinfo {author} {\bibfnamefont {K.}~\bibnamefont {Watanabe}},\ and\ \bibinfo {author} {\bibfnamefont {A.~F.}\ \bibnamefont {Young}},\ }\bibfield  {title} {\bibinfo {title} {Superconductivity in rhombohedral trilayer graphene},\ }\href {https://doi.org/10.1038/s41586-021-03926-0} {\bibfield  {journal} {\bibinfo  {journal} {Nature}\ }\textbf {\bibinfo {volume} {598}},\ \bibinfo {pages} {434} (\bibinfo {year} {2021}{\natexlab{a}})}\BibitemShut {NoStop}%
\bibitem [{\citenamefont {Patterson}\ \emph {et~al.}(2024)\citenamefont {Patterson}, \citenamefont {Sheekey}, \citenamefont {Arp}, \citenamefont {Holleis}, \citenamefont {Koh}, \citenamefont {Choi}, \citenamefont {Xie}, \citenamefont {Xu}, \citenamefont {Redekop}, \citenamefont {Babikyan}, \citenamefont {Zhou}, \citenamefont {Cheng}, \citenamefont {Taniguchi}, \citenamefont {Watanabe}, \citenamefont {Jin}, \citenamefont {Lantagne-Hurtubise}, \citenamefont {Alicea},\ and\ \citenamefont {Young}}]{patterson_superconductivity_2024}%
  \BibitemOpen
  \bibfield  {author} {\bibinfo {author} {\bibfnamefont {C.~L.}\ \bibnamefont {Patterson}}, \bibinfo {author} {\bibfnamefont {O.~I.}\ \bibnamefont {Sheekey}}, \bibinfo {author} {\bibfnamefont {T.~B.}\ \bibnamefont {Arp}}, \bibinfo {author} {\bibfnamefont {L.~F.~W.}\ \bibnamefont {Holleis}}, \bibinfo {author} {\bibfnamefont {J.~M.}\ \bibnamefont {Koh}}, \bibinfo {author} {\bibfnamefont {Y.}~\bibnamefont {Choi}}, \bibinfo {author} {\bibfnamefont {T.}~\bibnamefont {Xie}}, \bibinfo {author} {\bibfnamefont {S.}~\bibnamefont {Xu}}, \bibinfo {author} {\bibfnamefont {E.}~\bibnamefont {Redekop}}, \bibinfo {author} {\bibfnamefont {G.}~\bibnamefont {Babikyan}}, \bibinfo {author} {\bibfnamefont {H.}~\bibnamefont {Zhou}}, \bibinfo {author} {\bibfnamefont {X.}~\bibnamefont {Cheng}}, \bibinfo {author} {\bibfnamefont {T.}~\bibnamefont {Taniguchi}}, \bibinfo {author} {\bibfnamefont {K.}~\bibnamefont {Watanabe}}, \bibinfo {author} {\bibfnamefont {C.}~\bibnamefont {Jin}}, \bibinfo {author} {\bibfnamefont {E.}~\bibnamefont
  {Lantagne-Hurtubise}}, \bibinfo {author} {\bibfnamefont {J.}~\bibnamefont {Alicea}},\ and\ \bibinfo {author} {\bibfnamefont {A.~F.}\ \bibnamefont {Young}},\ }\bibfield  {title} {\bibinfo {title} {Superconductivity and spin canting in spin-orbit proximitized rhombohedral trilayer graphene},\ }\href {http://arxiv.org/abs/2408.10190} {\bibfield  {journal} {\bibinfo  {journal} {arXiv: 2408.10190}\ } (\bibinfo {year} {2024})}\BibitemShut {NoStop}%
\bibitem [{\citenamefont {Zhou}\ \emph {et~al.}(2022)\citenamefont {Zhou}, \citenamefont {Holleis}, \citenamefont {Saito}, \citenamefont {Cohen}, \citenamefont {Huynh}, \citenamefont {Patterson}, \citenamefont {Yang}, \citenamefont {Taniguchi}, \citenamefont {Watanabe},\ and\ \citenamefont {Young}}]{zhou2022isospin}%
  \BibitemOpen
  \bibfield  {author} {\bibinfo {author} {\bibfnamefont {H.}~\bibnamefont {Zhou}}, \bibinfo {author} {\bibfnamefont {L.}~\bibnamefont {Holleis}}, \bibinfo {author} {\bibfnamefont {Y.}~\bibnamefont {Saito}}, \bibinfo {author} {\bibfnamefont {L.}~\bibnamefont {Cohen}}, \bibinfo {author} {\bibfnamefont {W.}~\bibnamefont {Huynh}}, \bibinfo {author} {\bibfnamefont {C.~L.}\ \bibnamefont {Patterson}}, \bibinfo {author} {\bibfnamefont {F.}~\bibnamefont {Yang}}, \bibinfo {author} {\bibfnamefont {T.}~\bibnamefont {Taniguchi}}, \bibinfo {author} {\bibfnamefont {K.}~\bibnamefont {Watanabe}},\ and\ \bibinfo {author} {\bibfnamefont {A.~F.}\ \bibnamefont {Young}},\ }\bibfield  {title} {\bibinfo {title} {Isospin magnetism and spin-polarized superconductivity in bernal bilayer graphene},\ }\href {https://www.science.org/doi/abs/10.1126/science.abm8386} {\bibfield  {journal} {\bibinfo  {journal} {Science}\ }\textbf {\bibinfo {volume} {375}},\ \bibinfo {pages} {774} (\bibinfo {year} {2022})}\BibitemShut {NoStop}%
\bibitem [{\citenamefont {Zhang}\ \emph {et~al.}(2023)\citenamefont {Zhang}, \citenamefont {Polski}, \citenamefont {Thomson}, \citenamefont {Lantagne-Hurtubise}, \citenamefont {Lewandowski}, \citenamefont {Zhou}, \citenamefont {Watanabe}, \citenamefont {Taniguchi}, \citenamefont {Alicea},\ and\ \citenamefont {Nadj-Perge}}]{zhang2023spin}%
  \BibitemOpen
  \bibfield  {author} {\bibinfo {author} {\bibfnamefont {Y.}~\bibnamefont {Zhang}}, \bibinfo {author} {\bibfnamefont {R.}~\bibnamefont {Polski}}, \bibinfo {author} {\bibfnamefont {A.}~\bibnamefont {Thomson}}, \bibinfo {author} {\bibfnamefont {E.}~\bibnamefont {Lantagne-Hurtubise}}, \bibinfo {author} {\bibfnamefont {C.}~\bibnamefont {Lewandowski}}, \bibinfo {author} {\bibfnamefont {H.}~\bibnamefont {Zhou}}, \bibinfo {author} {\bibfnamefont {K.}~\bibnamefont {Watanabe}}, \bibinfo {author} {\bibfnamefont {T.}~\bibnamefont {Taniguchi}}, \bibinfo {author} {\bibfnamefont {J.}~\bibnamefont {Alicea}},\ and\ \bibinfo {author} {\bibfnamefont {S.}~\bibnamefont {Nadj-Perge}},\ }\bibfield  {title} {\bibinfo {title} {Enhanced superconductivity in spin{\textendash}orbit proximitized bilayer graphene},\ }\href {https://doi.org/10.1038/s41586-022-05446-x} {\bibfield  {journal} {\bibinfo  {journal} {Nature}\ }\textbf {\bibinfo {volume} {613}},\ \bibinfo {pages} {268} (\bibinfo {year} {2023})}\BibitemShut {NoStop}%
\bibitem [{\citenamefont {Holleis}\ \emph {et~al.}(2023)\citenamefont {Holleis}, \citenamefont {Patterson}, \citenamefont {Zhang}, \citenamefont {Yoo}, \citenamefont {Zhou}, \citenamefont {Taniguchi}, \citenamefont {Watanabe}, \citenamefont {Nadj-Perge},\ and\ \citenamefont {Young}}]{holleis2023SC}%
  \BibitemOpen
  \bibfield  {author} {\bibinfo {author} {\bibfnamefont {L.}~\bibnamefont {Holleis}}, \bibinfo {author} {\bibfnamefont {C.~L.}\ \bibnamefont {Patterson}}, \bibinfo {author} {\bibfnamefont {Y.}~\bibnamefont {Zhang}}, \bibinfo {author} {\bibfnamefont {H.~M.}\ \bibnamefont {Yoo}}, \bibinfo {author} {\bibfnamefont {H.}~\bibnamefont {Zhou}}, \bibinfo {author} {\bibfnamefont {T.}~\bibnamefont {Taniguchi}}, \bibinfo {author} {\bibfnamefont {K.}~\bibnamefont {Watanabe}}, \bibinfo {author} {\bibfnamefont {S.}~\bibnamefont {Nadj-Perge}},\ and\ \bibinfo {author} {\bibfnamefont {A.~F.}\ \bibnamefont {Young}},\ }\bibfield  {title} {\bibinfo {title} {Ising superconductivity and nematicity in bernal bilayer graphene with strong spin orbit coupling},\ }\href {https://arxiv.org/abs/2303.00742} {\bibfield  {journal} {\bibinfo  {journal} {arXiv: 2303.00742}\ } (\bibinfo {year} {2023})}\BibitemShut {NoStop}%
\bibitem [{\citenamefont {Han}\ \emph {et~al.}(2024)\citenamefont {Han}, \citenamefont {Lu}, \citenamefont {Yao}, \citenamefont {Shi}, \citenamefont {Yang}, \citenamefont {Seo}, \citenamefont {Ye}, \citenamefont {Wu}, \citenamefont {Zhou}, \citenamefont {Liu}, \citenamefont {Shi}, \citenamefont {Hua}, \citenamefont {Watanabe}, \citenamefont {Taniguchi}, \citenamefont {Xiong}, \citenamefont {Fu},\ and\ \citenamefont {Ju}}]{han_signatures_2024}%
  \BibitemOpen
  \bibfield  {author} {\bibinfo {author} {\bibfnamefont {T.}~\bibnamefont {Han}}, \bibinfo {author} {\bibfnamefont {Z.}~\bibnamefont {Lu}}, \bibinfo {author} {\bibfnamefont {Y.}~\bibnamefont {Yao}}, \bibinfo {author} {\bibfnamefont {L.}~\bibnamefont {Shi}}, \bibinfo {author} {\bibfnamefont {J.}~\bibnamefont {Yang}}, \bibinfo {author} {\bibfnamefont {J.}~\bibnamefont {Seo}}, \bibinfo {author} {\bibfnamefont {S.}~\bibnamefont {Ye}}, \bibinfo {author} {\bibfnamefont {Z.}~\bibnamefont {Wu}}, \bibinfo {author} {\bibfnamefont {M.}~\bibnamefont {Zhou}}, \bibinfo {author} {\bibfnamefont {H.}~\bibnamefont {Liu}}, \bibinfo {author} {\bibfnamefont {G.}~\bibnamefont {Shi}}, \bibinfo {author} {\bibfnamefont {Z.}~\bibnamefont {Hua}}, \bibinfo {author} {\bibfnamefont {K.}~\bibnamefont {Watanabe}}, \bibinfo {author} {\bibfnamefont {T.}~\bibnamefont {Taniguchi}}, \bibinfo {author} {\bibfnamefont {P.}~\bibnamefont {Xiong}}, \bibinfo {author} {\bibfnamefont {L.}~\bibnamefont {Fu}},\ and\ \bibinfo {author} {\bibfnamefont
  {L.}~\bibnamefont {Ju}},\ }\bibfield  {title} {\bibinfo {title} {Signatures of {Chiral} {Superconductivity} in {Rhombohedral} {Graphene}},\ }\href {http://arxiv.org/abs/2408.15233} {\bibfield  {journal} {\bibinfo  {journal} {arXiv: 2408.15233}\ } (\bibinfo {year} {2024})}\BibitemShut {NoStop}%
\bibitem [{\citenamefont {Choi}\ \emph {et~al.}(2024)\citenamefont {Choi}, \citenamefont {Choi}, \citenamefont {Valentini}, \citenamefont {Patterson}, \citenamefont {Holleis}, \citenamefont {Sheekey}, \citenamefont {Stoyanov}, \citenamefont {Cheng}, \citenamefont {Taniguchi}, \citenamefont {Watanabe},\ and\ \citenamefont {Young}}]{choi_electric_2024}%
  \BibitemOpen
  \bibfield  {author} {\bibinfo {author} {\bibfnamefont {Y.}~\bibnamefont {Choi}}, \bibinfo {author} {\bibfnamefont {Y.}~\bibnamefont {Choi}}, \bibinfo {author} {\bibfnamefont {M.}~\bibnamefont {Valentini}}, \bibinfo {author} {\bibfnamefont {C.~L.}\ \bibnamefont {Patterson}}, \bibinfo {author} {\bibfnamefont {L.~F.~W.}\ \bibnamefont {Holleis}}, \bibinfo {author} {\bibfnamefont {O.~I.}\ \bibnamefont {Sheekey}}, \bibinfo {author} {\bibfnamefont {H.}~\bibnamefont {Stoyanov}}, \bibinfo {author} {\bibfnamefont {X.}~\bibnamefont {Cheng}}, \bibinfo {author} {\bibfnamefont {T.}~\bibnamefont {Taniguchi}}, \bibinfo {author} {\bibfnamefont {K.}~\bibnamefont {Watanabe}},\ and\ \bibinfo {author} {\bibfnamefont {A.~F.}\ \bibnamefont {Young}},\ }\bibfield  {title} {\bibinfo {title} {Electric field control of superconductivity and quantized anomalous {Hall} effects in rhombohedral tetralayer graphene},\ }\href {http://arxiv.org/abs/2408.12584} {\bibfield  {journal} {\bibinfo  {journal} {arXiv: 2408.12584}\ } (\bibinfo {year}
  {2024})}\BibitemShut {NoStop}%
\bibitem [{\citenamefont {Lu}\ \emph {et~al.}(2024)\citenamefont {Lu}, \citenamefont {Han}, \citenamefont {Yao}, \citenamefont {Reddy}, \citenamefont {Yang}, \citenamefont {Seo}, \citenamefont {Watanabe}, \citenamefont {Taniguchi}, \citenamefont {Fu},\ and\ \citenamefont {Ju}}]{lu2024fractional}%
  \BibitemOpen
  \bibfield  {author} {\bibinfo {author} {\bibfnamefont {Z.}~\bibnamefont {Lu}}, \bibinfo {author} {\bibfnamefont {T.}~\bibnamefont {Han}}, \bibinfo {author} {\bibfnamefont {Y.}~\bibnamefont {Yao}}, \bibinfo {author} {\bibfnamefont {A.~P.}\ \bibnamefont {Reddy}}, \bibinfo {author} {\bibfnamefont {J.}~\bibnamefont {Yang}}, \bibinfo {author} {\bibfnamefont {J.}~\bibnamefont {Seo}}, \bibinfo {author} {\bibfnamefont {K.}~\bibnamefont {Watanabe}}, \bibinfo {author} {\bibfnamefont {T.}~\bibnamefont {Taniguchi}}, \bibinfo {author} {\bibfnamefont {L.}~\bibnamefont {Fu}},\ and\ \bibinfo {author} {\bibfnamefont {L.}~\bibnamefont {Ju}},\ }\bibfield  {title} {\bibinfo {title} {Fractional quantum anomalous hall effect in multilayer graphene},\ }\href@noop {} {\bibfield  {journal} {\bibinfo  {journal} {Nature}\ }\textbf {\bibinfo {volume} {626}},\ \bibinfo {pages} {759} (\bibinfo {year} {2024})}\BibitemShut {NoStop}%
\bibitem [{\citenamefont {Andrei}\ and\ \citenamefont {MacDonald}(2020)}]{Andrei2020Graphene}%
  \BibitemOpen
  \bibfield  {author} {\bibinfo {author} {\bibfnamefont {E.~Y.}\ \bibnamefont {Andrei}}\ and\ \bibinfo {author} {\bibfnamefont {A.~H.}\ \bibnamefont {MacDonald}},\ }\bibfield  {title} {\bibinfo {title} {Graphene bilayers with a twist},\ }\href {https://doi.org/10.1038/s41563-020-00840-0} {\bibfield  {journal} {\bibinfo  {journal} {Nature Materials}\ }\textbf {\bibinfo {volume} {19}},\ \bibinfo {pages} {1265} (\bibinfo {year} {2020})}\BibitemShut {NoStop}%
\bibitem [{\citenamefont {Balents}\ \emph {et~al.}(2020)\citenamefont {Balents}, \citenamefont {Dean}, \citenamefont {Efetov},\ and\ \citenamefont {Young}}]{Balents2020Superconductivity}%
  \BibitemOpen
  \bibfield  {author} {\bibinfo {author} {\bibfnamefont {L.}~\bibnamefont {Balents}}, \bibinfo {author} {\bibfnamefont {C.~R.}\ \bibnamefont {Dean}}, \bibinfo {author} {\bibfnamefont {D.~K.}\ \bibnamefont {Efetov}},\ and\ \bibinfo {author} {\bibfnamefont {A.~F.}\ \bibnamefont {Young}},\ }\bibfield  {title} {\bibinfo {title} {Superconductivity and strong correlations in moir{\'{e}} flat bands},\ }\href {https://doi.org/10.1038/s41567-020-0906-9} {\bibfield  {journal} {\bibinfo  {journal} {Nature Physics}\ }\textbf {\bibinfo {volume} {16}},\ \bibinfo {pages} {725} (\bibinfo {year} {2020})}\BibitemShut {NoStop}%
\bibitem [{\citenamefont {Ghazaryan}\ \emph {et~al.}(2021{\natexlab{a}})\citenamefont {Ghazaryan}, \citenamefont {Holder}, \citenamefont {Serbyn},\ and\ \citenamefont {Berg}}]{ghazaryan_unconventional_2021}%
  \BibitemOpen
  \bibfield  {author} {\bibinfo {author} {\bibfnamefont {A.}~\bibnamefont {Ghazaryan}}, \bibinfo {author} {\bibfnamefont {T.}~\bibnamefont {Holder}}, \bibinfo {author} {\bibfnamefont {M.}~\bibnamefont {Serbyn}},\ and\ \bibinfo {author} {\bibfnamefont {E.}~\bibnamefont {Berg}},\ }\bibfield  {title} {\bibinfo {title} {Unconventional superconductivity in systems with annular fermi surface: Application to rhombohedral trilayer graphene},\ }\href {https://doi.org/10.1103/PhysRevLett.127.247001} {\bibfield  {journal} {\bibinfo  {journal} {Physical Review Letters}\ }\textbf {\bibinfo {volume} {127}},\ \bibinfo {pages} {247001} (\bibinfo {year} {2021}{\natexlab{a}})}\BibitemShut {NoStop}%
\bibitem [{\citenamefont {You}\ and\ \citenamefont {Vishwanath}(2022)}]{you2022kohn}%
  \BibitemOpen
  \bibfield  {author} {\bibinfo {author} {\bibfnamefont {Y.-Z.}\ \bibnamefont {You}}\ and\ \bibinfo {author} {\bibfnamefont {A.}~\bibnamefont {Vishwanath}},\ }\bibfield  {title} {\bibinfo {title} {Kohn-luttinger superconductivity and intervalley coherence in rhombohedral trilayer graphene},\ }\href {https://doi.org/10.1103/physrevb.105.134524} {\bibfield  {journal} {\bibinfo  {journal} {Physical Review B}\ }\textbf {\bibinfo {volume} {105}} (\bibinfo {year} {2022})}\BibitemShut {NoStop}%
\bibitem [{\citenamefont {Chatterjee}\ \emph {et~al.}(2022)\citenamefont {Chatterjee}, \citenamefont {Wang}, \citenamefont {Berg},\ and\ \citenamefont {Zaletel}}]{chatterjee_inter-valley_2022}%
  \BibitemOpen
  \bibfield  {author} {\bibinfo {author} {\bibfnamefont {S.}~\bibnamefont {Chatterjee}}, \bibinfo {author} {\bibfnamefont {T.}~\bibnamefont {Wang}}, \bibinfo {author} {\bibfnamefont {E.}~\bibnamefont {Berg}},\ and\ \bibinfo {author} {\bibfnamefont {M.~P.}\ \bibnamefont {Zaletel}},\ }\bibfield  {title} {\bibinfo {title} {Inter-valley coherent order and isospin fluctuation mediated superconductivity in rhombohedral trilayer graphene},\ }\href {https://doi.org/10.1038/s41467-022-33561-w} {\bibfield  {journal} {\bibinfo  {journal} {Nature Communications}\ }\textbf {\bibinfo {volume} {13}},\ \bibinfo {pages} {6013} (\bibinfo {year} {2022})}\BibitemShut {NoStop}%
\bibitem [{\citenamefont {Cea}\ \emph {et~al.}(2022)\citenamefont {Cea}, \citenamefont {Pantale{\'o}n}, \citenamefont {Phong},\ and\ \citenamefont {Guinea}}]{cea2022superconductivity}%
  \BibitemOpen
  \bibfield  {author} {\bibinfo {author} {\bibfnamefont {T.}~\bibnamefont {Cea}}, \bibinfo {author} {\bibfnamefont {P.~A.}\ \bibnamefont {Pantale{\'o}n}}, \bibinfo {author} {\bibfnamefont {V.~T.}\ \bibnamefont {Phong}},\ and\ \bibinfo {author} {\bibfnamefont {F.}~\bibnamefont {Guinea}},\ }\bibfield  {title} {\bibinfo {title} {Superconductivity from repulsive interactions in rhombohedral trilayer graphene: A kohn-luttinger-like mechanism},\ }\href {https://doi.org/10.1103/physrevb.105.075432} {\bibfield  {journal} {\bibinfo  {journal} {Physical Review B}\ }\textbf {\bibinfo {volume} {105}} (\bibinfo {year} {2022})}\BibitemShut {NoStop}%
\bibitem [{\citenamefont {Cea}(2023)}]{Cea2023Superconductivity}%
  \BibitemOpen
  \bibfield  {author} {\bibinfo {author} {\bibfnamefont {T.}~\bibnamefont {Cea}},\ }\bibfield  {title} {\bibinfo {title} {Superconductivity induced by the intervalley coulomb scattering in a few layers of graphene},\ }\href {https://link.aps.org/doi/10.1103/PhysRevB.107.L041111} {\bibfield  {journal} {\bibinfo  {journal} {Physical Review B}\ }\textbf {\bibinfo {volume} {107}},\ \bibinfo {pages} {L041111} (\bibinfo {year} {2023})}\BibitemShut {NoStop}%
\bibitem [{\citenamefont {Jimeno-Pozo}\ \emph {et~al.}(2023)\citenamefont {Jimeno-Pozo}, \citenamefont {Sainz-Cruz}, \citenamefont {Cea}, \citenamefont {Pantale{\'{o}}n},\ and\ \citenamefont {Guinea}}]{JimenoPozo2023}%
  \BibitemOpen
  \bibfield  {author} {\bibinfo {author} {\bibfnamefont {A.}~\bibnamefont {Jimeno-Pozo}}, \bibinfo {author} {\bibfnamefont {H.}~\bibnamefont {Sainz-Cruz}}, \bibinfo {author} {\bibfnamefont {T.}~\bibnamefont {Cea}}, \bibinfo {author} {\bibfnamefont {P.~A.}\ \bibnamefont {Pantale{\'{o}}n}},\ and\ \bibinfo {author} {\bibfnamefont {F.}~\bibnamefont {Guinea}},\ }\bibfield  {title} {\bibinfo {title} {Superconductivity from electronic interactions and spin-orbit enhancement in bilayer and trilayer graphene},\ }\href {https://doi.org/10.1103/physrevb.107.l161106} {\bibfield  {journal} {\bibinfo  {journal} {Physical Review B}\ }\textbf {\bibinfo {volume} {107}} (\bibinfo {year} {2023})}\BibitemShut {NoStop}%
\bibitem [{\citenamefont {Li}\ \emph {et~al.}(2023)\citenamefont {Li}, \citenamefont {Kuang}, \citenamefont {Jimeno-Pozo}, \citenamefont {Sainz-Cruz}, \citenamefont {Zhan}, \citenamefont {Yuan},\ and\ \citenamefont {Guinea}}]{ZiyanLi2023}%
  \BibitemOpen
  \bibfield  {author} {\bibinfo {author} {\bibfnamefont {Z.}~\bibnamefont {Li}}, \bibinfo {author} {\bibfnamefont {X.}~\bibnamefont {Kuang}}, \bibinfo {author} {\bibfnamefont {A.}~\bibnamefont {Jimeno-Pozo}}, \bibinfo {author} {\bibfnamefont {H.}~\bibnamefont {Sainz-Cruz}}, \bibinfo {author} {\bibfnamefont {Z.}~\bibnamefont {Zhan}}, \bibinfo {author} {\bibfnamefont {S.}~\bibnamefont {Yuan}},\ and\ \bibinfo {author} {\bibfnamefont {F.}~\bibnamefont {Guinea}},\ }\bibfield  {title} {\bibinfo {title} {Charge fluctuations, phonons, and superconductivity in multilayer graphene},\ }\href {https://doi.org/10.1103/physrevb.108.045404} {\bibfield  {journal} {\bibinfo  {journal} {Physical Review B}\ }\textbf {\bibinfo {volume} {108}} (\bibinfo {year} {2023})}\BibitemShut {NoStop}%
\bibitem [{\citenamefont {Dong}\ \emph {et~al.}(2023{\natexlab{a}})\citenamefont {Dong}, \citenamefont {Levitov},\ and\ \citenamefont {Chubukov}}]{Dong2023Multilayer}%
  \BibitemOpen
  \bibfield  {author} {\bibinfo {author} {\bibfnamefont {Z.}~\bibnamefont {Dong}}, \bibinfo {author} {\bibfnamefont {L.}~\bibnamefont {Levitov}},\ and\ \bibinfo {author} {\bibfnamefont {A.~V.}\ \bibnamefont {Chubukov}},\ }\bibfield  {title} {\bibinfo {title} {Superconductivity near spin and valley orders in graphene multilayers},\ }\href {http://dx.doi.org/10.1103/PhysRevB.108.134503} {\bibfield  {journal} {\bibinfo  {journal} {Physical Review B}\ }\textbf {\bibinfo {volume} {108}} (\bibinfo {year} {2023}{\natexlab{a}})}\BibitemShut {NoStop}%
\bibitem [{\citenamefont {Qin}\ \emph {et~al.}(2023)\citenamefont {Qin}, \citenamefont {Huang}, \citenamefont {Wolf}, \citenamefont {Wei}, \citenamefont {Blinov},\ and\ \citenamefont {MacDonald}}]{qin_functional_2023}%
  \BibitemOpen
  \bibfield  {author} {\bibinfo {author} {\bibfnamefont {W.}~\bibnamefont {Qin}}, \bibinfo {author} {\bibfnamefont {C.}~\bibnamefont {Huang}}, \bibinfo {author} {\bibfnamefont {T.}~\bibnamefont {Wolf}}, \bibinfo {author} {\bibfnamefont {N.}~\bibnamefont {Wei}}, \bibinfo {author} {\bibfnamefont {I.}~\bibnamefont {Blinov}},\ and\ \bibinfo {author} {\bibfnamefont {A.~H.}\ \bibnamefont {MacDonald}},\ }\bibfield  {title} {\bibinfo {title} {Functional {Renormalization} {Group} {Study} of {Superconductivity} in {Rhombohedral} {Trilayer} {Graphene}},\ }\href {https://doi.org/10.1103/PhysRevLett.130.146001} {\bibfield  {journal} {\bibinfo  {journal} {Physical Review Letters}\ }\textbf {\bibinfo {volume} {130}},\ \bibinfo {pages} {146001} (\bibinfo {year} {2023})}\BibitemShut {NoStop}%
\bibitem [{\citenamefont {Shavit}\ and\ \citenamefont {Oreg}(2023)}]{shavit_inducing_2023}%
  \BibitemOpen
  \bibfield  {author} {\bibinfo {author} {\bibfnamefont {G.}~\bibnamefont {Shavit}}\ and\ \bibinfo {author} {\bibfnamefont {Y.}~\bibnamefont {Oreg}},\ }\bibfield  {title} {\bibinfo {title} {Inducing superconductivity in bilayer graphene by alleviation of the {Stoner} blockade},\ }\href {https://doi.org/10.1103/PhysRevB.108.024510} {\bibfield  {journal} {\bibinfo  {journal} {Physical Review B}\ }\textbf {\bibinfo {volume} {108}},\ \bibinfo {pages} {024510} (\bibinfo {year} {2023})}\BibitemShut {NoStop}%
\bibitem [{\citenamefont {Dong}\ \emph {et~al.}(2024)\citenamefont {Dong}, \citenamefont {Lantagne-Hurtubise},\ and\ \citenamefont {Alicea}}]{dong_superconductivity_2024}%
  \BibitemOpen
  \bibfield  {author} {\bibinfo {author} {\bibfnamefont {Z.}~\bibnamefont {Dong}}, \bibinfo {author} {\bibfnamefont {E.}~\bibnamefont {Lantagne-Hurtubise}},\ and\ \bibinfo {author} {\bibfnamefont {J.}~\bibnamefont {Alicea}},\ }\bibfield  {title} {\bibinfo {title} {Superconductivity from spin-canting fluctuations in rhombohedral graphene},\ }\href {http://arxiv.org/abs/2406.17036} {\bibfield  {journal} {\bibinfo  {journal} {arXiv: 2406.17036}\ } (\bibinfo {year} {2024})}\BibitemShut {NoStop}%
\bibitem [{\citenamefont {Long}\ \emph {et~al.}(2024)\citenamefont {Long}, \citenamefont {Jimeno-Pozo}, \citenamefont {Sainz-Cruz}, \citenamefont {Pantaleon},\ and\ \citenamefont {Guinea}}]{long_evolution_2024}%
  \BibitemOpen
  \bibfield  {author} {\bibinfo {author} {\bibfnamefont {M.}~\bibnamefont {Long}}, \bibinfo {author} {\bibfnamefont {A.}~\bibnamefont {Jimeno-Pozo}}, \bibinfo {author} {\bibfnamefont {H.}~\bibnamefont {Sainz-Cruz}}, \bibinfo {author} {\bibfnamefont {P.~A.}\ \bibnamefont {Pantaleon}},\ and\ \bibinfo {author} {\bibfnamefont {F.}~\bibnamefont {Guinea}},\ }\bibfield  {title} {\bibinfo {title} {Evolution of superconductivity in twisted graphene multilayers},\ }\href {https://doi.org/10.1073/pnas.2405259121} {\bibfield  {journal} {\bibinfo  {journal} {Proceedings of the National Academy of Sciences}\ }\textbf {\bibinfo {volume} {121}},\ \bibinfo {pages} {e2405259121} (\bibinfo {year} {2024})}\BibitemShut {NoStop}%
\bibitem [{\citenamefont {Vi\~nas Bostr\"om}\ \emph {et~al.}(2024)\citenamefont {Vi\~nas Bostr\"om}, \citenamefont {Fischer}, \citenamefont {Profe}, \citenamefont {Zhang}, \citenamefont {Kennes},\ and\ \citenamefont {Rubio}}]{vinas_bostrom_phonon-mediated_2024}%
  \BibitemOpen
  \bibfield  {author} {\bibinfo {author} {\bibfnamefont {E.}~\bibnamefont {Vi\~nas Bostr\"om}}, \bibinfo {author} {\bibfnamefont {A.}~\bibnamefont {Fischer}}, \bibinfo {author} {\bibfnamefont {J.~B.}\ \bibnamefont {Profe}}, \bibinfo {author} {\bibfnamefont {J.}~\bibnamefont {Zhang}}, \bibinfo {author} {\bibfnamefont {D.~M.}\ \bibnamefont {Kennes}},\ and\ \bibinfo {author} {\bibfnamefont {A.}~\bibnamefont {Rubio}},\ }\bibfield  {title} {\bibinfo {title} {Phonon-mediated unconventional superconductivity in rhombohedral stacked multilayer graphene},\ }\href {https://doi.org/10.1038/s41524-024-01345-z} {\bibfield  {journal} {\bibinfo  {journal} {npj Computational Materials}\ }\textbf {\bibinfo {volume} {10}},\ \bibinfo {pages} {163} (\bibinfo {year} {2024})}\BibitemShut {NoStop}%
\bibitem [{\citenamefont {Fischer}\ \emph {et~al.}(2024)\citenamefont {Fischer}, \citenamefont {Klebl}, \citenamefont {Profe}, \citenamefont {Rothstein}, \citenamefont {Waldecker}, \citenamefont {Beschoten}, \citenamefont {Wehling},\ and\ \citenamefont {Kennes}}]{fischer_spin_2024}%
  \BibitemOpen
  \bibfield  {author} {\bibinfo {author} {\bibfnamefont {A.}~\bibnamefont {Fischer}}, \bibinfo {author} {\bibfnamefont {L.}~\bibnamefont {Klebl}}, \bibinfo {author} {\bibfnamefont {J.~B.}\ \bibnamefont {Profe}}, \bibinfo {author} {\bibfnamefont {A.}~\bibnamefont {Rothstein}}, \bibinfo {author} {\bibfnamefont {L.}~\bibnamefont {Waldecker}}, \bibinfo {author} {\bibfnamefont {B.}~\bibnamefont {Beschoten}}, \bibinfo {author} {\bibfnamefont {T.~O.}\ \bibnamefont {Wehling}},\ and\ \bibinfo {author} {\bibfnamefont {D.~M.}\ \bibnamefont {Kennes}},\ }\bibfield  {title} {\bibinfo {title} {Spin and charge fluctuation induced pairing in {ABCB} tetralayer graphene},\ }\href {https://doi.org/10.1103/PhysRevResearch.6.L012003} {\bibfield  {journal} {\bibinfo  {journal} {Physical Review Research}\ }\textbf {\bibinfo {volume} {6}},\ \bibinfo {pages} {L012003} (\bibinfo {year} {2024})}\BibitemShut {NoStop}%
\bibitem [{\citenamefont {Vituri}\ \emph {et~al.}(2024)\citenamefont {Vituri}, \citenamefont {Xiao}, \citenamefont {Pareek}, \citenamefont {Holder},\ and\ \citenamefont {Berg}}]{vituri2024incommensurateintervalleycoherentstates}%
  \BibitemOpen
  \bibfield  {author} {\bibinfo {author} {\bibfnamefont {Y.}~\bibnamefont {Vituri}}, \bibinfo {author} {\bibfnamefont {J.}~\bibnamefont {Xiao}}, \bibinfo {author} {\bibfnamefont {K.}~\bibnamefont {Pareek}}, \bibinfo {author} {\bibfnamefont {T.}~\bibnamefont {Holder}},\ and\ \bibinfo {author} {\bibfnamefont {E.}~\bibnamefont {Berg}},\ }\bibfield  {title} {\bibinfo {title} {Incommensurate inter-valley coherent states in abc graphene: collective modes and superconductivity},\ }\href {https://arxiv.org/abs/2408.10309} {\bibfield  {journal} {\bibinfo  {journal} {arXiv: 2408.10309}\ } (\bibinfo {year} {2024})}\BibitemShut {NoStop}%
\bibitem [{\citenamefont {Braz}\ \emph {et~al.}(2024)\citenamefont {Braz}, \citenamefont {Nag},\ and\ \citenamefont {Black-Schaffer}}]{Braz2024_magneticABC}%
  \BibitemOpen
  \bibfield  {author} {\bibinfo {author} {\bibfnamefont {L.~B.}\ \bibnamefont {Braz}}, \bibinfo {author} {\bibfnamefont {T.}~\bibnamefont {Nag}},\ and\ \bibinfo {author} {\bibfnamefont {A.~M.}\ \bibnamefont {Black-Schaffer}},\ }\bibfield  {title} {\bibinfo {title} {Competing magnetic states on the surface of multilayer abc-stacked graphene},\ }\bibfield  {journal} {\bibinfo  {journal} {Physical Review B}\ }\textbf {\bibinfo {volume} {110}},\ \href {https://doi.org/10.1103/physrevb.110.l241401} {10.1103/physrevb.110.l241401} (\bibinfo {year} {2024})\BibitemShut {NoStop}%
\bibitem [{\citenamefont {Monkhorst}(1979)}]{monkhorst_hartree-fock_1979}%
  \BibitemOpen
  \bibfield  {author} {\bibinfo {author} {\bibfnamefont {H.~J.}\ \bibnamefont {Monkhorst}},\ }\bibfield  {title} {\bibinfo {title} {Hartree-{Fock} density of states for extended systems},\ }\href {https://doi.org/10.1103/PhysRevB.20.1504} {\bibfield  {journal} {\bibinfo  {journal} {Physical Review B}\ }\textbf {\bibinfo {volume} {20}},\ \bibinfo {pages} {1504} (\bibinfo {year} {1979})}\BibitemShut {NoStop}%
\bibitem [{\citenamefont {Chou}\ \emph {et~al.}(2024)\citenamefont {Chou}, \citenamefont {Zhu},\ and\ \citenamefont {Sarma}}]{Chou_Intravalley_tetralayer_2024}%
  \BibitemOpen
  \bibfield  {author} {\bibinfo {author} {\bibfnamefont {Y.-Z.}\ \bibnamefont {Chou}}, \bibinfo {author} {\bibfnamefont {J.}~\bibnamefont {Zhu}},\ and\ \bibinfo {author} {\bibfnamefont {S.~D.}\ \bibnamefont {Sarma}},\ }\bibfield  {title} {\bibinfo {title} {Intravalley spin-polarized superconductivity in rhombohedral tetralayer graphene},\ }\href {https://arxiv.org/abs/2409.06701} {\bibfield  {journal} {\bibinfo  {journal} {arXiv: 2409.06701}\ } (\bibinfo {year} {2024})}\BibitemShut {NoStop}%
\bibitem [{\citenamefont {Yang}\ and\ \citenamefont {Zhang}(2024)}]{yang_topological_2024}%
  \BibitemOpen
  \bibfield  {author} {\bibinfo {author} {\bibfnamefont {H.}~\bibnamefont {Yang}}\ and\ \bibinfo {author} {\bibfnamefont {Y.-H.}\ \bibnamefont {Zhang}},\ }\bibfield  {title} {\bibinfo {title} {Topological incommensurate {Fulde}-{Ferrell}-{Larkin}-{Ovchinnikov} superconductor and {Bogoliubov} {Fermi} surface in rhombohedral tetra-layer graphene},\ }\href {http://arxiv.org/abs/2411.02503} {\bibfield  {journal} {\bibinfo  {journal} {arXiv: 2411.02503}\ } (\bibinfo {year} {2024})}\BibitemShut {NoStop}%
\bibitem [{\citenamefont {Qin}\ and\ \citenamefont {Wu}(2024)}]{qin_chiral_2024}%
  \BibitemOpen
  \bibfield  {author} {\bibinfo {author} {\bibfnamefont {Q.}~\bibnamefont {Qin}}\ and\ \bibinfo {author} {\bibfnamefont {C.}~\bibnamefont {Wu}},\ }\bibfield  {title} {\bibinfo {title} {Chiral finite-momentum superconductivity in the tetralayer graphene},\ }\href {http://arxiv.org/abs/2412.07145} {\bibfield  {journal} {\bibinfo  {journal} {arXiv: 2412.07145}\ } (\bibinfo {year} {2024})}\BibitemShut {NoStop}%
\bibitem [{\citenamefont {Geier}\ \emph {et~al.}(2024)\citenamefont {Geier}, \citenamefont {Davydova},\ and\ \citenamefont {Fu}}]{Geier_isospin_tetralayer_2024}%
  \BibitemOpen
  \bibfield  {author} {\bibinfo {author} {\bibfnamefont {M.}~\bibnamefont {Geier}}, \bibinfo {author} {\bibfnamefont {M.}~\bibnamefont {Davydova}},\ and\ \bibinfo {author} {\bibfnamefont {L.}~\bibnamefont {Fu}},\ }\bibfield  {title} {\bibinfo {title} {Chiral and topological superconductivity in isospin polarized multilayer graphene},\ }\href {https://arxiv.org/abs/2409.13829} {\bibfield  {journal} {\bibinfo  {journal} {arXiv: 2409.13829}\ } (\bibinfo {year} {2024})}\BibitemShut {NoStop}%
\bibitem [{\citenamefont {Yoon}\ \emph {et~al.}(2025)\citenamefont {Yoon}, \citenamefont {Xu}, \citenamefont {Barlas},\ and\ \citenamefont {Zhang}}]{Yoon2025Quarter}%
  \BibitemOpen
  \bibfield  {author} {\bibinfo {author} {\bibfnamefont {C.}~\bibnamefont {Yoon}}, \bibinfo {author} {\bibfnamefont {T.}~\bibnamefont {Xu}}, \bibinfo {author} {\bibfnamefont {Y.}~\bibnamefont {Barlas}},\ and\ \bibinfo {author} {\bibfnamefont {F.}~\bibnamefont {Zhang}},\ }\href {https://arxiv.org/abs/2502.17555} {\bibinfo {title} {Quarter metal superconductivity}} (\bibinfo {year} {2025}),\ \Eprint {https://arxiv.org/abs/2502.17555} {arXiv:2502.17555 [cond-mat.supr-con]} \BibitemShut {NoStop}%
\bibitem [{\citenamefont {Jahin}\ and\ \citenamefont {Lin}(2024)}]{jahin_enhanced_2024}%
  \BibitemOpen
  \bibfield  {author} {\bibinfo {author} {\bibfnamefont {A.}~\bibnamefont {Jahin}}\ and\ \bibinfo {author} {\bibfnamefont {S.-Z.}\ \bibnamefont {Lin}},\ }\bibfield  {title} {\bibinfo {title} {Enhanced {Kohn}-{Luttinger} topological superconductivity in bands with nontrivial geometry},\ }\href {http://arxiv.org/abs/2411.09664} {\bibfield  {journal} {\bibinfo  {journal} {arXiv: 2411.09664}\ } (\bibinfo {year} {2024})}\BibitemShut {NoStop}%
\bibitem [{\citenamefont {Wang}\ \emph {et~al.}(2024)\citenamefont {Wang}, \citenamefont {Gao},\ and\ \citenamefont {Yang}}]{wang_chiral_2024}%
  \BibitemOpen
  \bibfield  {author} {\bibinfo {author} {\bibfnamefont {Y.-Q.}\ \bibnamefont {Wang}}, \bibinfo {author} {\bibfnamefont {Z.-Q.}\ \bibnamefont {Gao}},\ and\ \bibinfo {author} {\bibfnamefont {H.}~\bibnamefont {Yang}},\ }\href {http://arxiv.org/abs/2410.05384} {\bibinfo {title} {Chiral superconductivity from parent {Chern} band and its non-{Abelian} generalization}} (\bibinfo {year} {2024}),\ \bibinfo {note} {arXiv:2410.05384 [cond-mat]}\BibitemShut {NoStop}%
\bibitem [{\citenamefont {Li}\ \emph {et~al.}(2024)\citenamefont {Li}, \citenamefont {Ren}, \citenamefont {Li},\ and\ \citenamefont {Jiang}}]{Li2024_spontaneous}%
  \BibitemOpen
  \bibfield  {author} {\bibinfo {author} {\bibfnamefont {S.}~\bibnamefont {Li}}, \bibinfo {author} {\bibfnamefont {Y.-H.}\ \bibnamefont {Ren}}, \bibinfo {author} {\bibfnamefont {A.-L.}\ \bibnamefont {Li}},\ and\ \bibinfo {author} {\bibfnamefont {H.}~\bibnamefont {Jiang}},\ }\bibfield  {title} {\bibinfo {title} {Spontaneous spin superconductor state in {ABCA}-stacked tetralayer graphene},\ }\href {http://dx.doi.org/10.1103/PhysRevB.110.174512} {\bibfield  {journal} {\bibinfo  {journal} {Physical Review B}\ }\textbf {\bibinfo {volume} {110}} (\bibinfo {year} {2024})}\BibitemShut {NoStop}%
\bibitem [{SM()}]{SM}%
  \BibitemOpen
  \href@noop {} {}\bibinfo {note} {See supplementary information which includes Refs.~\cite{McClure1956, dresselhaus1981intercalation, guinea_electronic_2006, min_chiral_2008, koshino_trigonal_2009, heikkila_dimensional_2011, ho_evolution_2016, slizovskiy_films_2019, ghazaryan_multilayer_2023, Zhou2021HalfMetRTG, Zhang2010Band, Kohn1965, Chubukov1993KL2D, cea21Coulomb, phong2021band, long_evolution_2024, guerci_topological_2024, Ghazaryan2021Anular, cea2022superconductivity, JimenoPozo2023, Pantaleon2023ReviewSC, ZiyanLi2023, Dong2023spin, Dong2023Multilayer, Herrera2024Topological, monkhorst_hartree-fock_1979, Cea2023Superconductivity, Chou_Intravalley_tetralayer_2024, yang_topological_2024, Geier_isospin_tetralayer_2024, qin_chiral_2024, loos_uniform_2016, castro_neto_electron-phonon_2007, basko_theory_2008, calandra_electron-phonon_2007}}\BibitemShut {NoStop}%
\bibitem [{Note1()}]{Note1}%
  \BibitemOpen
  \bibinfo {note} {This rests on the assumption that although standard Hartree-Fock theory may not get the correct DOS at the Fermi level, it is sufficient for the comparison of relative total energies of various competing ground state candidates. Therefore, when determining the phase diagram of the normal states, we continue to use standard HF. It is only when the wavefunctions are needed to compute superconducting properties that those obtained from standard HF no longer suffice. In those cases, we turn to our new formalism of including internal screening in HF theory.}\BibitemShut {Stop}%
\bibitem [{Note2()}]{Note2}%
  \BibitemOpen
  \bibinfo {note} {Throughout, the term isospin is used to refer generically to both spin and valley.}\BibitemShut {Stop}%
\bibitem [{Note3()}]{Note3}%
  \BibitemOpen
  \bibinfo {note} {In this work, we have not considered more general ground states. However, we note that they are possible, including valley-coherent ground states, for example. Driven by experimental constraints, we focus here primarily on the quarter-metal phase and its competition with the half-metal and full-metal phases. We leave other possibilities to future work.}\BibitemShut {Stop}%
\bibitem [{\citenamefont {Kohn}\ and\ \citenamefont {Luttinger}(1965)}]{Kohn1965}%
  \BibitemOpen
  \bibfield  {author} {\bibinfo {author} {\bibfnamefont {W.}~\bibnamefont {Kohn}}\ and\ \bibinfo {author} {\bibfnamefont {J.~M.}\ \bibnamefont {Luttinger}},\ }\bibfield  {title} {\bibinfo {title} {New mechanism for superconductivity},\ }\href {http://dx.doi.org/10.1103/PhysRevLett.15.524} {\bibfield  {journal} {\bibinfo  {journal} {Physical Review Letters}\ }\textbf {\bibinfo {volume} {15}},\ \bibinfo {pages} {524–526} (\bibinfo {year} {1965})}\BibitemShut {NoStop}%
\bibitem [{\citenamefont {Chubukov}(1993)}]{Chubukov1993KL2D}%
  \BibitemOpen
  \bibfield  {author} {\bibinfo {author} {\bibfnamefont {A.~V.}\ \bibnamefont {Chubukov}},\ }\bibfield  {title} {\bibinfo {title} {Kohn-luttinger effect and the instability of a two-dimensional repulsive fermi liquid att=0},\ }\href {http://dx.doi.org/10.1103/PhysRevB.48.1097} {\bibfield  {journal} {\bibinfo  {journal} {Physical Review B}\ }\textbf {\bibinfo {volume} {48}},\ \bibinfo {pages} {1097–1104} (\bibinfo {year} {1993})}\BibitemShut {NoStop}%
\bibitem [{\citenamefont {Stoner}(1938)}]{stoner1938collective}%
  \BibitemOpen
  \bibfield  {author} {\bibinfo {author} {\bibfnamefont {E.~C.}\ \bibnamefont {Stoner}},\ }\bibfield  {title} {\bibinfo {title} {Collective electron ferromagnetism},\ }\href {https://doi.org/10.1098/rspa.1938.0066} {\bibfield  {journal} {\bibinfo  {journal} {Proceedings of the Royal Society of London. Series A. Mathematical and Physical Sciences}\ }\textbf {\bibinfo {volume} {165}},\ \bibinfo {pages} {372} (\bibinfo {year} {1938})}\BibitemShut {NoStop}%
\bibitem [{\citenamefont {Pantaleon}\ \emph {et~al.}(2023)\citenamefont {Pantaleon}, \citenamefont {Jimeno-Pozo}, \citenamefont {Sainz-Cruz}, \citenamefont {Phong}, \citenamefont {Cea},\ and\ \citenamefont {Guinea}}]{pantaleon_superconductivity_2023}%
  \BibitemOpen
  \bibfield  {author} {\bibinfo {author} {\bibfnamefont {P.~A.}\ \bibnamefont {Pantaleon}}, \bibinfo {author} {\bibfnamefont {A.}~\bibnamefont {Jimeno-Pozo}}, \bibinfo {author} {\bibfnamefont {H.}~\bibnamefont {Sainz-Cruz}}, \bibinfo {author} {\bibfnamefont {V.~T.}\ \bibnamefont {Phong}}, \bibinfo {author} {\bibfnamefont {T.}~\bibnamefont {Cea}},\ and\ \bibinfo {author} {\bibfnamefont {F.}~\bibnamefont {Guinea}},\ }\bibfield  {title} {\bibinfo {title} {Superconductivity and correlated phases in non-twisted bilayer and trilayer graphene},\ }\href {https://doi.org/10.1038/s42254-023-00575-2} {\bibfield  {journal} {\bibinfo  {journal} {Nature Reviews Physics}\ }\textbf {\bibinfo {volume} {5}},\ \bibinfo {pages} {304} (\bibinfo {year} {2023})}\BibitemShut {NoStop}%
\bibitem [{Note4()}]{Note4}%
  \BibitemOpen
  \bibinfo {note} {For clarity, it is worth emphasizing that by $p$-wave, we mean that the order parameter exhibits strong $p$-wave character. In other words, its phase in momentum space winds around the center-of-mass momentum once every $2\pi .$ However, unlike the usual $p$-wave order parameter, ours exhibits a strong dependence on the radial coordinate of momentum as well, as clearly demonstrated in Fig. \ref {fig:Figura3}. More precisely, by $p$-wave, we mean that the order parameter can be written as $\Delta (\protect \mathbf {k}) \approx \Delta (|\protect \mathbf {k}|) e^{\pm i \theta _\protect \mathbf {k}}$}\BibitemShut {NoStop}%
\bibitem [{\citenamefont {Kagan}\ and\ \citenamefont {Chubukov}(1988)}]{kagan_possibility_1988}%
  \BibitemOpen
  \bibfield  {author} {\bibinfo {author} {\bibfnamefont {M.~Y.}\ \bibnamefont {Kagan}}\ and\ \bibinfo {author} {\bibfnamefont {A.~V.}\ \bibnamefont {Chubukov}},\ }\bibfield  {title} {\bibinfo {title} {Possibility of a superfluid transition in a slightly nonideal {Fermi} gas with repulsion},\ }\href {https://ui.adsabs.harvard.edu/abs/1988JETPL..47..614K} {\bibfield  {journal} {\bibinfo  {journal} {Soviet Journal of Experimental and Theoretical Physics Letters}\ }\textbf {\bibinfo {volume} {47}},\ \bibinfo {pages} {614} (\bibinfo {year} {1988})}\BibitemShut {NoStop}%
\bibitem [{\citenamefont {Shavit}\ and\ \citenamefont {Alicea}(2024)}]{shavit_quantum_2024-1}%
  \BibitemOpen
  \bibfield  {author} {\bibinfo {author} {\bibfnamefont {G.}~\bibnamefont {Shavit}}\ and\ \bibinfo {author} {\bibfnamefont {J.}~\bibnamefont {Alicea}},\ }\bibfield  {title} {\bibinfo {title} {Quantum {Geometric} {Unconventional} {Superconductivity}},\ }\href {http://arxiv.org/abs/2411.05071} {\bibfield  {journal} {\bibinfo  {journal} {arXiv: 2411.05071}\ } (\bibinfo {year} {2024})}\BibitemShut {NoStop}%
\bibitem [{Note5()}]{Note5}%
  \BibitemOpen
  \bibinfo {note} {It is worth noting that in moire lattices, in addition to the geometric effects discussed in~\cite {shavit_quantum_2024-1,jahin_enhanced_2024,guerci_topological_2024}, the RPA dielectric function and the screened potential are matrices whose elements depend on moiré reciprocal lattice vectors. The resulting Umklapp processes can give a significant contribution to the calculation of the critical temperature~\cite {cea21Coulomb,long_evolution_2024}.}\BibitemShut {Stop}%
\bibitem [{\citenamefont {Abrikosov}\ and\ \citenamefont {Gor'kov}(1959)}]{AG59}%
  \BibitemOpen
  \bibfield  {author} {\bibinfo {author} {\bibfnamefont {A.~A.}\ \bibnamefont {Abrikosov}}\ and\ \bibinfo {author} {\bibfnamefont {L.~P.}\ \bibnamefont {Gor'kov}},\ }\bibfield  {title} {\bibinfo {title} {On the theory of superconducting alloys {I}. the electrodynamics of alloys at absolute zero},\ }\href {http://jetp.ras.ru/cgi-bin/e/index/e/8/6/p1090?a=list} {\bibfield  {journal} {\bibinfo  {journal} {Soviet Physics JETP}\ }\textbf {\bibinfo {volume} {35}},\ \bibinfo {pages} {1090} (\bibinfo {year} {1959})}\BibitemShut {NoStop}%
\bibitem [{\citenamefont {Ambegaokar}\ and\ \citenamefont {Griffin}(1965)}]{ambegaokar_theory_1965}%
  \BibitemOpen
  \bibfield  {author} {\bibinfo {author} {\bibfnamefont {V.}~\bibnamefont {Ambegaokar}}\ and\ \bibinfo {author} {\bibfnamefont {A.}~\bibnamefont {Griffin}},\ }\bibfield  {title} {\bibinfo {title} {Theory of the {Thermal} {Conductivity} of {Superconducting} {Alloys} with {Paramagnetic} {Impurities}},\ }\href {https://doi.org/10.1103/PhysRev.137.A1151} {\bibfield  {journal} {\bibinfo  {journal} {Physical Review}\ }\textbf {\bibinfo {volume} {137}},\ \bibinfo {pages} {A1151} (\bibinfo {year} {1965})}\BibitemShut {NoStop}%
\bibitem [{\citenamefont {Maki}(1969)}]{noauthor_gapless_1969}%
  \BibitemOpen
  \bibfield  {author} {\bibinfo {author} {\bibfnamefont {K.}~\bibnamefont {Maki}},\ }\bibfield  {title} {\bibinfo {title} {Gapless superconductivity},\ }in\ \href {https://doi.org/10.1201/9780203737965} {\emph {\bibinfo {booktitle} {Superconductivity}}}\ (\bibinfo  {publisher} {Marcel Dekker Inc.},\ \bibinfo {year} {1969})\BibitemShut {NoStop}%
\bibitem [{\citenamefont {Gonz{\'{a}}lez}\ and\ \citenamefont {Stauber}(2019)}]{Gonzlez2019}%
  \BibitemOpen
  \bibfield  {author} {\bibinfo {author} {\bibfnamefont {J.}~\bibnamefont {Gonz{\'{a}}lez}}\ and\ \bibinfo {author} {\bibfnamefont {T.}~\bibnamefont {Stauber}},\ }\bibfield  {title} {\bibinfo {title} {Kohn-luttinger superconductivity in twisted bilayer graphene},\ }\href {https://doi.org/10.1103/physrevlett.122.026801} {\bibfield  {journal} {\bibinfo  {journal} {Physical Review Letters}\ }\textbf {\bibinfo {volume} {122}} (\bibinfo {year} {2019})}\BibitemShut {NoStop}%
\bibitem [{\citenamefont {Roy}\ and\ \citenamefont {Juri{\v{c}}i{\'{c}}}(2019)}]{Roy2019}%
  \BibitemOpen
  \bibfield  {author} {\bibinfo {author} {\bibfnamefont {B.}~\bibnamefont {Roy}}\ and\ \bibinfo {author} {\bibfnamefont {V.}~\bibnamefont {Juri{\v{c}}i{\'{c}}}},\ }\bibfield  {title} {\bibinfo {title} {Unconventional superconductivity in nearly flat bands in twisted bilayer graphene},\ }\href {https://doi.org/10.1103/physrevb.99.121407} {\bibfield  {journal} {\bibinfo  {journal} {Physical Review B}\ }\textbf {\bibinfo {volume} {99}} (\bibinfo {year} {2019})}\BibitemShut {NoStop}%
\bibitem [{\citenamefont {Goodwin}\ \emph {et~al.}(2019)\citenamefont {Goodwin}, \citenamefont {Corsetti}, \citenamefont {Mostofi},\ and\ \citenamefont {Lischner}}]{Goodwin2019}%
  \BibitemOpen
  \bibfield  {author} {\bibinfo {author} {\bibfnamefont {Z.~A.~H.}\ \bibnamefont {Goodwin}}, \bibinfo {author} {\bibfnamefont {F.}~\bibnamefont {Corsetti}}, \bibinfo {author} {\bibfnamefont {A.~A.}\ \bibnamefont {Mostofi}},\ and\ \bibinfo {author} {\bibfnamefont {J.}~\bibnamefont {Lischner}},\ }\bibfield  {title} {\bibinfo {title} {Twist-angle sensitivity of electron correlations in moiré graphene bilayers},\ }\href {http://dx.doi.org/10.1103/PhysRevB.100.121106} {\bibfield  {journal} {\bibinfo  {journal} {Physical Review B}\ }\textbf {\bibinfo {volume} {100}} (\bibinfo {year} {2019})}\BibitemShut {NoStop}%
\bibitem [{\citenamefont {Cea}\ and\ \citenamefont {Guinea}(2021)}]{cea21Coulomb}%
  \BibitemOpen
  \bibfield  {author} {\bibinfo {author} {\bibfnamefont {T.}~\bibnamefont {Cea}}\ and\ \bibinfo {author} {\bibfnamefont {F.}~\bibnamefont {Guinea}},\ }\bibfield  {title} {\bibinfo {title} {Coulomb interaction, phonons, and superconductivity in twisted bilayer graphene},\ }\href {http://dx.doi.org/10.1073/pnas.2107874118} {\bibfield  {journal} {\bibinfo  {journal} {Proceedings of the National Academy of Sciences}\ }\textbf {\bibinfo {volume} {118}} (\bibinfo {year} {2021})}\BibitemShut {NoStop}%
\bibitem [{\citenamefont {Lewandowski}\ \emph {et~al.}(2021)\citenamefont {Lewandowski}, \citenamefont {Chowdhury},\ and\ \citenamefont {Ruhman}}]{Lewandowski2021}%
  \BibitemOpen
  \bibfield  {author} {\bibinfo {author} {\bibfnamefont {C.}~\bibnamefont {Lewandowski}}, \bibinfo {author} {\bibfnamefont {D.}~\bibnamefont {Chowdhury}},\ and\ \bibinfo {author} {\bibfnamefont {J.}~\bibnamefont {Ruhman}},\ }\bibfield  {title} {\bibinfo {title} {Pairing in magic-angle twisted bilayer graphene: Role of phonon and plasmon umklapp},\ }\href {https://doi.org/10.1103/physrevb.103.235401} {\bibfield  {journal} {\bibinfo  {journal} {Physical Review B}\ }\textbf {\bibinfo {volume} {103}} (\bibinfo {year} {2021})}\BibitemShut {NoStop}%
\bibitem [{\citenamefont {Sharma}\ \emph {et~al.}(2020)\citenamefont {Sharma}, \citenamefont {Trushin}, \citenamefont {Sushkov}, \citenamefont {Vignale},\ and\ \citenamefont {Adam}}]{Sharma2020}%
  \BibitemOpen
  \bibfield  {author} {\bibinfo {author} {\bibfnamefont {G.}~\bibnamefont {Sharma}}, \bibinfo {author} {\bibfnamefont {M.}~\bibnamefont {Trushin}}, \bibinfo {author} {\bibfnamefont {O.~P.}\ \bibnamefont {Sushkov}}, \bibinfo {author} {\bibfnamefont {G.}~\bibnamefont {Vignale}},\ and\ \bibinfo {author} {\bibfnamefont {S.}~\bibnamefont {Adam}},\ }\bibfield  {title} {\bibinfo {title} {Superconductivity from collective excitations in magic-angle twisted bilayer graphene},\ }\href {https://doi.org/10.1103/physrevresearch.2.022040} {\bibfield  {journal} {\bibinfo  {journal} {Physical Review Research}\ }\textbf {\bibinfo {volume} {2}} (\bibinfo {year} {2020})}\BibitemShut {NoStop}%
\bibitem [{\citenamefont {Samajdar}\ and\ \citenamefont {Scheurer}(2020)}]{Samajdar2020SC}%
  \BibitemOpen
  \bibfield  {author} {\bibinfo {author} {\bibfnamefont {R.}~\bibnamefont {Samajdar}}\ and\ \bibinfo {author} {\bibfnamefont {M.~S.}\ \bibnamefont {Scheurer}},\ }\bibfield  {title} {\bibinfo {title} {Microscopic pairing mechanism, order parameter, and disorder sensitivity in moir{\'{e}} superlattices: Applications to twisted double-bilayer graphene},\ }\href {https://doi.org/10.1103/physrevb.102.064501} {\bibfield  {journal} {\bibinfo  {journal} {Physical Review B}\ }\textbf {\bibinfo {volume} {102}} (\bibinfo {year} {2020})}\BibitemShut {NoStop}%
\bibitem [{\citenamefont {Pahlevanzadeh}\ \emph {et~al.}(2021)\citenamefont {Pahlevanzadeh}, \citenamefont {Sahebsara},\ and\ \citenamefont {Sénéchal}}]{Pahlevanzadeh2021DMFT}%
  \BibitemOpen
  \bibfield  {author} {\bibinfo {author} {\bibfnamefont {B.}~\bibnamefont {Pahlevanzadeh}}, \bibinfo {author} {\bibfnamefont {P.}~\bibnamefont {Sahebsara}},\ and\ \bibinfo {author} {\bibfnamefont {D.}~\bibnamefont {Sénéchal}},\ }\bibfield  {title} {\bibinfo {title} {Chiral $p$-wave superconductivity in twisted bilayer graphene from dynamical mean field theory},\ }\href {http://dx.doi.org/10.21468/SciPostPhys.11.1.017} {\bibfield  {journal} {\bibinfo  {journal} {SciPost Physics}\ }\textbf {\bibinfo {volume} {11}} (\bibinfo {year} {2021})}\BibitemShut {NoStop}%
\bibitem [{\citenamefont {Cr{\'{e}}pel}\ \emph {et~al.}(2022)\citenamefont {Cr{\'{e}}pel}, \citenamefont {Cea}, \citenamefont {Fu},\ and\ \citenamefont {Guinea}}]{Crepel2022Unconventional}%
  \BibitemOpen
  \bibfield  {author} {\bibinfo {author} {\bibfnamefont {V.}~\bibnamefont {Cr{\'{e}}pel}}, \bibinfo {author} {\bibfnamefont {T.}~\bibnamefont {Cea}}, \bibinfo {author} {\bibfnamefont {L.}~\bibnamefont {Fu}},\ and\ \bibinfo {author} {\bibfnamefont {F.}~\bibnamefont {Guinea}},\ }\bibfield  {title} {\bibinfo {title} {Unconventional superconductivity due to interband polarization},\ }\href {https://doi.org/10.1103/physrevb.105.094506} {\bibfield  {journal} {\bibinfo  {journal} {Physical Review B}\ }\textbf {\bibinfo {volume} {105}} (\bibinfo {year} {2022})}\BibitemShut {NoStop}%
\bibitem [{\citenamefont {González}\ and\ \citenamefont {Stauber}(2023)}]{Gonzlez2023TTG}%
  \BibitemOpen
  \bibfield  {author} {\bibinfo {author} {\bibfnamefont {J.}~\bibnamefont {González}}\ and\ \bibinfo {author} {\bibfnamefont {T.}~\bibnamefont {Stauber}},\ }\bibfield  {title} {\bibinfo {title} {Ising superconductivity induced from spin-selective valley symmetry breaking in twisted trilayer graphene},\ }\href {http://dx.doi.org/10.1038/s41467-023-38250-w} {\bibfield  {journal} {\bibinfo  {journal} {Nature Communications}\ }\textbf {\bibinfo {volume} {14}} (\bibinfo {year} {2023})}\BibitemShut {NoStop}%
\bibitem [{\citenamefont {Guerci}\ \emph {et~al.}(2024)\citenamefont {Guerci}, \citenamefont {Kaplan}, \citenamefont {Ingham}, \citenamefont {Pixley},\ and\ \citenamefont {Millis}}]{guerci_topological_2024}%
  \BibitemOpen
  \bibfield  {author} {\bibinfo {author} {\bibfnamefont {D.}~\bibnamefont {Guerci}}, \bibinfo {author} {\bibfnamefont {D.}~\bibnamefont {Kaplan}}, \bibinfo {author} {\bibfnamefont {J.}~\bibnamefont {Ingham}}, \bibinfo {author} {\bibfnamefont {J.~H.}\ \bibnamefont {Pixley}},\ and\ \bibinfo {author} {\bibfnamefont {A.~J.}\ \bibnamefont {Millis}},\ }\bibfield  {title} {\bibinfo {title} {Topological superconductivity from repulsive interactions in twisted {WSe}$_{2}$},\ }\href {http://arxiv.org/abs/2408.16075} {\bibfield  {journal} {\bibinfo  {journal} {arXiv: 2408.16075}\ } (\bibinfo {year} {2024})}\BibitemShut {NoStop}%
\bibitem [{\citenamefont {Yuan}\ and\ \citenamefont {Fu}(2022)}]{yuan2022supercurrent}%
  \BibitemOpen
  \bibfield  {author} {\bibinfo {author} {\bibfnamefont {N.~F.~Q.}\ \bibnamefont {Yuan}}\ and\ \bibinfo {author} {\bibfnamefont {L.}~\bibnamefont {Fu}},\ }\bibfield  {title} {\bibinfo {title} {Supercurrent diode effect and finite-momentum superconductors},\ }\href {http://dx.doi.org/10.1073/pnas.2119548119} {\bibfield  {journal} {\bibinfo  {journal} {Proceedings of the National Academy of Sciences}\ }\textbf {\bibinfo {volume} {119}} (\bibinfo {year} {2022})}\BibitemShut {NoStop}%
\bibitem [{\citenamefont {Kaplan}\ \emph {et~al.}(2025)\citenamefont {Kaplan}, \citenamefont {Lucht}, \citenamefont {Volkov},\ and\ \citenamefont {Pixley}}]{Kaplan2025}%
  \BibitemOpen
  \bibfield  {author} {\bibinfo {author} {\bibfnamefont {D.}~\bibnamefont {Kaplan}}, \bibinfo {author} {\bibfnamefont {K.~P.}\ \bibnamefont {Lucht}}, \bibinfo {author} {\bibfnamefont {P.~A.}\ \bibnamefont {Volkov}},\ and\ \bibinfo {author} {\bibfnamefont {J.~H.}\ \bibnamefont {Pixley}},\ }\href {https://arxiv.org/abs/2502.12265} {\bibinfo {title} {Quantum geometric photocurrents of quasiparticles in superconductors}} (\bibinfo {year} {2025}),\ \Eprint {https://arxiv.org/abs/2502.12265} {arXiv:2502.12265 [cond-mat.supr-con]} \BibitemShut {NoStop}%
\bibitem [{\citenamefont {Wang}\ \emph {et~al.}(2023)\citenamefont {Wang}, \citenamefont {Kaplan}, \citenamefont {Zhang}, \citenamefont {Holder}, \citenamefont {Cao}, \citenamefont {Wang}, \citenamefont {Zhou}, \citenamefont {Zhou}, \citenamefont {Jiang}, \citenamefont {Zhang}, \citenamefont {Ru}, \citenamefont {Cai}, \citenamefont {Watanabe}, \citenamefont {Taniguchi}, \citenamefont {Yan},\ and\ \citenamefont {Gao}}]{wang2023quantum}%
  \BibitemOpen
  \bibfield  {author} {\bibinfo {author} {\bibfnamefont {N.}~\bibnamefont {Wang}}, \bibinfo {author} {\bibfnamefont {D.}~\bibnamefont {Kaplan}}, \bibinfo {author} {\bibfnamefont {Z.}~\bibnamefont {Zhang}}, \bibinfo {author} {\bibfnamefont {T.}~\bibnamefont {Holder}}, \bibinfo {author} {\bibfnamefont {N.}~\bibnamefont {Cao}}, \bibinfo {author} {\bibfnamefont {A.}~\bibnamefont {Wang}}, \bibinfo {author} {\bibfnamefont {X.}~\bibnamefont {Zhou}}, \bibinfo {author} {\bibfnamefont {F.}~\bibnamefont {Zhou}}, \bibinfo {author} {\bibfnamefont {Z.}~\bibnamefont {Jiang}}, \bibinfo {author} {\bibfnamefont {C.}~\bibnamefont {Zhang}}, \bibinfo {author} {\bibfnamefont {S.}~\bibnamefont {Ru}}, \bibinfo {author} {\bibfnamefont {H.}~\bibnamefont {Cai}}, \bibinfo {author} {\bibfnamefont {K.}~\bibnamefont {Watanabe}}, \bibinfo {author} {\bibfnamefont {T.}~\bibnamefont {Taniguchi}}, \bibinfo {author} {\bibfnamefont {B.}~\bibnamefont {Yan}},\ and\ \bibinfo {author} {\bibfnamefont {W.}~\bibnamefont {Gao}},\ }\bibfield  {title}
  {\bibinfo {title} {Quantum-metric-induced nonlinear transport in a topological antiferromagnet},\ }\href {http://dx.doi.org/10.1038/s41586-023-06363-3} {\bibfield  {journal} {\bibinfo  {journal} {Nature}\ }\textbf {\bibinfo {volume} {621}},\ \bibinfo {pages} {487–492} (\bibinfo {year} {2023})}\BibitemShut {NoStop}%
\bibitem [{\citenamefont {Gao}\ \emph {et~al.}(2023)\citenamefont {Gao}, \citenamefont {Liu}, \citenamefont {Qiu}, \citenamefont {Ghosh}, \citenamefont {V.~Trevisan}, \citenamefont {Onishi}, \citenamefont {Hu}, \citenamefont {Qian}, \citenamefont {Tien}, \citenamefont {Chen}, \citenamefont {Huang}, \citenamefont {Bérubé}, \citenamefont {Li}, \citenamefont {Tzschaschel}, \citenamefont {Dinh}, \citenamefont {Sun}, \citenamefont {Ho}, \citenamefont {Lien}, \citenamefont {Singh}, \citenamefont {Watanabe}, \citenamefont {Taniguchi}, \citenamefont {Bell}, \citenamefont {Lin}, \citenamefont {Chang}, \citenamefont {Du}, \citenamefont {Bansil}, \citenamefont {Fu}, \citenamefont {Ni}, \citenamefont {Orth}, \citenamefont {Ma},\ and\ \citenamefont {Xu}}]{gao2023quantum}%
  \BibitemOpen
  \bibfield  {author} {\bibinfo {author} {\bibfnamefont {A.}~\bibnamefont {Gao}}, \bibinfo {author} {\bibfnamefont {Y.-F.}\ \bibnamefont {Liu}}, \bibinfo {author} {\bibfnamefont {J.-X.}\ \bibnamefont {Qiu}}, \bibinfo {author} {\bibfnamefont {B.}~\bibnamefont {Ghosh}}, \bibinfo {author} {\bibfnamefont {T.}~\bibnamefont {V.~Trevisan}}, \bibinfo {author} {\bibfnamefont {Y.}~\bibnamefont {Onishi}}, \bibinfo {author} {\bibfnamefont {C.}~\bibnamefont {Hu}}, \bibinfo {author} {\bibfnamefont {T.}~\bibnamefont {Qian}}, \bibinfo {author} {\bibfnamefont {H.-J.}\ \bibnamefont {Tien}}, \bibinfo {author} {\bibfnamefont {S.-W.}\ \bibnamefont {Chen}}, \bibinfo {author} {\bibfnamefont {M.}~\bibnamefont {Huang}}, \bibinfo {author} {\bibfnamefont {D.}~\bibnamefont {Bérubé}}, \bibinfo {author} {\bibfnamefont {H.}~\bibnamefont {Li}}, \bibinfo {author} {\bibfnamefont {C.}~\bibnamefont {Tzschaschel}}, \bibinfo {author} {\bibfnamefont {T.}~\bibnamefont {Dinh}}, \bibinfo {author} {\bibfnamefont {Z.}~\bibnamefont {Sun}}, \bibinfo
  {author} {\bibfnamefont {S.-C.}\ \bibnamefont {Ho}}, \bibinfo {author} {\bibfnamefont {S.-W.}\ \bibnamefont {Lien}}, \bibinfo {author} {\bibfnamefont {B.}~\bibnamefont {Singh}}, \bibinfo {author} {\bibfnamefont {K.}~\bibnamefont {Watanabe}}, \bibinfo {author} {\bibfnamefont {T.}~\bibnamefont {Taniguchi}}, \bibinfo {author} {\bibfnamefont {D.~C.}\ \bibnamefont {Bell}}, \bibinfo {author} {\bibfnamefont {H.}~\bibnamefont {Lin}}, \bibinfo {author} {\bibfnamefont {T.-R.}\ \bibnamefont {Chang}}, \bibinfo {author} {\bibfnamefont {C.~R.}\ \bibnamefont {Du}}, \bibinfo {author} {\bibfnamefont {A.}~\bibnamefont {Bansil}}, \bibinfo {author} {\bibfnamefont {L.}~\bibnamefont {Fu}}, \bibinfo {author} {\bibfnamefont {N.}~\bibnamefont {Ni}}, \bibinfo {author} {\bibfnamefont {P.~P.}\ \bibnamefont {Orth}}, \bibinfo {author} {\bibfnamefont {Q.}~\bibnamefont {Ma}},\ and\ \bibinfo {author} {\bibfnamefont {S.-Y.}\ \bibnamefont {Xu}},\ }\bibfield  {title} {\bibinfo {title} {Quantum metric nonlinear hall effect in a topological
  antiferromagnetic heterostructure},\ }\href {http://dx.doi.org/10.1126/science.adf1506} {\bibfield  {journal} {\bibinfo  {journal} {Science}\ }\textbf {\bibinfo {volume} {381}},\ \bibinfo {pages} {181–186} (\bibinfo {year} {2023})}\BibitemShut {NoStop}%
\bibitem [{\citenamefont {Fulde}\ and\ \citenamefont {Ferrell}(1964)}]{fulde_superconductivity_1964}%
  \BibitemOpen
  \bibfield  {author} {\bibinfo {author} {\bibfnamefont {P.}~\bibnamefont {Fulde}}\ and\ \bibinfo {author} {\bibfnamefont {R.~A.}\ \bibnamefont {Ferrell}},\ }\bibfield  {title} {\bibinfo {title} {Superconductivity in a {Strong} {Spin}-{Exchange} {Field}},\ }\href {https://doi.org/10.1103/PhysRev.135.A550} {\bibfield  {journal} {\bibinfo  {journal} {Physical Review}\ }\textbf {\bibinfo {volume} {135}},\ \bibinfo {pages} {A550} (\bibinfo {year} {1964})}\BibitemShut {NoStop}%
\bibitem [{\citenamefont {Larkin}\ and\ \citenamefont {Ovchinnikov}(1964)}]{larkin_nonuniform_64}%
  \BibitemOpen
  \bibfield  {author} {\bibinfo {author} {\bibfnamefont {A.~I.}\ \bibnamefont {Larkin}}\ and\ \bibinfo {author} {\bibfnamefont {Y.~N.}\ \bibnamefont {Ovchinnikov}},\ }\bibfield  {title} {\bibinfo {title} {Nonuniform state of superconductors},\ }\bibfield  {journal} {\bibinfo  {journal} {Zh. Eksperim. i Teor. Fiz.}\ }\textbf {\bibinfo {volume} {47}},\ \href {https://www.osti.gov/biblio/4653415} {} (\bibinfo {year} {1964})\BibitemShut {NoStop}%
\bibitem [{\citenamefont {McClure}(1956)}]{McClure1956}%
  \BibitemOpen
  \bibfield  {author} {\bibinfo {author} {\bibfnamefont {J.~W.}\ \bibnamefont {McClure}},\ }\bibfield  {title} {\bibinfo {title} {Diamagnetism of graphite},\ }\href {https://doi.org/10.1103/physrev.104.666} {\bibfield  {journal} {\bibinfo  {journal} {Physical Review}\ }\textbf {\bibinfo {volume} {104}},\ \bibinfo {pages} {666} (\bibinfo {year} {1956})}\BibitemShut {NoStop}%
\bibitem [{\citenamefont {Dresselhaus}\ and\ \citenamefont {Dresselhaus}(1981)}]{dresselhaus1981intercalation}%
  \BibitemOpen
  \bibfield  {author} {\bibinfo {author} {\bibfnamefont {M.~S.}\ \bibnamefont {Dresselhaus}}\ and\ \bibinfo {author} {\bibfnamefont {G.}~\bibnamefont {Dresselhaus}},\ }\bibfield  {title} {\bibinfo {title} {Intercalation compounds of graphite},\ }\href {https://doi.org/10.1080/00018730110113644} {\bibfield  {journal} {\bibinfo  {journal} {Advances in Physics}\ }\textbf {\bibinfo {volume} {30}},\ \bibinfo {pages} {139} (\bibinfo {year} {1981})}\BibitemShut {NoStop}%
\bibitem [{\citenamefont {Guinea}\ \emph {et~al.}(2006)\citenamefont {Guinea}, \citenamefont {Castro~Neto},\ and\ \citenamefont {Peres}}]{guinea_electronic_2006}%
  \BibitemOpen
  \bibfield  {author} {\bibinfo {author} {\bibfnamefont {F.}~\bibnamefont {Guinea}}, \bibinfo {author} {\bibfnamefont {A.~H.}\ \bibnamefont {Castro~Neto}},\ and\ \bibinfo {author} {\bibfnamefont {N.~M.~R.}\ \bibnamefont {Peres}},\ }\bibfield  {title} {\bibinfo {title} {Electronic states and {Landau} levels in graphene stacks},\ }\href {https://doi.org/10.1103/PhysRevB.73.245426} {\bibfield  {journal} {\bibinfo  {journal} {Physical Review B}\ }\textbf {\bibinfo {volume} {73}},\ \bibinfo {pages} {245426} (\bibinfo {year} {2006})}\BibitemShut {NoStop}%
\bibitem [{\citenamefont {Min}\ and\ \citenamefont {MacDonald}(2008)}]{min_chiral_2008}%
  \BibitemOpen
  \bibfield  {author} {\bibinfo {author} {\bibfnamefont {H.}~\bibnamefont {Min}}\ and\ \bibinfo {author} {\bibfnamefont {A.~H.}\ \bibnamefont {MacDonald}},\ }\bibfield  {title} {\bibinfo {title} {Chiral decomposition in the electronic structure of graphene multilayers},\ }\href {https://doi.org/10.1103/PhysRevB.77.155416} {\bibfield  {journal} {\bibinfo  {journal} {Physical Review B}\ }\textbf {\bibinfo {volume} {77}},\ \bibinfo {pages} {155416} (\bibinfo {year} {2008})}\BibitemShut {NoStop}%
\bibitem [{\citenamefont {Koshino}\ and\ \citenamefont {McCann}(2009)}]{koshino_trigonal_2009}%
  \BibitemOpen
  \bibfield  {author} {\bibinfo {author} {\bibfnamefont {M.}~\bibnamefont {Koshino}}\ and\ \bibinfo {author} {\bibfnamefont {E.}~\bibnamefont {McCann}},\ }\bibfield  {title} {\bibinfo {title} {Trigonal warping and {Berry}’s phase {N} $\pi$ in {ABC}-stacked multilayer graphene},\ }\href {https://doi.org/10.1103/PhysRevB.80.165409} {\bibfield  {journal} {\bibinfo  {journal} {Physical Review B}\ }\textbf {\bibinfo {volume} {80}},\ \bibinfo {pages} {165409} (\bibinfo {year} {2009})}\BibitemShut {NoStop}%
\bibitem [{\citenamefont {Heikkilä}\ and\ \citenamefont {Volovik}(2011)}]{heikkila_dimensional_2011}%
  \BibitemOpen
  \bibfield  {author} {\bibinfo {author} {\bibfnamefont {T.~T.}\ \bibnamefont {Heikkilä}}\ and\ \bibinfo {author} {\bibfnamefont {G.~E.}\ \bibnamefont {Volovik}},\ }\bibfield  {title} {\bibinfo {title} {Dimensional crossover in topological matter: {Evolution} of the multiple {Dirac} point in the layered system to the flat band on the surface},\ }\href {https://doi.org/10.1134/S002136401102007X} {\bibfield  {journal} {\bibinfo  {journal} {JETP Letters}\ }\textbf {\bibinfo {volume} {93}},\ \bibinfo {pages} {59} (\bibinfo {year} {2011})}\BibitemShut {NoStop}%
\bibitem [{\citenamefont {Ho}\ \emph {et~al.}(2016)\citenamefont {Ho}, \citenamefont {Chang},\ and\ \citenamefont {Lin}}]{ho_evolution_2016}%
  \BibitemOpen
  \bibfield  {author} {\bibinfo {author} {\bibfnamefont {C.-H.}\ \bibnamefont {Ho}}, \bibinfo {author} {\bibfnamefont {C.-P.}\ \bibnamefont {Chang}},\ and\ \bibinfo {author} {\bibfnamefont {M.-F.}\ \bibnamefont {Lin}},\ }\bibfield  {title} {\bibinfo {title} {Evolution and dimensional crossover from the bulk subbands in {ABC}-stacked graphene to a three-dimensional {Dirac} cone structure in rhombohedral graphite},\ }\href {https://doi.org/10.1103/PhysRevB.93.075437} {\bibfield  {journal} {\bibinfo  {journal} {Physical Review B}\ }\textbf {\bibinfo {volume} {93}},\ \bibinfo {pages} {075437} (\bibinfo {year} {2016})}\BibitemShut {NoStop}%
\bibitem [{\citenamefont {Slizovskiy}\ \emph {et~al.}(2019)\citenamefont {Slizovskiy}, \citenamefont {McCann}, \citenamefont {Koshino},\ and\ \citenamefont {Fal’ko}}]{slizovskiy_films_2019}%
  \BibitemOpen
  \bibfield  {author} {\bibinfo {author} {\bibfnamefont {S.}~\bibnamefont {Slizovskiy}}, \bibinfo {author} {\bibfnamefont {E.}~\bibnamefont {McCann}}, \bibinfo {author} {\bibfnamefont {M.}~\bibnamefont {Koshino}},\ and\ \bibinfo {author} {\bibfnamefont {V.~I.}\ \bibnamefont {Fal’ko}},\ }\bibfield  {title} {\bibinfo {title} {Films of rhombohedral graphite as two-dimensional topological semimetals},\ }\href {https://www.nature.com/articles/s42005-019-0268-8} {\bibfield  {journal} {\bibinfo  {journal} {Communications Physics}\ }\textbf {\bibinfo {volume} {2}},\ \bibinfo {pages} {164} (\bibinfo {year} {2019})}\BibitemShut {NoStop}%
\bibitem [{\citenamefont {Ghazaryan}\ \emph {et~al.}(2023)\citenamefont {Ghazaryan}, \citenamefont {Holder}, \citenamefont {Berg},\ and\ \citenamefont {Serbyn}}]{ghazaryan_multilayer_2023}%
  \BibitemOpen
  \bibfield  {author} {\bibinfo {author} {\bibfnamefont {A.}~\bibnamefont {Ghazaryan}}, \bibinfo {author} {\bibfnamefont {T.}~\bibnamefont {Holder}}, \bibinfo {author} {\bibfnamefont {E.}~\bibnamefont {Berg}},\ and\ \bibinfo {author} {\bibfnamefont {M.}~\bibnamefont {Serbyn}},\ }\bibfield  {title} {\bibinfo {title} {Multilayer graphenes as a platform for interaction-driven physics and topological superconductivity},\ }\href {https://doi.org/10.1103/PhysRevB.107.104502} {\bibfield  {journal} {\bibinfo  {journal} {Physical Review B}\ }\textbf {\bibinfo {volume} {107}},\ \bibinfo {pages} {104502} (\bibinfo {year} {2023})}\BibitemShut {NoStop}%
\bibitem [{\citenamefont {Zhou}\ \emph {et~al.}(2021{\natexlab{b}})\citenamefont {Zhou}, \citenamefont {Xie}, \citenamefont {Ghazaryan}, \citenamefont {Holder}, \citenamefont {Ehrets}, \citenamefont {Spanton}, \citenamefont {Taniguchi}, \citenamefont {Watanabe}, \citenamefont {Berg}, \citenamefont {Serbyn},\ and\ \citenamefont {Young}}]{Zhou2021HalfMetRTG}%
  \BibitemOpen
  \bibfield  {author} {\bibinfo {author} {\bibfnamefont {H.}~\bibnamefont {Zhou}}, \bibinfo {author} {\bibfnamefont {T.}~\bibnamefont {Xie}}, \bibinfo {author} {\bibfnamefont {A.}~\bibnamefont {Ghazaryan}}, \bibinfo {author} {\bibfnamefont {T.}~\bibnamefont {Holder}}, \bibinfo {author} {\bibfnamefont {J.~R.}\ \bibnamefont {Ehrets}}, \bibinfo {author} {\bibfnamefont {E.~M.}\ \bibnamefont {Spanton}}, \bibinfo {author} {\bibfnamefont {T.}~\bibnamefont {Taniguchi}}, \bibinfo {author} {\bibfnamefont {K.}~\bibnamefont {Watanabe}}, \bibinfo {author} {\bibfnamefont {E.}~\bibnamefont {Berg}}, \bibinfo {author} {\bibfnamefont {M.}~\bibnamefont {Serbyn}},\ and\ \bibinfo {author} {\bibfnamefont {A.~F.}\ \bibnamefont {Young}},\ }\bibfield  {title} {\bibinfo {title} {Half- and quarter-metals in rhombohedral trilayer graphene},\ }\href {https://doi.org/10.1038/s41586-021-03938-w} {\bibfield  {journal} {\bibinfo  {journal} {Nature}\ }\textbf {\bibinfo {volume} {598}},\ \bibinfo {pages} {429} (\bibinfo {year}
  {2021}{\natexlab{b}})}\BibitemShut {NoStop}%
\bibitem [{\citenamefont {Zhang}\ \emph {et~al.}(2010)\citenamefont {Zhang}, \citenamefont {Sahu}, \citenamefont {Min},\ and\ \citenamefont {MacDonald}}]{Zhang2010Band}%
  \BibitemOpen
  \bibfield  {author} {\bibinfo {author} {\bibfnamefont {F.}~\bibnamefont {Zhang}}, \bibinfo {author} {\bibfnamefont {B.}~\bibnamefont {Sahu}}, \bibinfo {author} {\bibfnamefont {H.}~\bibnamefont {Min}},\ and\ \bibinfo {author} {\bibfnamefont {A.~H.}\ \bibnamefont {MacDonald}},\ }\bibfield  {title} {\bibinfo {title} {Band structure of $abc$-stacked graphene trilayers},\ }\href {https://doi.org/10.1103/PhysRevB.82.035409} {\bibfield  {journal} {\bibinfo  {journal} {Phys. Rev. B}\ }\textbf {\bibinfo {volume} {82}},\ \bibinfo {pages} {035409} (\bibinfo {year} {2010})}\BibitemShut {NoStop}%
\bibitem [{\citenamefont {Phong}\ \emph {et~al.}(2021)\citenamefont {Phong}, \citenamefont {Pantale{\'{o}}n}, \citenamefont {Cea},\ and\ \citenamefont {Guinea}}]{phong2021band}%
  \BibitemOpen
  \bibfield  {author} {\bibinfo {author} {\bibfnamefont {V.~T.}\ \bibnamefont {Phong}}, \bibinfo {author} {\bibfnamefont {P.~A.}\ \bibnamefont {Pantale{\'{o}}n}}, \bibinfo {author} {\bibfnamefont {T.}~\bibnamefont {Cea}},\ and\ \bibinfo {author} {\bibfnamefont {F.}~\bibnamefont {Guinea}},\ }\bibfield  {title} {\bibinfo {title} {Band structure and superconductivity in twisted trilayer graphene},\ }\href {https://doi.org/10.1103/physrevb.104.l121116} {\bibfield  {journal} {\bibinfo  {journal} {Physical Review B}\ }\textbf {\bibinfo {volume} {104}} (\bibinfo {year} {2021})}\BibitemShut {NoStop}%
\bibitem [{\citenamefont {Ghazaryan}\ \emph {et~al.}(2021{\natexlab{b}})\citenamefont {Ghazaryan}, \citenamefont {Holder}, \citenamefont {Serbyn},\ and\ \citenamefont {Berg}}]{Ghazaryan2021Anular}%
  \BibitemOpen
  \bibfield  {author} {\bibinfo {author} {\bibfnamefont {A.}~\bibnamefont {Ghazaryan}}, \bibinfo {author} {\bibfnamefont {T.}~\bibnamefont {Holder}}, \bibinfo {author} {\bibfnamefont {M.}~\bibnamefont {Serbyn}},\ and\ \bibinfo {author} {\bibfnamefont {E.}~\bibnamefont {Berg}},\ }\bibfield  {title} {\bibinfo {title} {Unconventional superconductivity in systems with annular fermi surfaces: Application to rhombohedral trilayer graphene},\ }\href {https://link.aps.org/doi/10.1103/PhysRevLett.127.247001} {\bibfield  {journal} {\bibinfo  {journal} {Physical Review Letters}\ }\textbf {\bibinfo {volume} {127}},\ \bibinfo {pages} {247001} (\bibinfo {year} {2021}{\natexlab{b}})}\BibitemShut {NoStop}%
\bibitem [{\citenamefont {Pantale{\'{o}}n}\ \emph {et~al.}(2023)\citenamefont {Pantale{\'{o}}n}, \citenamefont {Jimeno-Pozo}, \citenamefont {Sainz-Cruz}, \citenamefont {Phong}, \citenamefont {Cea},\ and\ \citenamefont {Guinea}}]{Pantaleon2023ReviewSC}%
  \BibitemOpen
  \bibfield  {author} {\bibinfo {author} {\bibfnamefont {P.~A.}\ \bibnamefont {Pantale{\'{o}}n}}, \bibinfo {author} {\bibfnamefont {A.}~\bibnamefont {Jimeno-Pozo}}, \bibinfo {author} {\bibfnamefont {H.}~\bibnamefont {Sainz-Cruz}}, \bibinfo {author} {\bibfnamefont {V.~T.}\ \bibnamefont {Phong}}, \bibinfo {author} {\bibfnamefont {T.}~\bibnamefont {Cea}},\ and\ \bibinfo {author} {\bibfnamefont {F.}~\bibnamefont {Guinea}},\ }\bibfield  {title} {\bibinfo {title} {Superconductivity and correlated phases in non-twisted bilayer and trilayer graphene},\ }\href {https://doi.org/10.1038/s42254-023-00575-2} {\bibfield  {journal} {\bibinfo  {journal} {Nature Reviews Physics}\ } (\bibinfo {year} {2023})}\BibitemShut {NoStop}%
\bibitem [{\citenamefont {Dong}\ \emph {et~al.}(2023{\natexlab{b}})\citenamefont {Dong}, \citenamefont {Chubukov},\ and\ \citenamefont {Levitov}}]{Dong2023spin}%
  \BibitemOpen
  \bibfield  {author} {\bibinfo {author} {\bibfnamefont {Z.}~\bibnamefont {Dong}}, \bibinfo {author} {\bibfnamefont {A.~V.}\ \bibnamefont {Chubukov}},\ and\ \bibinfo {author} {\bibfnamefont {L.}~\bibnamefont {Levitov}},\ }\bibfield  {title} {\bibinfo {title} {Transformer spin-triplet superconductivity at the onset of isospin order in bilayer graphene},\ }\href {https://doi.org/10.1103/physrevb.107.174512} {\bibfield  {journal} {\bibinfo  {journal} {Physical Review B}\ }\textbf {\bibinfo {volume} {107}} (\bibinfo {year} {2023}{\natexlab{b}})}\BibitemShut {NoStop}%
\bibitem [{\citenamefont {Herrera}\ \emph {et~al.}(2024)\citenamefont {Herrera}, \citenamefont {Parra-Martinez}, \citenamefont {Rosenzweig}, \citenamefont {Matta}, \citenamefont {Polley}, \citenamefont {K\"{u}ster}, \citenamefont {Starke}, \citenamefont {Guinea}, \citenamefont {Silva-Guillen}, \citenamefont {Naumis},\ and\ \citenamefont {Pantaleon}}]{Herrera2024Topological}%
  \BibitemOpen
  \bibfield  {author} {\bibinfo {author} {\bibfnamefont {S.~A.}\ \bibnamefont {Herrera}}, \bibinfo {author} {\bibfnamefont {G.}~\bibnamefont {Parra-Martinez}}, \bibinfo {author} {\bibfnamefont {P.}~\bibnamefont {Rosenzweig}}, \bibinfo {author} {\bibfnamefont {B.}~\bibnamefont {Matta}}, \bibinfo {author} {\bibfnamefont {C.~M.}\ \bibnamefont {Polley}}, \bibinfo {author} {\bibfnamefont {K.}~\bibnamefont {K\"{u}ster}}, \bibinfo {author} {\bibfnamefont {U.}~\bibnamefont {Starke}}, \bibinfo {author} {\bibfnamefont {F.}~\bibnamefont {Guinea}}, \bibinfo {author} {\bibfnamefont {J.~A.}\ \bibnamefont {Silva-Guillen}}, \bibinfo {author} {\bibfnamefont {G.~G.}\ \bibnamefont {Naumis}},\ and\ \bibinfo {author} {\bibfnamefont {P.~A.}\ \bibnamefont {Pantaleon}},\ }\bibfield  {title} {\bibinfo {title} {Topological superconductivity in heavily doped single-layer graphene},\ }\href {http://dx.doi.org/10.1021/acsnano.4c12532} {\bibfield  {journal} {\bibinfo  {journal} {ACS Nano}\ }\textbf {\bibinfo {volume} {18}},\ \bibinfo
  {pages} {34842–34857} (\bibinfo {year} {2024})}\BibitemShut {NoStop}%
\bibitem [{\citenamefont {Loos}\ and\ \citenamefont {Gill}(2016)}]{loos_uniform_2016}%
  \BibitemOpen
  \bibfield  {author} {\bibinfo {author} {\bibfnamefont {P.}~\bibnamefont {Loos}}\ and\ \bibinfo {author} {\bibfnamefont {P.~M.~W.}\ \bibnamefont {Gill}},\ }\bibfield  {title} {\bibinfo {title} {The uniform electron gas},\ }\href {https://doi.org/10.1002/wcms.1257} {\bibfield  {journal} {\bibinfo  {journal} {WIREs Computational Molecular Science}\ }\textbf {\bibinfo {volume} {6}},\ \bibinfo {pages} {410} (\bibinfo {year} {2016})}\BibitemShut {NoStop}%
\bibitem [{\citenamefont {Castro~Neto}\ and\ \citenamefont {Guinea}(2007)}]{castro_neto_electron-phonon_2007}%
  \BibitemOpen
  \bibfield  {author} {\bibinfo {author} {\bibfnamefont {A.~H.}\ \bibnamefont {Castro~Neto}}\ and\ \bibinfo {author} {\bibfnamefont {F.}~\bibnamefont {Guinea}},\ }\bibfield  {title} {\bibinfo {title} {Electron-phonon coupling and {Raman} spectroscopy in graphene},\ }\href {https://doi.org/10.1103/PhysRevB.75.045404} {\bibfield  {journal} {\bibinfo  {journal} {Physical Review B}\ }\textbf {\bibinfo {volume} {75}},\ \bibinfo {pages} {045404} (\bibinfo {year} {2007})}\BibitemShut {NoStop}%
\bibitem [{\citenamefont {Basko}(2008)}]{basko_theory_2008}%
  \BibitemOpen
  \bibfield  {author} {\bibinfo {author} {\bibfnamefont {D.~M.}\ \bibnamefont {Basko}},\ }\bibfield  {title} {\bibinfo {title} {Theory of resonant multiphonon {Raman} scattering in graphene},\ }\href {https://doi.org/10.1103/PhysRevB.78.125418} {\bibfield  {journal} {\bibinfo  {journal} {Physical Review B}\ }\textbf {\bibinfo {volume} {78}},\ \bibinfo {pages} {125418} (\bibinfo {year} {2008})}\BibitemShut {NoStop}%
\bibitem [{\citenamefont {Calandra}\ and\ \citenamefont {Mauri}(2007)}]{calandra_electron-phonon_2007}%
  \BibitemOpen
  \bibfield  {author} {\bibinfo {author} {\bibfnamefont {M.}~\bibnamefont {Calandra}}\ and\ \bibinfo {author} {\bibfnamefont {F.}~\bibnamefont {Mauri}},\ }\bibfield  {title} {\bibinfo {title} {Electron-phonon coupling and electron self-energy in electron-doped graphene: {Calculation} of angular-resolved photoemission spectra},\ }\href {https://doi.org/10.1103/PhysRevB.76.205411} {\bibfield  {journal} {\bibinfo  {journal} {Physical Review B}\ }\textbf {\bibinfo {volume} {76}},\ \bibinfo {pages} {205411} (\bibinfo {year} {2007})}\BibitemShut {NoStop}%
\end{thebibliography}
\end{document}